\documentclass[lettersize,journal]{IEEEtran}
\usepackage{cite}
\usepackage{comment}
\usepackage{amsmath,amssymb,amsfonts}
\usepackage{graphicx}
\usepackage{textcomp}
\usepackage{xcolor}
\usepackage{colortbl}
\usepackage{subcaption}
\usepackage[most]{tcolorbox}
\usepackage{booktabs,tabularx,multirow}
\usepackage{pythonhighlight}
\usepackage{hyperref}
\usepackage{lipsum} 
\usepackage{microtype} 
\usepackage{array}
\usepackage{amsthm}
\usepackage{listings}
\usepackage{color}
\usepackage{pgfplots}
\pgfplotsset{compat=1.18}
\usepackage[vlined,ruled,linesnumbered]{algorithm2e}
\usepackage{algpseudocode}
\usepackage{soul}
\usepackage{enumitem}
\usepackage{verbatim}
\usepackage{etoolbox} 
\usepackage{glossaries}
\makeglossaries
\usepackage{marginnote}
\usepackage[normalem]{ulem}

\definecolor{dkgreen}{rgb}{0,0.6,0}
\definecolor{gray}{rgb}{0.5,0.5,0.5}
\definecolor{mauve}{rgb}{0.58,0,0.82}

\theoremstyle{definition} 
\newtheorem{definition}{Definition} 
\def\algosize{\small}

\newtcolorbox[auto counter]{finding}{enhanced,
  attach boxed title to top text left={yshift=-2mm},
  fonttitle=\bfseries, title=Answer to RQ\thetcbcounter}
\newcommand{\Finding}[1]{\begin{finding}#1\end{finding}}

\SetKw{Continue}{continue}
\newif\ifannotated%
\annotatedtrue
\newcommand{\annotate}[2][0cm]{\ifannotated\marginnote{\scriptsize\bfseries\color{black}#2}[#1]\fi}
\newcommand{\addedtext}[1]{%
 \ifannotated{%
 \begingroup%
 \renewcommand*{\glstextformat}[1]{\textcolor{blue}{##1}}%
 \hypersetup{citecolor=blue,linkcolor=blue}%
 \color{blue}#1%
 \hypersetup{citecolor=black,linkcolor=black}%
 \endgroup%
 }%
 \else#1\fi%
}
\newcommand{\deletedtext}[1]{%
 \ifannotated%
 \begingroup%
 \renewcommand*{\glstextformat}[1]{\textcolor{red}{##1}}%
 \hypersetup{citecolor=red,linkcolor=red}%
 \color{red}\sout{#1}%
 \hypersetup{citecolor=black,linkcolor=black}%
 \endgroup%
 \fi%
}

\renewcommand{\addedtext}[1]{#1}
\renewcommand{\deletedtext}[1]{}
\renewcommand{\annotate}[2][0cm]{}

\newif\ifnewannotated
\newannotatedfalse 

\newcommand{\newadded}[1]{%
  \ifnewannotated
    {\color{blue}#1}
  \else
    #1
  \fi
}

\newcommand{\newdeleted}[1]{%
  \ifnewannotated
    {\color{red}\sout{#1}}
  \else
  \fi
}

\newcommand{\newannotate}[2][0cm]{%
  \ifnewannotated
    \marginnote{\scriptsize\bfseries\color{black}#2}[#1]
  \fi
}
\begin{document}
\title{\annotate{C3.1}\addedtext{Efficient Black-Box Fault Localization for System-Level Test Code Using Large Language Models}}
\author{
    Ahmadreza Saboor Yaraghi\textsuperscript{*},
    Golnaz Gharachorlu\textsuperscript{*},
    Sakina Fatima,
    Lionel C. Briand, \IEEEmembership{Fellow, IEEE},
    Ruiyuan Wan,
    and Ruifeng Gao
\IEEEcompsocitemizethanks{
\IEEEcompsocthanksitem \textsuperscript{*}A. Saboor Yaraghi and G. Gharachorlu contributed equally to this work.
\IEEEcompsocthanksitem A. Saboor Yaraghi, G. Gharachorlu, and L. C. Briand are with the School of Electrical Engineering and Computer Science, University of Ottawa, Ottawa, Ontario, Canada, K1N 5N6. L. C. Briand also holds an appointment with the Lero, the Research Ireland Centre for Software, University of Limerick, Ireland.
\protect\\
E-mail: \{a.saboor, ggharach, lbriand\}@uottawa.ca
\IEEEcompsocthanksitem S. Fatima is currently with Fujitsu Research, Canada. She contributed to this work while affiliated with the School of Electrical Engineering and Computer Science at the University of Ottawa, Ottawa, Ontario, Canada, K1N 5N6.
\protect\\
E-mail: sakina.fatima@fujitsu.com
\IEEEcompsocthanksitem R. Wan and R. Gao are with Huawei Technologies Co., Ltd. \protect\\
E-mail: \{wanruiyuan, gaoruifeng1\}@huawei.com}
}
\maketitle

\begin{abstract}
Fault localization (FL) is a critical step in debugging, which typically relies on repeated executions to pinpoint faulty code regions. However, repeated executions can be impractical in the presence of non-deterministic failures or high execution costs. While recent efforts have leveraged Large Language Models (LLMs) to aid execution-free FL, these have primarily focused on identifying faults in the system-under-test (SUT) rather than in the often complex system\addedtext{-level} test code\annotate{C3.3}\addedtext{, i.e., the test script for system-level testing}. However, the latter is also important, as in practice, many failures are triggered by faulty test code. To overcome these challenges, we introduce a fully static, LLM-driven approach for system-level test code fault localization (TCFL) that does not require executing the test case. Our method uses a single failure execution log to estimate the test's execution trace through three novel algorithms that identify only code statements likely involved in the failure. This pruned trace, combined with the error message, is used to prompt the LLM to rank potential faulty locations. Our black-box, system-level approach requires no access to the SUT source code and is applicable to \annotate{C3.2}\deletedtext{large}\addedtext{complex} test scripts that assess full system behavior. We evaluate our technique at the function, block, and line levels using an industrial dataset of faulty Python test cases that were not used in pre-training LLMs. Results show that our best-estimated traces closely match the actual traces, with an F1 score of around 90\%.  Additionally, pruning the complex system-level test code reduces the LLM's inference time by up to 34\% without any loss in FL performance. Our results further suggest that block-level TCFL offers a practical balance, narrowing the search space while preserving useful context, achieving \annotate{C1.2 \& C2.3}an 81\% hit rate at top-3 (Hit@3). 
\addedtext{Finally, our TCFL method achieves equal or higher FL accuracy, requiring over 85\% less average inference time per test case and 93\% fewer tokens than the latest LLM-guided FL method.}

\end{abstract}

\begin{IEEEkeywords}
Fault Localization, Large Language Models, Execution Trace Estimation, Software Testing and Debugging.
\end{IEEEkeywords}

\section{Introduction}\label{Introduction}
Fault localization (FL) plays a crucial role in software testing and debugging, serving as one of the first steps in addressing an observed failure\cite{DBLP:journals/tse/ZouLXEZ21, wong2016survey, DBLP:conf/icse/PearsonCJFAEPK17}. Given a program that fails to
meet an expected behavior, FL aims to identify and pinpoint the underlying causes of the failure, referred to as \emph{faults}.
Manual FL is a tedious and cumbersome task.
As a result, there has been a continuous interest in developing
effective and efficient automated FL techniques within both the research and industrial practice.

Recently, like
other software engineering tasks, automated FL has benefited from the rise of Large Language Models (LLMs), which help
automate the identification of problematic sections in the code\cite{DBLP:journals/corr/abs-2403-16362, DBLP:conf/icse/YangGMH24, 10.1109/TSE.2025.3553363}.
Although they have shown promising results, such FL efforts primarily target the system-under-test (SUT), with less emphasis on identifying issues in the system\addedtext{-level} test code itself\annotate{C3.3}\addedtext{, i.e., the test script for system-level testing}. \addedtext{This can lead to unaddressed faults in the test code, resulting in misleading system quality assessments and wasted debugging effort.} \annotate{C1.1}\addedtext{In such cases, when a test fails, the debugging process may target the SUT under the assumption that it is faulty, even though the SUT is functioning correctly and the invalid behavior actually stems from the test code itself\cite{vahabzadeh2015empirical}. \autoref{fig::motiv} depicts two faulty \texttt{assertEqual()} statements and their fixed variants for testing the \texttt{hypot()} function from the \texttt{math} module in CPython,\footnote{\url{https://github.com/python/cpython}} which computes the Euclidean norm of a vector. These assertions fail on certain distributions and architectures when comparing two floating points that are not exactly equal in all decimal places.\footnote{\url{https://github.com/python/cpython/issues/124040}} Failing to recognize such nuanced faults in the test code may lead to unnecessary investigation of the larger SUT, which can be time-consuming. For example, the \texttt{hypot} function spans roughly 40 lines of code with multiple branches and loops. In this case, the issue was resolved by using test arguments that match exactly in all decimal places, regardless of the testing environment.
 
This real-world example highlights the importance of localizing faults within the test code. The issue becomes even more significant in industrial practice, particularly during system-level testing, where test code is typically more complex
and the cause of many apparent failures.} \annotate{Part of C2.1 \& C3.2}\addedtext{This complexity arises from the large number of interactions between the SUT and the system-level test code, which is a common characteristic of these types of tests\cite{DBLP:conf/iolts/AppelloC0PBR21}.
}

\begin{figure}[t]
\centering
\begin{minipage}{\columnwidth}
\definecolor{diffadd}{RGB}{220,255,220} 
\definecolor{diffdel}{RGB}{255,220,220} 
\definecolor{lightgray}{RGB}{245,245,245}

\lstdefinestyle{diff}{
    basicstyle=\color{blue}\ttfamily\scriptsize,
     numbers=left,                   
     numberstyle=\tiny\color{gray},    
     stepnumber=1,                    
     numbersep=5pt,                   
     backgroundcolor=\color{lightgray}, 
     showspaces=false,               
     showstringspaces=false,         
     showtabs=false,                  
     tabsize=4,                       
     captionpos=b,                    
     breaklines=true,                 
     breakatwhitespace=true,
     escapeinside={(*@}{@*)}
}
\annotate{C1.1}
\begin{lstlisting}[style=diff]
(*@\sethlcolor{diffdel}\hl{- self.assertEqual(hypot(1, -1), math.sqrt(2))}@*)
(*@\sethlcolor{diffdel}\hl{- self.assertEqual(hypot(1, FloatLike(-1.)), math.sqrt(2))}@*)
(*@\sethlcolor{diffadd}\hl{+ self.assertEqual(hypot(0.75, -1), 1.25)}@*) 
(*@\sethlcolor{diffadd}\hl{+ self.assertEqual(hypot(-1, 0.75), 1.25)}@*)
(*@\sethlcolor{diffadd}\hl{+ self.assertEqual(hypot(0.75, FloatLike(-1.)), 1.25)}@*) 
(*@\sethlcolor{diffadd}\hl{+ self.assertEqual(hypot(FloatLike(-1.), 0.75), 1.25)}@*)
\end{lstlisting}
\caption{\protect\addedtext{Two faulty test statements in the CPython project, causing assertion failure at commit \protect\href{https://github.com/python/cpython/blob/7628f67d55cb65bad9c9266e0457e468cd7e3775/Lib/test/test_math.py/\#L815}{7628f67}, along with their fixed variants.}}
\label{fig::motiv}
\end{minipage}
\end{figure}

In addition to using LLMs for fault localization in the SUT, several studies have explored applying LLMs to repair faulty test code, including flaky tests\cite{DBLP:journals/tse/FatimaHB24, ke2025niodebugger} and broken tests within test suites\cite{yaraghi2025automated, rahman2025utfix}. \annotate{Part of C1.2}\addedtext{However, these techniques do not perform fault localization and focus solely on the repair phase, either by taking already identified faulty locations as input or by initiating the repair process on the entire test code without localization. We believe this latter design choice is influenced by the relatively small size of the test cases they target, which are mostly simple unit tests rather than larger system-level test scripts, making explicit localization less necessary.}

Therefore, despite the investigation of LLM-based solutions for system-under-test fault localization (SUTFL) and test code repair, solutions for \emph{system-level test code fault localization (TCFL)} have remained largely unexplored.
Additionally, automated FL can still be challenging, particularly because it typically requires the repeated \emph{execution} of test cases to compare passing and failing runs to pinpoint suspicious fault locations, as seen in well-known techniques such as spectrum-based fault localization (SBFL)\cite{DBLP:journals/ieicetd/ZhengHCYFX24, DBLP:journals/jss/RaselimoF24, DBLP:journals/access/SarhanB22, de2016spectrum, zakari2020spectrum}. \newannotate{C2.2}\newadded{Furthermore, since the ultimate goal of fault localization is fault repair, these multiple execution iterations accumulate, substantially increasing the overall repair cost.}As a result, these automated FL techniques may become impractical or even infeasible in situations where collecting multiple execution traces is hindered by non-deterministic, irreproducible failures\cite{DBLP:conf/issta/ChoiZ02, DBLP:conf/icst/Feldt14} or high execution costs\cite{DBLP:journals/spe/Larus90, DBLP:journals/tse/MarchettoSS19, DBLP:conf/wosp/HorkyKLT16}.

To investigate the potential of LLMs in TCFL while avoiding execution-related limitations, this paper proposes a fully static, LLM-driven TCFL technique that localizes faults in system-level test code using a single failure execution log generated after the test failure. \newannotate{C2.2}\newadded{The goal of this research is twofold: (1) to address the lack of automated techniques for system-level test code fault localization, and (2) to improve the efficiency of this FL process by reducing token overhead and LLM's inference time without sacrificing accuracy, thereby making the approach more practical for CI pipelines, where faults are frequently localized and repaired over multiple iterations. To this end,}we propose three novel algorithms that can \emph{estimate} the execution trace of a faulty test case with sufficient accuracy while pruning irrelevant information, enabling effective and efficient FL without requiring the test case to be executed. Using the estimated trace and other information, including the error message, we prompt the LLM to perform ranking-based FL, which outputs several faulty locations ordered by their likelihood of being faulty. Our TCFL technique is \emph{black-box}, meaning it does not require access to the SUT's source code and operates at the \emph{system level} by performing FL on complex, faulty test scripts spanning multiple layers and designed to assess the entire system's functionality. To the best of our knowledge, this work is the first to investigate fault localization in test code.

In summary, we make the following contributions:
\begin{enumerate}
\newannotate{C2.2}
    \newadded{\item We characterize the problem of test code fault localization (TCFL) as a distinct research problem and propose an automated solution leveraging large language models (LLMs).}
    \newadded{\item We enhance the efficiency of our TCFL technique to facilitate its integration into fault localization and repair pipelines. To achieve this, we:}
    \begin{enumerate}
    \item \newadded{propose novel algorithms to estimate a faulty test case's execution trace by mapping its code statements to corresponding failure log messages. The goal is to identify only those code statements that are potentially relevant to the failure---those likely executed at the time---without executing the test itself. We further propose a \emph{masking strategy} inspired by the token masking approach used in LLMs}\cite{DBLP:conf/naacl/DevlinCLT19} \newadded{to statically evaluate the accuracy of our estimated traces.}
    \item \newadded{use the pruned set of test code statements identified as relevant to the failure to prompt the LLM, reducing the search space compared to using the entire test code, thus enabling more efficient FL.}
    \end{enumerate}
\item We evaluate our TCFL technique at different levels of test code granularity, i.e., function, block, and line levels, using an industrial dataset of faulty system-level test cases.\annotate{C1.2 \& C2.3}\addedtext{\item We evaluate our technique against three state-of-the-art methods, including the most recent LLM-guided fault localization approach.}
\annotate{C1.3}\addedtext{\item We make our tool available online, including the implementations of all trace estimation algorithms and the LLM-based fault localization component.\footnote{
\underline{\url{https://github.com/Ahmadreza-SY/TCFL}}
}}
\end{enumerate}

Our results indicate that our estimated traces closely approximate actual traces, with the best prediction F1 score reaching around 90\%, and support effective and efficient FL across different levels of granularity. Specifically, our TCFL approach achieves the highest precision of 82.1\% when prompting at the coarser function level and requesting a single faulty function, and the highest recall of 95.7\% when requesting three functions. We also find that using the estimated trace with maximal pruning enables the most efficient FL, reducing inference time by up to 34\% while maintaining the same performance as using the full test code. Our results suggest that block-level TCFL offers a more practical trade-off, narrowing the search space more than function-level TCFL while providing more context than individual lines. \annotate{C1.2 \& C2.3}\addedtext{Finally, compared to state-of-the-art techniques, our TCFL method achieves equal or higher FL accuracy while being significantly more efficient and scalable, requiring over 85\% less average inference time per test case and 93\% fewer tokens than the latest LLM-guided FL method.}

\section{Problem Definition and Challenges}\label{Problem_Def}
This section defines the system-level test code fault localization (TCFL) problem at a high level, highlights its differences with system-under-test fault localization (SUTFL), and explores the challenges from different perspectives.

\subsection{\newadded{System-Level Test Code Fault Localization: Definition and Example}}\label{tcfl::example}

Fault localization (FL)\cite{DBLP:journals/tse/ZouLXEZ21}, in general, refers to the task of identifying the root cause of an observed failure.
In SUTFL~\cite{DBLP:journals/ieicetd/ZhengHCYFX24, DBLP:journals/jss/RaselimoF24, DBLP:journals/access/SarhanB22,de2016spectrum,zakari2020spectrum, DBLP:conf/icse/YangGMH24, DBLP:journals/corr/abs-2403-16362}, while the root cause is in the SUT, failures are either identified within the SUT or occur in the test code. In contrast, TCFL involves identifying the root cause of a failure observed during SUT testing, with both the failure and its cause residing within the test code itself. 

\begin{figure}[tbp]
\centering
\includegraphics[width=\columnwidth, keepaspectratio, trim=25mm 135mm 40mm 58mm, clip]{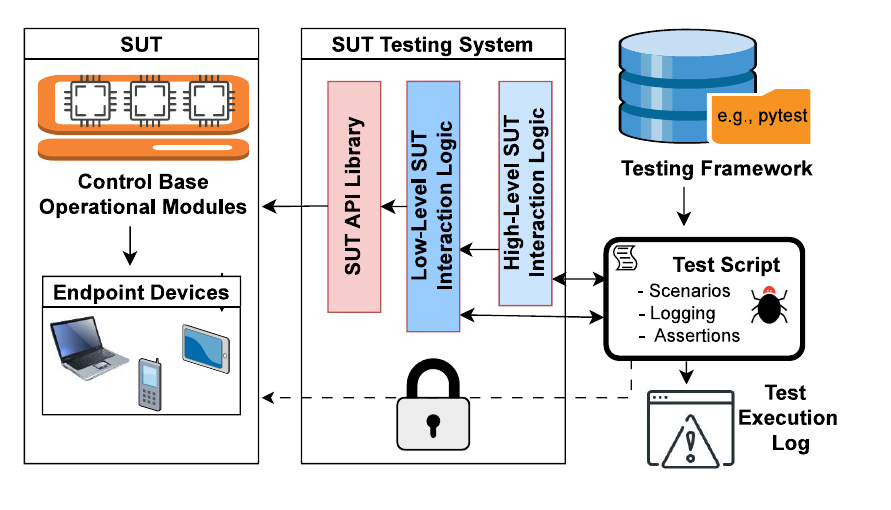}
    \caption{An illustration of black-box system-level testing, where a fault in the test script leads to a failure observed in the test execution log. The test script interacts with the SUT without having direct access to its source code.
}
    \label{fig:problem_def_overview}
\end{figure}

Consider the simplified example in \autoref{fig:problem_def_overview}, illustrating a testing environment for a video conferencing setup. Here, the SUT consists of a control base, several operational modules, and various endpoint devices that serve as terminal nodes for receiving conference signals. As shown, multiple components interact to assess the system's functionality by evaluating requirements, such as signal quality. Since the test script does not have direct access to the SUT's source code, this assessment is performed through several layers that connect the SUT and the testing environment. These layers include the SUT API library as well as both low-level and high-level interaction logic between the SUT and the testing environment. While these layers facilitate interaction between the SUT and the test script, they also introduce additional complexity into the test script\cite{DBLP:conf/iolts/AppelloC0PBR21,DBLP:journals/tosem/TangZZZGLGLMXL23}. This increased complexity can result in faulty test scripts that produce failed execution logs.

Examples of faults in the test code of the system shown in \autoref{fig:problem_def_overview} include incorrectly setting the parameters of an endpoint device in the test script or failing to specify the correct minimum threshold for signal quality. \newdeleted{Below, we discuss several challenges encountered when attempting to localize such faults using a TCFL technique.}

\subsection{\newadded{General Challenges in Fault Localization}}\label{challenges::general}
\newannotate{C2.1}\newadded{Fault localization, even when automated, remains a challenging task. The following list highlights challenges common to both SUTFL and TCFL.}

\newdeleted{\noindent\textbf{Black-Box Setting.} 
Understanding the SUT is often necessary to determine the root cause of a failure in the test code. For example, detecting the faults described above can be challenging without access to the SUT's source code, which provides critical information such as valid parameter ranges or acceptable signal quality thresholds.
However, in many contexts, such as black-box testing or within industrial Quality Assurance (QA) teams, testers typically do not have access to the production code}\newdeleted{, making TCFL significantly more difficult. The use of proprietary and third-party code}\newdeleted{further exacerbates this challenge. As a result, there is a need for a black-box TCFL approach that can effectively localize faults in system-level test code without requiring access to the SUT's source code.}

\subsubsection{Execution Costs}\label{subsub::exec::cost}
Effective and efficient FL, whether SUTFL or TCFL, often requires additional data such as multiple execution traces from both passing and failing runs\cite{DBLP:journals/ieicetd/ZhengHCYFX24, DBLP:journals/jss/RaselimoF24, DBLP:journals/access/SarhanB22,de2016spectrum,zakari2020spectrum}.
However, the complexity of large configuration files, along with the high cost of setup, execution, and performance overhead in large codebases\cite{DBLP:journals/spe/Larus90, DBLP:journals/tse/MarchettoSS19, DBLP:conf/wosp/HorkyKLT16}, makes it challenging, and at times impossible, to repeatedly run the faulty code across different environmental settings to replicate test failures or successes. \annotate{Part of C2.1 \& C3.2}\addedtext{This also applies to industrial system-level testing, where test cases can be complex despite their size due to involving different components of the system.} Additionally, the non-deterministic nature of some failures further complicates this challenge by making them non-reproducible\cite{DBLP:conf/issta/ChoiZ02, DBLP:conf/pldi/RegehrCCEEY12, DBLP:journals/tse/FatimaHB24}.

While both SUTFL and TCFL share this limitation, the challenge is further intensified in TCFL because, to optimize test cases for efficiency\cite{DBLP:journals/software/BaudryFJT05, DBLP:journals/access/GuptaSP19}, test execution traces may lack detailed data, such as the precise statement IDs collected during code instrumentation, which are crucial for thorough FL analysis\cite{DBLP:journals/ese/WhiteK22}. 
For the reasons above, obtaining precise test execution traces may not always be feasible, necessitating the development of an execution-free FL approach based on effective execution trace estimators.

\subsubsection{Scalability} 
As LLM-based FL techniques gain wider adoption, ensuring their scalability has become increasingly important. The limited context window of LLMs in general, along with the token-based pricing of commercial models, makes it preferable to optimize the input provided to the model and the output requested from it\cite{DBLP:journals/corr/abs-2402-01742}.
    
\subsubsection{Efficiency} While automated techniques can significantly accelerate the manual inspection of faulty code, they may still suffer from slow performance, particularly when processing large amounts of data for numerous faulty cases. Such situations often arise in automated test generation scenarios\cite{DBLP:journals/corr/abs-2305-04764, DBLP:journals/infsof/AlagarsamyTA24, lemieux2023codamosa} or when there is a substantial backlog of faulty test cases\cite{IMTIAZ20191} \newadded{or failing programs.}Slow performance can disrupt developer workflow and reduce efficiency\cite{DBLP:journals/sqj/ParninR11}. Moreover, it may hinder the scalability of emerging FL applications, including its use in models, neural networks, and hardware specifications\cite{DBLP:journals/tim/MohammedMCS21, DBLP:conf/icse/WardatLR21, DBLP:conf/iccd/Wu0YMHML22}.
    
In LLM-based FL, the model's inference time is a crucial factor in determining the overall speed of the localization process. This inference time is influenced by various factors, including the model architecture (e.g., the number of parameters), hardware resources (e.g., available GPUs), and the model's input and output sizes\cite{liu2025bag,jain2024effllm}. Consequently, a technique that reduces input size without losing essential information would be highly beneficial for improving FL efficiency.

\subsubsection{Practicality} The ultimate goal of effective automated FL techniques should be to generate outputs that are practical and useful for debugging purposes. Achieving this requires a precise problem formulation that produces results in the desired format. For example, listing the names of the faulty \emph{functions} as potential locations to investigate can yield a broad search space with redundant information, making the debugging process inefficient. On the other hand, outputting only the faulty \emph{lines} could be too limited, offering no surrounding context to consider for repairs. As a result, an effective middle-ground approach is required. 
Moreover, offering a range of candidate faulty locations in the appropriate format enables users to explore potential faulty locations and reduces the tendency to focus solely on a limited scope, thereby enhancing the practicality of the technique. For example, returning a \emph{larger} number of faulty locations increases the chance of finding the actual fault but requires more investigation, whereas a \emph{smaller} number narrows the scope but reduces the likelihood of including the actual fault. A comprehensive analysis of these variants can help tailor the FL technique to meet different requirements.
    
\subsubsection{Data Leakage} An important characteristic of an effective FL technique is its FL accuracy, which can be evaluated by thorough empirical evaluation on real datasets. Many existing FL techniques lack adequate evaluation on these datasets, undermining the validity of their results. This issue is especially concerning when adopting LLMs and evaluating them on open-source datasets, as the model may have been previously trained on this data\cite{DBLP:journals/corr/abs-2310-16253, DBLP:conf/eacl/BalloccuSLD24}. Evaluating FL techniques on datasets that LLMs have not been exposed to before can greatly help mitigate this challenge.
\subsubsection{Multiple Faulty Locations}\newannotate{C2.1}\newadded{Many scenarios involve more than one fault, which makes localizing issues significantly more challenging than identifying a single fault. Interactions between faults can obscure their individual effects on failures, making it harder to accurately pinpoint the exact faulty locations. This complexity makes manual debugging considerably more tedious, even when the codebase is relatively small, and renders many automated techniques less effective in multiple-fault scenarios}\cite{zakari2018simultaneous, zakari2020multiple}.

\subsubsection{Large Codebases}
\newadded{Large codebases introduce significant challenges for fault localization. As code size increases, the number of possible execution paths, code elements, and interactions grows rapidly, making it harder to pinpoint the source of failures. Dependencies on external or third-party modules can further complicate the task, as faults may arise in or propagate through these modules, obscuring their origin. This growth in scale and complexity not only increases the computational effort required by FL techniques but also exacerbates the risk of misdiagnosis.}

\subsection{\newadded{Challenges Unique to Test Code Fault Localization}}\label{challenges::unique}
\newannotate{C2.1}\newadded{Here, we discuss challenges that are unique to the TCFL setting, with a focus on system-level TCFL.}
\subsubsection{Black-Box Setting} 
\newadded{Understanding the SUT is often necessary to determine the root cause of a failure in the test code. For example, detecting the faults described in \autoref {fig:problem_def_overview} in \autoref{tcfl::example} can be challenging without access to the SUT's source code, which provides critical information, such as valid parameter ranges and acceptable signal quality thresholds.
However, in many contexts, such as black-box testing or within industrial Quality Assurance (QA) teams, testers typically do not have access to the production code}\cite{nidhra2012black, sandhu1998role}\newadded{, making TCFL significantly more difficult. The use of proprietary and third-party code}\cite{katyal2018paradox, raemaekers2011exploring, DBLP:conf/icsm/Wang0HSX0WL20} \newadded{further exacerbates this challenge. As a result, there is a need for a black-box TCFL approach that can effectively localize faults in system-level test code without requiring access to the SUT's source code.}

\subsubsection{Absence of Test Cases for Test Code}
\newadded{SUTFL techniques such as spectrum-based fault localization (SBFL)}\cite{DBLP:journals/ieicetd/ZhengHCYFX24, DBLP:journals/jss/RaselimoF24, DBLP:journals/access/SarhanB22,de2016spectrum,zakari2020spectrum} \newadded{can exploit rich information from multiple test cases. By executing these tests and monitoring which parts of the program under test are exercised, each code element can be assigned a suspiciousness score based on its presence in passing and failing runs. In contrast, TCFL often lacks test cases for the test code itself. As a result, such dynamic information is unavailable, making it much more difficult to estimate which lines or elements of the test code are likely to be faulty. This absence of test cases introduces a unique challenge for TCFL that is not present in standard SUTFL scenarios.}

\subsubsection{Limited Execution Detail}\label{subsub::limited::detail}
\newadded{As further discussed in \autoref{subsub::exec::cost}, the execution information of the faulty code itself can provide valuable insights for fault localization, such as which statements were executed, the sequence of method or function calls, and the interactions between different components of the code\cite{DBLP:journals/ese/WhiteK22}.}\newadded{However, such fine-grained information is often unavailable in the TCFL setting because the testing process is optimized to use lightweight tests, rather than tests instrumented to collect precise execution data\cite{DBLP:journals/software/BaudryFJT05, DBLP:journals/access/GuptaSP19}. This introduces a challenge that further complicates TCFL.}

\subsubsection{Frequent SUT Changes}
\newadded{Changes in the SUT can significantly complicate TCFL because even small modifications in system behavior, interfaces, or configurations may cause previously valid tests to fail. As a result, test failures may frequently originate not from faults in the SUT, but from outdated or incorrect logic in the test code itself. The problem is further exacerbated by the black-box nature of system-level testing\cite{nidhra2012black, sandhu1998role}.}

\subsubsection{Extensive SUT Interactions}
\newadded{Unlike unit tests that target isolated components, system-level tests typically interact with multiple subsystems and rely on complex sequences of operations, timing assumptions, and environment configurations (see \autoref{tcfl::example} for an example). These interactions involve invoking multiple system components, which can make the test script complex even with relatively few lines of code. We discuss this complexity in more detail in the following subsection.}

\begin{figure*}
\centering
\resizebox{!}{0.5\textheight}{%
  \begin{minipage}{\textwidth}
    \input{Images/problem_def_motivation_code}
    \input{Images/problem_def_motivation_log}
     \end{minipage}
}
\end{figure*}

\subsection{\newadded{Importance of Automation in Localizing System-Level Test Code Faults}}\label{sub::problem::importance}
\newannotate{C2.1}\newadded{To illustrate the challenges of black-box system-level fault localization and the need for an automated technique, a system-level test script and its execution log are shown in \autoref{lst:faulty:system::test} and \autoref{lst:faulty:system::test:log}, respectively. This example is inspired by our industrial test script data but has been altered for privacy and simplified for brevity. The test script, intended to evaluate the signal quality of endpoint devices in a video conference, fails due to an assertion error on line 32. In real-world scenarios, when such failures occur, the FL search space consists of three components: (1) the potentially large SUT codebase, which is absent in black-box settings; (2) the potentially large test codebase, including testing-related modules and dependencies; and (3) the failing test script itself. The earlier the fault in the test script is identified, the less time and effort are spent investigating the larger SUT and test codebase. In this example, method-level FL offers little benefit because \texttt{test\_signal\_interactions} is the main test method that spans most of the script. Simply identifying this method as faulty provides no more insight than localizing the fault at the file level or not localizing it at all. Therefore, a finer-grained analysis at the code-block or line level is necessary.}

\newadded{To localize the faults causing this failure, a human developer begins by examining the execution log. However, aside from a few \texttt{DEBUG} and \texttt{INFO} messages and the assertion error itself, the log provides little useful information due to its \emph{limited execution details}. Looking closely at line 32, the developer focuses on the \texttt{quality} variable and the valid signal range \texttt{[0, 100]}. The failure clearly occurs because the \texttt{quality} value falls outside this range. Two immediate possibilities could explain this: 1) the \texttt{quality} value computed at line 31 by \texttt{get\_signal\_quality()} is incorrect due to a fault in the function's implementation (i.e., a fault in the SUT), or 2) the valid range itself has changed from \texttt{[0, 100]} because of a modification in the SUT configuration. In either case, further reasoning is hindered by the \emph{black-box setting}, which prevents access to the SUT source code. Continuing the analysis, the developer encounters an \emph{extensive set of interactions} between the SUT and the test script, each of which could directly or indirectly affect the \texttt{quality} value. These interactions exist to cover various system-level testing scenarios, such as dynamic module switching, enabling or disabling the video stream, network degradation, or an endpoint device joining late or leaving the conference unexpectedly. Even the sequence of these interactions can influence the signal quality of the endpoint under test. However, in the black-box setting, the developer cannot determine whether any of these interactions contribute to the failure. In addition to SUT interactions, the test script also contains callsites that invoke functions defined not in the SUT or test script itself but in the test codebase, e.g., \texttt{enable\_network\_simulation} on line 11. This potentially \emph{large codebase} contains dependencies, such as utility functions and modules available to the testing team. In practice, the developer cannot know whether the fault lies in the test script or in the codebase, which significantly increases the complexity of TCFL.}

\newadded{A possible approach would be to compare passing and failing runs across different executions, assigning high suspiciousness scores to code lines that appear more frequently in failing runs, as is done in SBFL techniques. However, this is not feasible in TCFL, since there is only a single failing execution of the test script and no additional passing or failing runs due to the \emph{lack of test cases} for the test code itself. Many developers may halt here, assuming the fault lies within the SUT. This can lead to misleading assessments of system quality and wasted debugging effort.}

\newadded{In contrast, LLMs can automatically and effectively localize the faults causing this failure by leveraging their understanding of code semantics and analyzing patterns in the test code, identifying lines 22 and 25 as inconsistent with preceding lines and marking them as suspicious. Line 22, which verifies whether the type of the endpoint device under test is included in a specified list, is similar to line 14 with one subtle difference: the type \texttt{wireless} is present in the list on line 14 but missing from the list on line 22. The length of the two lists, combined with the differing item orders, makes the missing type harder for a human to notice. Note that in real-world test scripts, these lists are generally longer, with element names more difficult to remember, such as identifiers in hexadecimal form. Furthermore, with the increasing prevalence of LLM-generated code---including LLM-generated test code\cite{DBLP:journals/corr/abs-2305-04764, DBLP:journals/infsof/AlagarsamyTA24, lemieux2023codamosa}---the repetition of code elements has become common~\cite{dong2025rethinking, liu2025code}, which can make manual FL more challenging even when the script contains only a few hundred lines.}

\newadded{In addition to line 22, which is marked as suspicious due to its difference from line 14, the LLM also flags line 25 as suspicious because it differs from lines 20 and 23 by assigning the endpoint to the wrong module through the use of an incorrect index (i.e., \((i + 1) \%
\text{len}(\text{self.modules})\) instead of \(i \%
\text{len}(\text{self.modules})\)). Note that the execution log messages may not be helpful and can even be misleading in this case, as the endpoint-to-module assignment information is printed earlier in the code (line 21) before the lines that actually perform the assignment (lines 23 and 25). Additionally, localizing \emph{multiple faults}, which are common in practice\cite{zakari2020multiple}, is more cumbersome and time-consuming because interactions between faults hinder debugging\cite{6681344}. As a result, deeper code pattern analysis, such as that enabled by automated LLM reasoning, is required.}

\newadded{Once these two lines are identified as suspicious, the cause of the failure becomes evident. An endpoint of type \texttt{wireless} executes the \texttt{else} branch because its type is missing from the condition of the \texttt{if} statement (line 23), resulting in an incorrect module being assigned to it (line 25), which subsequently leads to an incorrect signal quality range.}

\noindent\textbf{\emph{\newadded{How Execution Traces Help in This Example.}}} \newadded{Knowing the execution trace would help narrow the FL search space under the practical assumption that faults causing a failure occur on executed lines\cite{DBLP:conf/ACMse/FrancelR92, DBLP:conf/issre/AgrawalHLW95, DBLP:conf/icse/JonesHS02, DBLP:journals/tse/RuthruffBR06, DBLP:conf/kbse/RenierisR03, Pan-Heuristics}. This reduced search space can enable more effective, or at least more efficient, fault localization by decreasing the number of lines of code that need to be investigated. In this example, knowing that the \texttt{if} branch is not taken would help focus attention immediately on the \texttt{False} condition and the statements in the \texttt{else} branch. In general, focusing on executed lines helps exclude a significant number of function invocations---whether defined within the script or in the test codebase---as well as many SUT interactions from the search space. Since obtaining actual execution traces is costly and may even be infeasible for non-deterministic failures, as described in \autoref{subsub::exec::cost}, an automated trace estimation technique becomes essential.}

\subsection{\newadded{Prevalence of System-Level Test Code Faults}}\label{sub::problem::prevalence}
\newannotate{C2.1}\newadded{Faults in system-level test scripts are a widespread and common issue. Jordan et al.\cite{10.1145/3510457.3513069} report that 27\% to 53\% of test failures in five automotive electrical control units responsible for system-level customer functions are caused by errors in test implementation or incorrect test specifications. Herzig and Nagappan\cite{7202948} classify a substantial number of faults in Windows 8.1 and Microsoft Dynamics AX as originating from integration and system-level test scripts. They refer to these as false test alarms, since they do not indicate faults in the SUT but instead reveal faults in the test code itself. Jiang et al.\cite{7985707} report that 27\% to 46\% of test failures in two Huawei system-level test datasets are attributed to defects in the test scripts or to obsolete tests. While these studies highlight the prevalence of such faults, they do not attempt to localize them. Instead, their focus is on classifying test execution failures by likely cause, such as faults in the SUT or in the test code.}

\subsection{\newadded{Measuring Size and Complexity of System-Level Test Code}}\label{sub::problem::complexity}

\newannotate{C2.4}\newadded{A smaller number of lines in system-level test code does not necessarily imply lower complexity. As illustrated in \autoref{sub::problem::importance}, the complexity of system-level tests primarily arises from the extensive interactions of test scripts with both the SUT and the test codebase. Consequently, some system-level test scripts actively used for testing and verifying large industrial products may consist of only a few hundred lines of code\cite{7985707}, which is fewer than the number of lines in other test cases, such as fuzzer-generated tests, that are designed to reveal faults in the SUT\cite{10.1145/1993498.1993532}. Analyzing hundreds of lines of code---primarily consisting of calls to the SUT and test codebase components, each potentially affecting the expected behavior---is considerably more challenging for both humans and LLMs than analyzing larger test code, such as fuzzer-generated code, which may consist of thousands of lines but is more likely to contain duplicates due to randomness. To better capture the complexity of system-level test scripts, the literature often uses two alternative metrics: (1) the number of test steps, defined as the count of high-level interactions with the SUT, such as commands or code snippets that display or verify specific SUT behaviors\cite{7985707, 9978993} (e.g., \texttt{endpoint.join\_conference()} in \autoref{lst:faulty:system::test}), and (2) the size of the test execution log files in bytes\cite{7985707}.}

\subsection{\newadded{Problem Formulation}}\label{sub::formulation}

\newannotate{C2.2}\newadded{Given the prevalence of faults in system-level test code (see \autoref{sub::problem::prevalence}) and the importance of automated techniques for localizing them (see \autoref{sub::problem::importance}), we propose the first automated LLM-based technique for test code fault localization. Since the complexity and nuances of system-level test code---making fault localization subtle---are not solely determined by the size of these scripts (see \autoref{sub::problem::complexity}), we evaluate our technique on a set of real-world system-level test scripts actively used by our industrial partner to verify their SUTs.}

\newadded{Since the ultimate goal of fault localization is fault repair, which typically occurs over multiple iterative rounds, we incorporate an additional measure to improve the efficiency of our localization technique and make it more practical in real-world settings, including CI pipelines. To this end, we reduce the FL search space by excluding test code lines that are not executed and are therefore unlikely to be faulty. To determine whether a line of code has been executed, we rely on the single available execution trace of the failing test. Because reobtaining this trace is often infeasible---due to cost or non-determinism, as discussed in \autoref{subsub::exec::cost}---and because the trace itself lacks complete execution details (see \autoref{subsub::limited::detail}), we leverage existing logging statements in the test code together with the log messages recorded in the log file to estimate the execution trace as accurately as possible.}

\newadded{This research pursues two main goals: (1) addressing the lack of automated techniques for system-level test code fault localization, and (2) making the localization process practical by reducing token overhead and inference time while preserving accuracy. To work toward these goals, we aim to address the challenges outlined in \autoref{challenges::unique} and \autoref{challenges::general}.

We believe our technique can serve as a foundational baseline for future research in LLM-based TCFL, paving the way for further exploration in this area. It can also help minimize the developer's debugging time and effort by accurately localizing faults in system-level test scripts, thereby reducing the need to examine the potentially extensive SUT or test codebase.

The next section presents the background material.}

\newdeleted{This research aims to develop and evaluate an effective and efficient system-level black-box TCFL technique that leverages LLMs to address the challenges described above. To achieve this, we first minimize the LLM's input by using only the faulty test code and a single failed execution log. This minimization phase is part of our black-box approach, which does not access the SUT. The input reduction is performed to improve the scalability and efficiency of LLM-guided FL. Using this reduced input, we guide LLMs to produce a range of practical outputs for debugging.}

\newdeleted{While the fault lies in the test code, and the lack of SUT access does not prevent us from locating it, understanding the SUT could help interpret faulty test behavior for better localization. Furthermore, because the faulty test code interacts extensively with SUT components in system-level testing, access to those components could improve the precision of reducing the LLM's input using the execution trace estimation algorithms in}\newdeleted{ as it would provide additional information.}

\newdeleted{Given the lack of access to the SUT and the benefits such access could provide, our setting further highlights the importance of a fully black-box technique like ours.
To the best of our knowledge, this work is the first to investigate fault localization in test code and leverages an industrial context, providing a reliable environment for evaluation.}

\section{Background}\label{Background}
In this section, we provide the essential definitions and illustrations that form the foundation for the rest of the paper. 

Consider the running example in \autoref{lst:T0}, simplified for illustration purposes, which is a faulty test case written in Python, along with its corresponding execution log in \autoref{lst:running_example_log}. In this case, initializing the set \texttt{values} on line 3 causes a division-by-zero error on line 17. Here, line 3 contains the fault (root cause), which leads to a failure (error) on line 17. To identify the lines responsible for the failure in this example, an effective FL technique would take the faulty test case as input and output line 3 as the faulty line.
FL is not limited to localizing faults at the line level. It can also be performed at other levels, using \emph{elements} such as files, functions, and code blocks\cite{cfgbook} as distinct units. In this context, a basic block of code is described as follows.

\begin{definition}[\emph{\textbf{Basic Block of Code}}]\label{def:block}
A basic block of code, or simply a block, is a sequence of consecutive lines of code that has a single entry point, a single exit point, and no branches or jumps in between. Formally, a block is defined as a sequence of code lines \(\{l_1, l_2, \dots, l_n\}\), where \(l_1\) and \(l_n\) represent the entry and exit points, respectively. 
Furthermore, assuming the function \emph{Next} takes \(l_i\) as input and returns subsequent executable lines after \(l_i\), for all lines within a block, we have the following:
\[
Next(l_i) = l_{i+1}, \,\,\,\forall i,1 \leq i \leq n-1
\]
\end{definition}

In the example shown in \autoref{lst:T0}, the code blocks are indicated by their IDs. For instance, the first basic block (\(B1\)) begins at line 1 and ends at line 3, while the entry and exit points of the third block (\(B3\)) are at lines 8 and 12, respectively. Regardless of the level of detail at which the faulty code is analyzed, referred to as \emph{granularity}, the following formal definition provides a rigorous description of ranking-based FL.

\begin{definition}[\emph{\textbf{Ranking-Based Fault Localization}}]\label{def:fault_localization}
Given a faulty test case \( \tau_f \) and the set of its input elements \( E_g \) at a granularity level \( g \), a ranking-based fault localization technique, \( \text{RBFL} \), returns a list of the \( n \) most suspicious elements, ranked in descending order of suspiciousness, based on the probability function \( P: E_g \rightarrow [0, 1] \), as follows:

\[
\text{RBFL}(\tau_f) = \{ e_1, e_2, \dots, e_n \}
\]

where \( e_i \in E_g \) for \( i = 1, 2, \dots, n \), and \( P(e_n) \leq P(e_{n-1}) \leq \dots \leq P(e_2) \leq P(e_1) \).
\end{definition}

In our example, where line 3 is identified as faulty, the granularity level \( g \) is defined at the line level, and \( E_g \) is a set whose elements are the lines numbered from 1 to 24.

Upon further examination of this example, we observe that not every line of code is executed to trigger the failure. For example, when the \texttt{if condition} evaluates to true and the true branch is taken (line 13), the \texttt{else} branch (lines 14--15) will not be executed. Consequently, we define the execution trace of \(\tau_f\) as a subset of elements in \(E_g\) as follows.

\begin{definition}[\emph{\textbf{Execution Trace}}]\label{def:exe_state}
Given a faulty test \( \tau_f \) and the set of its input elements \( E_g \), the execution trace of \(\tau_f\) is defined as $X_{\tau_f} \subseteq E_g$, where $X_{\tau_f} = \{e \in E_g \mid e \text{ is executed}\}$

\end{definition}

The execution trace is a set of elements that were executed, without regard to their execution \emph{order}.

Intuitively, the executed elements of the faulty test case are the most likely locations where the fault or root cause of the failure resides. This idea underlies 
many FL techniques in the literature\cite{DBLP:conf/ACMse/FrancelR92, DBLP:conf/issre/AgrawalHLW95, DBLP:conf/icse/JonesHS02, DBLP:journals/tse/RuthruffBR06, DBLP:conf/kbse/RenierisR03, Pan-Heuristics}. Consequently, by focusing on the execution trace, the \emph{search space} of a FL technique will be reduced, as there would be fewer elements to select from. In our example, line 13 is part of \(X_{\tau_f}\) and is thus included in the FL search space, whereas lines 14 and 15 are not.

If the actual execution trace of the test case is unavailable, for instance when obtaining this information is costly or infeasible as discussed in \autoref{Problem_Def}, the execution trace needs to be \emph{estimated}\cite{DBLP:conf/msr/SchipperAD19, DBLP:conf/racs/BushongSCDCFBTS20, DBLP:conf/ism2/Jurgensen20}.
Various sources of information, such as execution logs and control flow graphs (CFGs)\cite{cfgbook} derived from test code functions, can be helpful in this estimation process.
For each function in the faulty test case, we can define a unique CFG as follows.

\begin{figure}[tbp]
    \centering
    \scriptsize
    \begin{minipage}{\columnwidth}
        \centering
        \input{Images/running_example_T0}
        \input{Images/running_example_log}
    \end{minipage}
    \caption{A faulty test code and its corresponding execution log.}
    \label{fig:running}
\end{figure}

\begin{definition}[\emph{\textbf{Per-Function Control Flow Graph}}]\label{def:cfg}
Given a faulty test \( \tau_f \), \( CFG_F \) is the directed graph representation for function \( F \) in \( \tau_f \), where nodes correspond to basic blocks of code in \( F \), and edges show the execution flow between them. 
\end{definition}

For example, after constructing the CFG for the function \texttt{execute\_tests()} and utilizing the execution log from \autoref{lst:running_example_log}, we can eliminate the basic block spanning lines 14 and 15 from the execution trace. Notably, this conclusion is drawn from the log message \texttt{Test 1 executed!} in line 4 of the execution log, which corresponds to a log statement in line 2 of \autoref{lst:T0}. This implies that function \texttt{test\_1}, and therefore lines 1--3 and 13, were executed.

Note that the execution log in this example does not provide a precise mapping between each log message and the corresponding line in the faulty code. Therefore, this scenario falls under cases where the actual execution trace is unavailable, necessitating an estimation of the execution. In this simplified example, because there is only one log statement with the message \texttt{Test 1 executed!}, the mapping process is straightforward, and the pruning of the \texttt{else} branch is clear. In real-world scenarios with more complex test cases, this estimation process may not be as straightforward. We will address this further in \autoref{Approach}.

The execution estimate above can be further refined by accounting for call sites. A call site is a location (or line) in the code where a function or method is invoked. In \autoref{lst:T0}, there are three call sites on lines 13, 15, and 23. Removing lines 14 and 15 eliminates the call site for \texttt{test\_2}, which in turn enables the removal of its definition (lines 5 and 6), since a function that is never called is not executed. This example demonstrates further pruning of the execution trace by removing function definitions when their call sites no longer exist in the execution trace. 

The following section describes how we leverage the information sources presented above to develop new algorithms for estimating the execution trace of a faulty test case.

\section{Approach}\label{Approach}
Our approach consists of two phases. In the first phase, based on static analysis, we attempt to determine the execution trace of the faulty test case using available software artifacts, such as the test code and the test execution log. Since this cannot be perfectly achieved without executing the test cases, we refer to this task as \emph{estimating} the execution trace. In the second phase, we use the estimated trace to prune the faulty test code and provide it, along with the error message, to the LLM using a prompt template that includes FL instructions and supports various FL granularity levels (line, block, and function). The idea is that a smaller, pruned test code that still contains the faulty locations can reduce the LLM's search space, potentially improving its ability to identify faults in the input code. Moreover, a pruned test case contains fewer tokens, which not only helps prevent exceeding the LLM's context window size but also potentially decreases the inference time required for the LLM to generate an answer\cite{DBLP:journals/corr/abs-2407-00434, DBLP:conf/mlsys/AgrawalKMPKGRT24}.

\subsection{Execution Trace Estimation}\label{approach::estimation}

This section outlines our methodology for estimating the execution trace of a faulty test case based on the test code and the test failure's execution log. Our trace estimation approach is completely static and does not execute the test case. In general, the approach includes two key steps. Initially, we utilize the execution log to match log messages with test code lines, aiming for a \emph{unique} and \emph{precise} match when feasible. This step identifies the initial set of \emph{log-based} executed and non-executed lines in the test code. Examples of such log-based lines in a typical Python script include \texttt{print()}, \texttt{logging.info()}, \texttt{logging.debug()}, and similar constructs. Subsequently, we propose a \emph{fill-in-the-gaps} algorithm and a CFG-based algorithm to infer the execution states of the \emph{remaining} lines (i.e., whether they were executed). This inference is further refined by leveraging the test code's function calls. In the following, we describe each step with precise definitions and explanations.

\subsubsection{Source-to-Log Matching}\label{src::to::log}
Each log message in the test's execution log originates from the execution of a line of code that outputs the message, whether it resides in the test code, the SUT, or a library. This section presents an effective, automated method for matching each log message to a corresponding line in the \emph{test code} when a unique, accurate match is possible. 
We refer to such lines in the test code that produce log messages as \emph{log statements}.
Our matching methodology is based on \emph{static} log statements, as defined below.

\begin{definition}[\emph{\textbf{Static Log Statement}}]\label{def:static_log_stmt}
We define a static log statement \(s:= p(a_1, a_2, \dots, a_n)\) in a faulty test case \(\tau_f\) as a call to a function \(p\) with arguments \(a_1, a_2, \dots, a_n\), where: 
\begin{enumerate}
    \item \(p\) is designed to print a message to the output (e.g., the console or a saved log file) during code execution, and
    \item each argument \(a_i\), for \(1 \leq i \leq n\), can be decomposed into one or more components, and
    \item there exists at least one component whose value can be determined without executing \(\tau_f\). We refer to such components as \emph{static parts}.
\end{enumerate} 
\end{definition}

Based on the above definition, in our running example in \autoref{lst:T0}, lines 2, 10, 11, 16, 18, 22, and 24 are static log statements, whereas line 6 is not. 
Specifically, lines 2 and 22 print a string literal when executed. Lines 10, 11, 16, 18, and 24 print a string literal concatenated with a dynamic (non-static) part whose value is determined at runtime. On the other hand, line 6 is a non-static log statement where the value of argument \texttt{var} is fully determined at runtime.

To identify and collect static log statements from the test code, we build and traverse its abstract syntax tree (AST). In our set of Python test cases, which we will describe in more detail in \autoref{Evaluation}, we initially identify all log statements by locating nodes of type \texttt{attribute} that start with \texttt{"Log."} and have a parent node of type \texttt{function\_call}. 
These criteria, proposed by domain experts, are generalizable to accommodate log statements from test code of other products with different languages and testing frameworks. We then refine the collection by keeping only those log statements that include at least one string literal part. This is achieved through static analysis of the identified log statements, where we examine the types of their arguments and define a set of rules for extracting the string literal parts. \autoref{lst:T0} depicts these argument types in a comment following each log statement. For instance, line 18 is a log statement that has a \texttt{binary\_operator} argument type, containing two operands: one string literal part, \texttt{"var\_2: "}, and a dynamic part, \texttt{str(var\_2)}. Our approach is generalizable and can be adapted to incorporate any set of heuristic rules to identify static log statements.

After collecting all static log statements in the test code, we need to match them with execution log messages to determine their execution state. To this end, we generate a regular expression (\emph{regex}) for each static log statement, marking the dynamic parts using the \texttt{.*} quantifier. For instance, the regexes for lines 10 and 11 are generated as \texttt{/Current time: .*/} and \texttt{/.* var\_1: .*/}, respectively. 

\newannotate{C1.2}We define a static log statement as executed if there exists a log message that \emph{exclusively} matches that statement. \newadded{Under this definition, the actual content of the log message is irrelevant and only the uniqueness criterion is verified.}This uniqueness is essential for precise matching, as a log message that matches multiple log statements cannot be accurately linked to any one of them without knowing the execution order, which is unavailable in our static approach since it does not execute the code. However, there is no such requirement for the reverse. Multiple log messages can be matched to a single log statement.
This is because our trace estimation algorithms do not require knowing how many times each statement was executed. They only need to determine whether a statement was executed. As a result, similar to state-of-the-art work\cite{DBLP:journals/corr/abs-2408-02816}, we treat statements inside loops the same as those inside conditional branches. 

Our matching algorithm is shown in \autoref{algorithm0}. \annotate{C2.4 (b)}Given the sets of faulty test code \addedtext{static log} statements (\(S\)), the execution log messages (\(Log\)), and the error statements (\(E\))\annotate{C2.4 (a)}\addedtext{, which are the statements in the test code that trigger the failure exception}\deletedtext{as inputs}, the matching algorithm produces two sets of statements as outputs. The first set contains the error statements combined with the static log statements that match the log messages and are therefore executed (\(L_{exe}\)). Error statements are always assumed to be executed because they can only fail if they are reached. The second output set contains the static log statements that do not match any messages and are therefore not executed (\(L_{nexe}\)). 

\begin{algorithm}[tbp]\algosize
\caption{Matching statements to log messages}
\label{algorithm0}
\KwIn{$S$ -- \annotate{C2.4 (b)}\addedtext{Static log}\deletedtext{All} statements of the faulty test code}
\KwIn{$Log$ --  Test execution log messages}
\KwIn{$E$ --  Error statements}
\KwOut{$L_{exe}$ --  Executed static log and error statements}
\KwOut{$L_{nexe}$ --  Non-executed static log statements}
$L_{exe}, L_{nexe}, L_{mm} \gets E, \emptyset, \emptyset$

\For{$m \in Log$}{
$ M_s \gets \{s \in S \mid \texttt{match}(\texttt{regex}(s), m)\}$

\If{$ \left| M_s\right| = 1$}
{$L_{exe} \gets L_{exe} \cup M_s$}
\Else{$L_{mm} \gets L_{mm} \cup (M_s \setminus L_{exe})$}
}
\For{$s \in S$}{
\If{$s \notin (L_{exe} \cup L_{mm})$}{
$L_{nexe} \gets L_{nexe} \cup \{s\}$
}
}
\KwRet $L_{exe}, L_{nexe}$

\end{algorithm}

We begin by initializing \(L_{exe}\) as \(E\) and set both \(L_{nexe}\) and \(L_{mm}\) to empty sets (line 1). The \(L_{mm}\) set stores statements whose matched messages also match other statements (\emph{multi-matched} statements). For each message \(m\) in the set of execution log messages, we find all static log statements in \(S\) that match \(m\) using their regex (lines 2 and 3). If exactly one statement is found, it indicates that the statement \emph{uniquely} matches the message, so we add it to the list of executed statements, \(L_{exe}\) (lines 4 and 5). Otherwise, we add each matched statement not previously identified as executed to \(L_{mm}\) (line 7). After processing all log messages, we iterate over the static log statements and add those that are neither executed nor multi-matched to \(L_{nexe}\) (lines 8--10). Finally, we return the two sets of executed and non-executed statements as outputs (line 11), which will serve as inputs to our execution trace estimation algorithms in the next step. 

\begin{algorithm}[tbp]
\algosize  
\caption{Fill-in-the-gaps execution trace estimation}
\label{algorithm1}
\KwIn{$F$ -- All functions of the faulty test code}
\KwIn{$L_{exe}$ --  Executed static log and error statements}
\KwIn{$L_{nexe}$ --  Non-executed static log statements}
\KwOut{$T_1$ -- Estimated execution trace}
$T_1 \gets \emptyset$

\For{$f \in F$}{
    $S_f \gets \texttt{collect\_statements}(f)$
    
    \If{$\left| S_f\right| = 1$}{
        $T_1 \gets T_1 \cup (S_f \setminus L_{nexe})$
        
        $\Continue$
    }
    
    $K_f \gets S_f \cap (L_{exe} \cup L_{nexe})$
    
    \If{$K_f.\texttt{empty}()$}{
        $T_1 \gets T_1 \cup S_f$
    
        $\Continue$
    }
    $G_f \gets \{ (b, I, e) \mid b, e \in S_f,\, I \subseteq (S_f \setminus K_f),\;$\\
    $\forall\, s \in I,\; b < s < e \}$
    
    \For{$(b, I, e) \in G_f$}{
        \If{($b \notin K_f$ \textbf{or} $b \in L_{exe}$) \textbf{and} $e \in L_{exe}$}{
            $\forall\, s \in \{b, e\} \cup I,\, s.{state} \gets \text{\scriptsize{EXE}}$
        }
        \ElseIf{$b \in L_{nexe}$ \textbf{and} ($e \notin K_f$ \textbf{or} $e \in L_{nexe}$)}{
            $\forall\, s \in \{b, e\} \cup I,\, s.{state} \gets \text{\scriptsize{NEXE}}$
        }
        \Else{
            $\forall\, s \in \{b, e\} \cup I,\, s.{state} \gets dbt \in \{\text{\scriptsize{EXE}}, \text{\scriptsize{NEXE}}\}$
        }
        $T_1 \gets T_1 \cup \{s \in \{b, e\} \cup I \mid s.state = \text{\scriptsize{EXE}}\}$
    }
}
\KwRet $T_1$
\end{algorithm}

\subsubsection{Estimation Algorithms}
We propose three trace-estimation algorithms for the test code, each operating at a different code-granularity level. We begin with a line-level estimation, proceed to a block-level estimation that accounts for conditional paths, and conclude with a function-level estimation based on function-call relationships.

\paragraph{Fill-in-the-Gaps Algorithm (\(T_1\))}

Once each static log statement's execution state is determined by \autoref{algorithm0}, we infer the execution state of the remaining lines in the test code to estimate the execution trace.
To this end, we first develop a simple \emph{fill-in-the-gaps} algorithm under the initial assumption that the test code consists of a sequence of statements without branches. We then gradually refine this algorithm as we further explore the details and complexities of the problem. The initial algorithm is presented in \autoref{algorithm1}.

The algorithm takes as inputs the functions of the faulty test code (\(F\)) and the sets of executed and non-executed statements (\(L_{exe}\) and \(L_{nexe}\), respectively), as determined by \autoref{algorithm0}. The output (\(T_1\)) is the set of estimated executed lines in the faulty test case. 
For each function, we extract its body statements (lines 2 and 3). 
Statements in the global scope are also collected as a dummy function called \texttt{module}.
If the function contains only a single statement and it is in \(L_{nexe}\), the function is excluded from the trace.
Otherwise, it is included, and the algorithm continues to the next function (lines 4--6). \(K_f\), defined on line 7, is a subset of \(S_f\), including statements with a \emph{known} execution state. An empty \(K_f\) indicates that we have no information about the execution state of the statements in the function. Consequently, we cannot perform any pruning and must include the entire function in the trace and continue (lines 8--10). 
On lines 11 and 12, we define a set of \emph{gaps} in the function, represented as tuples, where the first and last elements of each tuple are individual statements denoting the beginning (\(b\)) and end (\(e\)) of a gap, and the second element is a set of statements inside the gap, forming its body (\(I\)). The beginning and end statements, referred to as \emph{gap bounds}, can have either known or unknown execution states, whereas the statements in the body are unknown and must have their execution states determined.
We infer the execution state of each gap's body based on its bounds.
When the start statement is either unknown or known to be executed, and the end is executed (lines 14 and 15), all statements within the gap are marked as executed (\texttt{EXE}). The rationale is that, without control-flow branches, if the endpoint is executed, then all preceding statements---regardless of the unknown start---must also be executed. Conversely, if the start is known to be non-executed and the end is either non-executed or unknown (lines 16 and 17), all statements in the gap are marked as non-executed (\texttt{NEXE}). This is based on the assumption that execution stops at the start point, and thus, the subsequent statements are not executed. For all other configurations where a clear determination is not possible, the execution state is marked as \(dbt\) (doubtful) (line 19). These cases represent uncertainty, where the statements may be either executed or non-executed. For instance, when the start is non-executed, and the end is executed, it creates an inconsistency within a linear execution model that lacks branches, as execution cannot logically continue after it has already terminated. Because the trace estimator must assign a definite state to every line, it cannot leave any statement unknown. As a result, we consider two straightforward strategies in \autoref{Evaluation}, where one treats all \(dbt\) cases as \texttt{EXE} and the other treats them as \texttt{NEXE}. Finally, all statements marked as executed are added to the estimated trace \(T_1\) (line 20). This process is repeated for all functions.

\(T_1\) is an estimate of the test code's execution trace, performed at the line level. This estimation assumes the absence of branches and treats all lines of code as if executed sequentially for simplicity. Next, we leverage the detailed information from CFGs\cite{cfgbook} to propose a more precise estimate by accounting for branches in the test code. The algorithm is presented below.

\paragraph{CFG-Based Algorithm (\(T_2\))}

Similar to \autoref{algorithm1}, \autoref{algorithm2} takes as input the functions of the faulty test code (\(F\)), along with \(L_{exe}\) and \(L_{nexe}\). The output is an execution trace estimated by CFG analysis (\(T_2\)). The goal of this algorithm is to identify all potentially executed paths, referred to as \emph{pexe}, within the CFGs of functions in the test code, given \(L_{exe}\) and \(L_{nexe}\). Any node---and its corresponding code block---along these paths is estimated to be executed. Since iterating over all paths in the CFGs comes at an exponential cost, \autoref{algorithm2} introduces an optimized and practical approach for efficiently finding \emph{pexes}.

To this end, we begin by initializing \(T_2\) as an empty set (line 1). 
For each function \(f\) in \(F\), we collect its body statements and construct a set of statements with known execution states (lines 2--4). If no such statements are found, the entire function is assumed to have been executed, and the algorithm proceeds to the next function (lines 5--7). Otherwise, we then build an acyclic CFG, \(cfg\), for function \(f\) (line 8). Next, we identify executed and non-executed nodes within the \(cfg\) (lines 9 and 10). A node is classified as executed (\(n \in V_{exe}\)) if it contains at least one executed statement, i.e., a statement belonging to \(L_{exe}\). Otherwise, if the node contains at least one non-executed statement and no executed statements, it is added to the set of non-executed nodes, \(V_{nexe}\). We refer to a node in \(V_{exe}\) as an \emph{exe} node, and a node in \(V_{nexe}\) as a \emph{nexe} node.
A node may contain both executed and non-executed statements and still be considered entirely executed. This can occur, for example, when the node includes an error statement: the error statement itself is executed, but any statements that follow it within the same node may not be, as the program may terminate at the error point.
This limitation is inherent to this algorithm, which, by design, does not operate at a granularity finer than the node level. As a result, the estimated trace may include the entire node rather than pruning the non-executed statements that follow the executed line. In \autoref{Evaluation}, we demonstrate how \(T_1\) and \(T_2\) can be combined to mitigate this problem by leveraging both \(T_1\)'s fine-grained line-level estimation and \(T_2\)'s block-level estimation.

\begin{algorithm}[!t]\algosize
\caption{Execution trace estimation using CFGs}
\label{algorithm2}
\KwIn{$F$ -- All functions of the faulty test code}
\KwIn{$L_{exe}$ --  Executed static log and error statements}
\KwIn{$L_{nexe}$ --  Non-executed static log statements}
\KwOut{$T_2$ -- Estimated execution trace}

$T_2 \gets \emptyset$

\For{$f \in F$}{
    $S_f \gets \texttt{collect\_statements}(f)$
 
    $K_f \gets S_f \cap (L_{exe} \cup L_{nexe})$

    \If{$K_f.\texttt{empty}()$}{
        $T_2 \gets T_2 \cup S_f$
        
        $\Continue$
    }
    $cfg \gets \texttt{build\_CFG}(f)$

    $V_{exe} \gets \{n \in cfg \mid n \cap L_{exe} \neq \emptyset \}$

    $V_{nexe} \gets \{n \in cfg \mid n \cap L_{nexe} \neq \emptyset$ \textbf{and} $n \cap L_{exe} = \emptyset\}$

    $\forall\, n \in V_{nexe},\, n.{state} \gets \text{\scriptsize{NEXE}}$

    $stack \gets [(cfg.root,[cfg.root])]$

    \While{$not\,\, stack.\texttt{empty}()$}{
        $n, path \gets stack.\texttt{pop}()$
    
        \If{$n.\text{children} = \emptyset$ or $n.state = \texttt{\scriptsize{NEXE}}$ }{
            \If{$n \in V_{nexe}$}{
                $path \gets path[:n]$
            }
            $\texttt{set\_state}(path, V_{exe}, V_{nexe})$
        }
        \Else{
            \For{$n_c \in n.children$}{
                \If{$n_c.state = \texttt{\scriptsize{EXE}}$}{
                    $path_c \gets n_c.pexe\_path[n_c:]$
                    
                    $\texttt{set\_state}(path + path_{c}, V_{exe}, V_{nexe})$
                }
                \Else{
                    $stack.\texttt{push}((n_c,path + [n_c]))$
                }
            }
        }
    }
    $\forall\, n \in cfg,\,T_2 \gets T_2 \cup \{s \in n \mid n.state = \texttt{\scriptsize{EXE}}\}$
}
\KwRet $T_2$

\SetKwFunction{Fn}{set\_state}
\SetKwProg{FnDef}{Function}{:}{}
\FnDef{\Fn{$path, V_{exe}, V_{nexe}$}}{
    \If{$path\cap V_{exe} = V_{exe}\,\,\,\text{and}\,\,\, path\cap V_{nexe} = \emptyset$}{
        $\forall\, n \in path,\,n.{state} \gets \texttt{\scriptsize{EXE}},\,\,n.pexe\_path \gets path$
    }
    \Else{
        $n_{l}\gets last\{n \in path \mid n.state = \texttt{\scriptsize{EXE}}\}$
        
        \If{$n_{l}$ \text{exists}}{
         $\forall\, n \in path[n_{l}+1 :\,],\,n.{state} \gets \texttt{\scriptsize{NEXE}}$
        }
    }
}

\end{algorithm}

Once all \emph{exe} and \emph{nexe} nodes in the function's CFG are identified, we set the execution state of all \emph{nexe} nodes, i.e., the execution state of their statements, as \texttt{NEXE} (line 11). Next, we perform a traversal similar to a depth-first search (DFS) approach, starting from the \(cfg\) root node, to collect all \emph{pexes} within \(cfg\) (lines 12--25). Unlike standard DFS, our traversal allows nodes to be revisited. This is necessary for nodes in the CFG with multiple incoming paths, each of which we aim to explore to determine whether it is a \emph{pexe}. We use the \texttt{set\_state()} function (lines 28--34) to determine whether a traversed path is a \emph{pexe} and to attempt marking its nodes accordingly.
We consider a path a \emph{pexe} if it includes all \emph{exe} nodes and excludes all \emph{nexe} nodes (line 29). 

During the traversal, when we encounter an \emph{end} node, we verify whether the traversed path qualifies as a \emph{pexe}. An end node is either a leaf node with no children or a node with a \texttt{NEXE} execution state (line 15). The traversed path excludes the last node if it belongs to \(V_{nexe}\), as including it would violate the definition of a \emph{pexe} (lines 16 and 17).
If the traversed path is a \emph{pexe}, we set the execution state of its nodes to executed and save the traversed path as the \emph{pexe} of all nodes along it (lines 29 and 30). If the path is not a \emph{pexe}, we identify the last \emph{exe} node along it. If such a node exists, we mark all nodes that come after it as \emph{nexe} nodes (lines 31--34). This step optimizes the traversal by marking the execution state of nodes that do not belong to any \emph{pexes} as \texttt{NEXE}, preventing them from being revisited during subsequent rounds.

When we encounter a node that is not an end node, we iterate through its child nodes. If the child's execution state is \texttt{EXE} (line 21), this indicates that the child is already part of a \emph{pexe}. We then invoke \texttt{set\_state()} on the current traversed path extended with the child's \emph{pexe}, starting from the child node (lines 22 and 23). Note that the \emph{pexe} associated with each node is saved on line 30 and can be retrieved for the child here. By augmenting the traversed path with the child's \emph{pexe}, we aim to determine whether the resulting longer path also qualifies as a \emph{pexe}. This verification is performed via \texttt{set\_state()} in the same manner as before. 
If the child's execution state is not \texttt{EXE}, the child and its traversed path are pushed to the DFS stack for later visitation (line 25). 

After traversing all nodes in the function's CFG, we collect the statements from any node whose execution state is \texttt{EXE} and add them to the estimated trace, $T_2$, excluding the rest from the output (lines 26). This process is repeated for all functions.

\paragraph{Call Site Refinement Algorithm (CSR)}
Finally, the estimated traces \(T_1\) and \(T_2\) can be enhanced by leveraging execution information from call sites. 
A call site is the location in the code where a function or method is invoked.
In a trace where the call site for a particular function is absent, the function's body can be safely considered as non-executed and removed from the estimated trace.
This is particularly beneficial because when execution details for log statements within a function are missing, the entire function body is included in \(T_1\) and \(T_2\), as shown in \autoref{algorithm1} (lines 8 and 9) and~\autoref{algorithm2} (lines 5 and 6). Leveraging call site information helps address execution overestimation by removing unused function definitions from \(T_1\) and \(T_2 \).

To enable the removal of unused function definitions, we propose a \emph{call site refinement} algorithm, presented in \autoref{algorithm3}. This algorithm takes as input an estimated execution trace (\(T\)) and the helper functions of the faulty test code (\(F_h\)). A helper function is generally intended to simplify and assist the main logic by performing a specific task. We classify a function as a helper if 1) its definition appears within the test code, and 2) it is not a function that is executed before or after the test, such as \texttt{setup} or \texttt{teardown} in the testing framework. The output trace \(T'\) is a refined version of the input trace \(T\).

\begin{algorithm}[t]\algosize
\caption{Call Site Refinement}
\label{algorithm3}
\KwIn{$F_h$ -- All helper functions of the faulty test code}
\KwIn{$T$ -- Estimated execution trace}
\KwOut{$T'$ -- Refined estimated execution trace}

$T' \gets T$

\For{$f \in F_h$}{
$C \gets \texttt{get\_call\_sites}(f) $

\If{$f \in T'$ and $C \cap T' = \emptyset$}{
$T' \gets T' \setminus \{f\}$
}
}
\KwRet $T'$
\end{algorithm}

The algorithm begins by initializing \(T'\) to \(T\) (line 1). For each function definition in \(F_h\), we collect the set of its call sites (lines 2 and 3). If the function definition appears in the estimated trace but no corresponding call site is found within the trace, the function definition is removed from the trace (lines 4 and 5). This happens either because the function has no call sites in the test case or because none of its call sites are included in the trace. Once all helper functions are examined, the refined execution trace is returned (line 6).

The proposed trace estimation algorithms in this section can be used individually or in combination to eliminate non-executed elements from the test code, thereby reducing the input provided to the LLM. This reduction can directly impact the search space, potentially improving FL accuracy and inference time. Next, we outline our prompt engineering technique for TCFL and explain how the estimated execution trace of the faulty test is incorporated into the LLM's prompt, along with other relevant information.

\subsection{Prompt Engineering}\label{approach::prompt}

Prompt engineering is the practice of designing structured inputs (prompts) to query LLMs to produce accurate and relevant outputs\cite{shin2023prompt}. To identify the most suspicious code elements in faulty test scripts (files), we design a prompt template, as described in \autoref{fig:prmpt_template}. The template follows a general format, making it applicable to faulty test scripts written in various programming languages with different error messages as input. It allows for requests to the LLM to generate a variable number of faulty elements  \(k \in \{1, 2, ...\}\) at different output granularity levels, such as function, block, or line.

\begin{figure}[!t]
  \centering
  \scriptsize
  \framebox[\columnwidth][l]{\parbox{0.95\columnwidth}{
  \Large \textbf{Task Description} \vspace{0.4em} \\
  \normalsize As an expert software engineer and tester, your mission is to localize faults in \{\emph{programming\_language\}} test scripts at the \emph{\{element\}} level. You will be provided with the test scripts and the error message caused by the test failure. Your goal is to identify \emph{\{k\}} \emph{\{element\}}s that are most likely responsible for the failure and require modification.\vspace{0.6em}\\
 \Large \textbf{Inputs} \vspace{0.4em} \\
\large \textbf{Error Message}
 \vspace{0.4em}\\
\normalsize Here is the error message caused by the test failure:\vspace{0.4em}\\
  \emph{\{err\_msg\}}\vspace{0.4em}\\
\large \textbf{Code}\vspace{0.4em}\\
\normalsize Below are the \emph{\{programming\_language\}} test scripts:\vspace{0.4em}\\
   \emph{\{test\_code\}}\vspace{0.6em}\\
\Large\textbf{Task Instructions}\vspace{0.4em}
\normalsize
\begin{enumerate}[leftmargin=15pt]
\item Carefully examine the provided test scripts and the associated error message.
    \item Identify the \emph{\{k\}} \emph{\{element\}}s that are most likely to contain the faults.
    \item Return a list of faulty \emph{\{element\}}s and their \emph{\{ID\}}s, without any additional explanation. Note that the list of \emph{\{element\}}s and their \emph{\{ID\}}s should be within the range 1 to \emph{\{max\_element\_id\}} and the size of the list must be exactly \emph{\{k\}}. The list should be also in descending order of likelihood of containing the fault, with the most suspicious \emph{\{element\}} first and the least suspicious \emph{\{element\}} last. Ensure that your response is strictly in the specified format. The output should follow this format: \emph{\{output\_template\}}
\end{enumerate}
  }}
  \caption{Our prompt template for test code fault localization. Text enclosed in curly braces (\{ \}) represents variable placeholders dynamically filled during the prompting process.}
  \label{fig:prmpt_template}
\end{figure}
\begin{figure}[htbp]
    \centering
    \scriptsize
    \begin{minipage}[]{\columnwidth}
        \centering
        \subfloat[\scriptsize Function-Level]{
        \fbox{
        \parbox{6.1cm}{
        \texttt{\{"faulty\_functions": ["foo", "bar", \texttt{...}]\}}
        }
        }
        }\\
        \vspace{1.5em}
        \subfloat[\scriptsize Block-Level]{
           \fbox{
        \parbox{7cm}{
        \texttt{\{"faulty\_blocks": ["BLOCK 10", "BLOCK 7", \texttt{...}]\}}
        }
        }
       }\\
      \vspace{1.5em}
    \subfloat[\scriptsize Line-Level]{
        \fbox{
        \parbox{5.4cm}{
        \texttt{10: print("This line is faulty!")}\\
        \texttt{5: print("This line is faulty too!")}\\
        \texttt{...}
        }
        }
        }
    \end{minipage}
  
    \caption{Examples of the requested output format for function, block, and line-level fault localization.}
    \label{fig:outputs}
\end{figure}

Here are the details of each component of the prompt.

\noindent\textbf{Task Description.} This part involves role-based prompting~\cite{DBLP:conf/naacl/KongZCLQSZWD24, han2024rethinking}, assigning a specific role to the LLM (i.e., an expert software engineer and tester), and outlining the FL objective. It specifies the language of the faulty test scripts, the element type requested (function, block, or line), and the number of faulty elements, \emph{k}, to be identified.

\noindent\textbf{Inputs.} 
This part provides the LLM with two inputs: the error message and the faulty test code. The test code is constructed based on the estimated execution traces generated by algorithms discussed in \autoref{approach::estimation}. When the entire faulty test code is considered for FL (i.e., no elements are excluded based on execution trace estimation), it is simply a concatenation of the test scripts. When trace estimation is enabled, the input test code includes only the code present in the trace, excluding all non-executed lines. As shown in our running example in \autoref{lst:T0}, for easier reference, we assign unique integer IDs in the test code for line-level and block-level localization, where each ID corresponds to a specific line or block. When providing input prompts to the LLM, we use these unique integer IDs to label each block or line. Function IDs are the unique names in the test scripts.  The error message is extracted from the test execution log and contains the location and reason of the test failure.

\noindent\textbf{Task Instructions.} This section provides step-by-step instructions, prompting the LLM to examine the inputs and identify and rank the root causes of the failure in a descending order of suspiciousness. It instructs the LLM to generate valid responses by requesting it to return \emph{exactly k} values, avoid including any IDs outside the input range, and refrain from adding additional information to the output. These requests enhance the output's verifiability and help prevent the LLM from hallucinating\cite{DBLP:journals/corr/abs-2310-01469}.

This part also provides a structured output template for the LLM to follow.
Examples of these templates are shown in \autoref{fig:outputs}.  
For all levels, the output elements are expected to be ranked in descending order of suspiciousness, as specified in the prompt. We use the widely adopted JSON format\footnote{\underline{\url{https://platform.openai.com/docs/guides/structured-outputs}}} for representing outputs at both the function and block levels.  The LLM is prompted to include the names of suspicious functions and the IDs of suspicious blocks (assigned during preprocessing) in the output template.
In line-level FL, along with the line number as the ID, we also request the line's content. This method ensures that each line number matches its content, corresponding to a single line in the labeled input code. It further enables the detection and handling of cases during post-processing in which the content is correctly identified as faulty, but the associated line number is incorrect (see \autoref{discuss::output::validity} for details). 
Since certain characters, such as quotation marks and backslashes, require specific escaping to comply with JSON formatting rules, and these characters may appear within a line, we define and use a custom output template for line-level FL. This template represents each faulty line by indicating its line number, followed by a colon and the line content.
We do not request content at the block or function level in FL because including entire code blocks and function bodies significantly increases output size, thereby increasing token usage. This not only risks exceeding the LLM's context window when combined with input tokens but also increases inference time, reducing efficiency. Moreover, LLMs are less likely to struggle with block and function label mismatches compared to line labels since a block or a function is typically longer and more distinctive than a single line.

Next, we present a comprehensive evaluation of our approach on an industrial dataset, using various configurations of estimated execution traces, \emph{k} values, and granularity levels.

\section{Evaluation} \label{Evaluation}
In this section, we address the following research questions.

\textbf{RQ1:} How accurately do the various versions of our estimated execution traces reflect the actual execution states of different elements in the test code, and how effectively do they prune the test code for fault localization (FL)?

\annotate{C2.2}\addedtext{\textbf{RQ2:} How does pruning the input prompt code with estimated traces impact the effectiveness of LLM-based FL in terms of FL accuracy, and its efficiency and scalability in terms of inference time and total token count?}

\annotate{C1.2 \& C2.3}\addedtext{
\textbf{RQ3:} How does our LLM-based test code fault localization (TCFL) approach perform at the function, block, and line levels, compared to state-of-the-art techniques?
}

\deletedtext{
\textbf{RQ2:} What is the overall performance of LLM-based test code fault localization (TCFL) at the function, block, and line levels, using various parameter configurations?

\textbf{RQ3:} How does pruning the input prompt code with estimated traces impact the efficiency and scalability of LLM-based FL in terms of inference time and total token count?
}

\subsection{Experimental Setup}\label{setup}
\subsubsection{Benchmarks}\label{benchs} We conduct our experiments on a set of 785 real-world faulty Python system test cases, provided by our industrial partner. Prior to applying our execution trace estimators or LLM-based FL techniques, we preprocess each test case by removing blank lines. This step helps eliminate unnecessary complexity, as blank lines do not contribute meaningful information about the code's logic, structure, or behavior\cite{DBLP:journals/jss/ZhangZKTYYH24, galiullin_2024_14132684}. Comments, however, are retained because they have been shown to aid in code comprehension\cite{6613836}. Following preprocessing, our test cases range from 56 to 1,070 lines of code, with an average of 244 lines per test case. \autoref{table::benchmark} depicts the size of these test cases across various levels of code granularity, including the number of files, functions, blocks, and lines of code. \newannotate{C2.4}\newadded{To better reflect the complexity of the test cases in our benchmark, in addition to these numbers, we report three additional metrics that have also been used in prior work\cite{7985707, 9978993}: (1) the number of test steps, defined as the number of high-level interactions between the test script and the SUT (see \autoref{fig:problem_def_overview}); (2) the size of the failed execution log file, measured in bytes; and (3) the number of lines in the test script when accounting for calls to dependencies defined within the test codebase. The values of these metrics are presented in \autoref{table::benchmark::additional::characteristics}. On average, each test case in our benchmark performs 479 interactions with the SUT, with a maximum of 13,935 for a single test case. Execution logs average 458.4 KB, and test cases average 1,087 lines of code. The line counts in \autoref{table::benchmark::additional::characteristics} reflect the total number of lines obtained by replacing calls to functions in the test scripts with the corresponding function definitions from our test codebase, which contains 5,159,790 lines in total, excluding the test scripts themselves. In contrast, the line counts in \autoref{table::benchmark} include only the lines of code in the test scripts themselves, without inlining code from their dependencies. While our technique is applied only to the line counts reported in \autoref{table::benchmark}, it is important to note that, in practice, the actual search space for FL includes the total number of lines in both the test script code and the test codebase (presented in \autoref{table::benchmark::additional::characteristics}), along with the SUT code. Our approach aims to minimize the developer's effort by identifying faults in the test script early in the debugging process. This helps prevent the developer from wasting time on the broader test codebase or making incorrect assumptions that the fault lies in the larger or unavailable SUT code.}
\begin{table}[tbp]
    \centering
    \resizebox{\columnwidth}{!}{
    \begin{tabular}{l cccc cccc}
        \toprule
         \multirow{2.5}{*}{\shortstack{Code \\ Granularity}} & \multicolumn{4}{c}{\# Total} &  \multicolumn{4}{c}{Faulty Ratio (\%)} \\ 
        \cmidrule(lr){2-5} \cmidrule(lr){6-9} 
        & Min & Median & Mean & Max & Min & Median & Mean & Max \\  
        \midrule
        File  & 1 & 2 & 2 & 2 & 50.0 & 50.0 & 54.2 & 100.0  \\  
        Function  & 2  & 5 & 6 & 26 & 3.8 & 25.0 & 24.5 & 100.0 \\    
        Block  & 3 & 31 & 40 & 280 & 0.4 & 6.7 & 9.5 & 70.6\\   
        Line & 56 & 202 & 244 & 1,070 & 0.1 & 1.2 & 3.2 & 33.1 \\  
        \bottomrule
    \end{tabular}
    }
    \caption{Distribution of total and faulty levels of code granularity in our benchmark of 785 faulty Python test cases \protect\newadded{from a test codebase of 5,159,790 lines.}}
    \label{table::benchmark}
\end{table}

\begin{table}[tbp]
    \centering
    \resizebox{\columnwidth}{!}{
     \newannotate{C2.4}\newadded{
    \begin{tabular}{l cccc}
        \toprule
        Test Characteristics & Min & Median & Mean & Max \\ 
        \midrule
        Test Step (\#)  & 1 & 161 & 479 & 13,935 \\ 
        Log File Size (KB)  & 3.9 & 162.5 & 458.4 & 43,281.7 \\  
        Test Script + Codebase Deps (LOC) & 124 & 807 & 1,087 & 4,721 \\
        \bottomrule
    \end{tabular}
    }
    }
    \caption{\protect\newadded{Distribution of the total test steps (high-level SUT interactions), execution log file sizes, and lines of code in test scripts---including dependencies from the test codebase---across our benchmark of 785 faulty Python test cases from a test codebase of 5,159,790 lines.}}
\label{table::benchmark::additional::characteristics}
\end{table}

To identify the faulty elements in each test case, we initially compute the code \emph{diff} between the preprocessed faulty version and its corresponding preprocessed repaired version. The repaired version is selected as the first subsequent commit in the repository history where the failure no longer occurs. Any line in the faulty version that is either removed or modified in the repaired version is marked as faulty. The associated code blocks, functions, and files containing that line are also marked as faulty. When the diff includes new lines added in the repaired version that were not present in the faulty one, we follow prior work\cite{DBLP:journals/tse/ZouLXEZ21} and treat the line immediately following the inserted segment as the faulty line. 

Since bug-fixing commits often include unrelated changes, such as refactorings, that are not directly tied to the root cause of the failure\cite{DBLP:conf/msr/BagheriH22}, our automated diff-based approach may overestimate the faulty locations. To mitigate this, we further employ a semi-automated refinement process. We begin by identifying outliers, defined as test cases where the ratio of faulty lines of code \newannotate{C1.4}\newadded{(i.e., the number of faulty lines divided by the total number of lines per test case)}exceeds \(mean + 3 * \sigma\), based on the \emph{3-\(\sigma\) rule}\cite{pukelsheim1994}. Here, \(\sigma\) represents the standard deviation, and \(mean\) is the average faulty line ratio across all test cases. We then manually inspect this small set of outliers using expert knowledge to confirm the true fault locations. If a test case exercises only refactoring changes, it is excluded from our dataset. Otherwise, we retain the test case and update its ground truth solely based on the manually verified faulty lines. Our benchmark, consisting of 785 test cases, is the result of applying this refinement process. \newadded{Specifically, the \(mean + 3 * \sigma\) threshold computed for our dataset is 52\%, leading to the identification of 19 outliers from an original set of 794 test cases (i.e., those exhibiting a line-change ratio exceeding 52\%).}\newdeleted{We identified 19 outliers from an original set of 794 test cases, each exhibiting a line change ratio exceeding 52\% (i.e., the \(mean + 3 * \sigma\) threshold for our dataset).}\newadded{Note that computing the threshold in this manner follows standard statistical practice in the literature for identifying outliers\cite{6520712, https://doi.org/10.1155/2021/1899225}. 

The 52\% threshold computed automatically by the formula on our dataset is not so high as to produce many false negatives (i.e., failing to detect outlier test cases), nor so low as to result in many false positives. A subsequent manual investigation by industry engineers further demonstrated the 3-\(\sigma\) rule's effectiveness for our task.}
The manual inspection confirmed that nine out of 19 outliers involved only refactoring changes and were thus excluded. The remaining 10 were retained, with updated fault labels determined by manual analysis. These updates added only a few newly identified faulty lines (ranging from 0 to 3 per case) while removing an average of 59 incorrectly flagged faulty lines. \newannotate{C1.4}\newadded{The use of expert knowledge during manual examination of outliers ensures that legitimate complex faults that deviate from typical diff patterns are not excluded, and that false positives in the ground-truth labels are effectively reduced.}Note that manual inspection of all test cases would be prohibitively labor-intensive. Therefore, we limited our manual inspection to the much smaller set of outliers automatically identified, which are more likely to include changes unrelated to the actual failure, \newadded{as outliers primarily serve as red flags for cases that may require further investigation.}\autoref{table::benchmark} also presents the ratio of faulty elements to the total number of elements at various levels of granularity. On average, 24.5\% of functions, 9.5\% of blocks, and 3.2\% of lines in our benchmark are faulty. 

To match test code statements to log messages, we collect a total of 42,121 static log statements from all 785 test cases. Among these, we uniquely match 13,643 statements to execution log messages. In addition, 1,078 error statements are included, bringing the total to 14,721 statements with execution state set to executed.
Of the remaining static log statements, 12,159 do not match any log message and are therefore marked as non-executed. \annotate{Part of C1.4}\addedtext{This non-execution is genuine because for each one of these statements, all static string literals are absent from the log.}

The remaining static log statements match multiple log messages. As a result, their execution state is considered unknown. \addedtext{Besides multi-match static log statements, there are 1,437 fully dynamic log statements whose execution state our algorithm cannot infer. These statements lack static content and therefore do not produce log information that could help our matching algorithm, for example, when they print only the runtime value of a variable without a string literal.}

To construct CFGs, we use the open-source tool Joern\footnote{\underline{\url{https://github.com/joernio/joern}}} and optimize the graphs by merging chains of nodes without branches or nodes that point to the same line of code into single nodes.

\begin{table*}[tbp]
    \centering
    \resizebox{\textwidth}{!}{
     \newannotate{C1.1}\newadded{
    \begin{tabular}{l|l|l|l}
        \hline
        Baseline  & Adaptation Description & Adaptation Rationale  & Replication Guide \\
        \hline
        \multirow{24}{*}{FlexFL} & \multirow{2}{*}{Excluding execution-based FL methods such as SBFL} & 1. No test cases for the test code & \multirow{2}{*}{\parbox{6.7cm}{Among the traditional techniques, only IRFL is enabled in the space reduction (SR) phase.}}\\
        & & 2. High execution cost for system-level test scripts~\cite{DBLP:journals/spe/Larus90, DBLP:journals/tse/MarchettoSS19, DBLP:conf/wosp/HorkyKLT16} & \\
        \cline{2-4}
        & \multirow{12}{*}{\parbox{5.7cm}{Expanding FL support across multiple granularity levels beyond the method level}} & \multirow{12}{*}{1. A more comprehensive and fine-grained FL analysis} & \multirow{12}{*}{\parbox{6.7cm}{Implement six additional agent function calls and use them in the same manner as method-level calls when prompting the LLMs for faulty lines or blocks: \texttt{find\_line(keyword)} and \texttt{find\_block(keyword)} to retrieve lines and blocks in the test code that contain the specified keyword, \texttt{get\_lines\_of\_path(path\_name)} and \texttt{get\_blocks\_of\_path(path\_name)} to get the range of line and block numbers in file \texttt{path\_name}, and \texttt{get\_code\_snippet\_of\_line(line\_num)} and \texttt{get\_code\_snippet\_of\_block(block\_num)} to retrieve the code snippet of a specific line or block.}} \\
        & & \multirow{12}{*}{2. A like-for-like comparison with our technique in terms of granularity levels} &  \\
         & & &  \\
        & & &  \\
        & & &  \\
        & & &  \\
        & & &  \\
        & & &  \\
         & & &  \\
        & & &  \\
        & & &  \\
        & & &  \\
        \cline{2-4}
        & Deleting class-related agent function calls & No class hierarchy in the source code of test cases in our dataset & \parbox{6.7cm}{Delete the agent function calls \texttt{get\_classes\_of\_path}, \texttt{get\_methods\_of\_class}, and \texttt{find\_class}, and add \texttt{get\_methods\_of\_path} instead.}\\ 
        \cline{2-4}
        & Excluding bug reports from the prompt &  \parbox{8cm}{Limited or no access to internal bug-tracking systems in commercial proprietary codebases, for both public users and internal teams~\cite{sandhu1998role}} & \parbox{6.7cm}{Remove the \texttt{<bug report>} component from the prompt template while leaving the rest unchanged.}\\ 
        \cline{2-4}
        & \multirow{5}{*}{Revising the prompt terminology} & \multirow{5}{*}{1. Our dataset comprising Python test cases rather than Java} & \multirow{5}{*}{\parbox{6.7cm}{Replace the term \texttt{Java} with \texttt{Python} and specify the FL granularity level in the prompt (e.g., \texttt{You are a debugging assistant for Python test code. Your task is to locate the top-5 most likely faulty test code blocks.}).}} \\
        & & \multirow{5}{*}{2. Enabling FL at three granularity levels: function, block, and line.}  &  \\
        & & & \\
        & & & \\
        & & & \\
        \hline
        \multirow{9}{*}{IRFL} & \multirow{6}{*}{\parbox{5.7cm}{Expanding FL support across multiple granularity levels beyond the method level}} &  \multirow{6}{*}{1. A more comprehensive and fine-grained FL analysis} & \multirow{6}{*}{\parbox{6.7cm}{In addition to the default corpus containing the methods' content, generate and query two additional corpora: a block-level corpus, where each document corresponds to the code content of a CFG block, and a line-level corpus, where each element represents the content of an individual line of the test code.}} \\ 
         & & \multirow{6}{*}{2. A like-for-like comparison with our technique in terms of granularity levels} &  \\
           & & &  \\
        & & &  \\
         & & &  \\
        & & &  \\
         \cline{2-4}
         & \parbox{5.7cm}{Using error tracebacks instead of bug reports when querying the data corpus} & \parbox{8cm}{Limited or no access to internal bug-tracking systems in commercial proprietary codebases, for both public users and internal teams~\cite{sandhu1998role}} & \parbox{6.7cm}{Query the corpus using error traceback messages with failure details (e.g., error locations) rather than bug report descriptions.} \\
        \hline
        \multirow{6}{*}{FauxPy} &   \multirow{2}{*}{Excluding execution-based FL methods  such as SBFL} & 1. No test cases for the test code  & \multirow{2}{*}{\parbox{6.7cm}{Among the seven FL techniques, only stack-trace-based FL is enabled.}}  \\
        & & 2. High execution cost for system-level test scripts~\cite{DBLP:journals/spe/Larus90, DBLP:journals/tse/MarchettoSS19, DBLP:conf/wosp/HorkyKLT16} &  \\
        \cline{2-4}
        & \multirow{4}{*}{\parbox{5.7cm}{Expanding FL support across multiple granularity levels beyond the method level}} & \multirow{4}{*}{1. A more comprehensive and fine-grained FL analysis} & \multirow{4}{*}{\parbox{6.7cm}{For line-level FL, consider the error lines from the traceback as faulty. For block-level FL, extract CFG blocks by parsing the test code, and mark as faulty any blocks containing error lines from the traceback.}} \\
        & & \multirow{4}{*}{2. A like-for-like comparison with our technique in terms of granularity levels} &  \\
         & & &  \\
         & & & \\
        \hline
    \end{tabular}
    }
    }
    \caption{\protect\newadded{List of adaptations applied to the SUTFL baselines for applicability to the TCFL context, along with their rationale and guidance for reimplementation.}}
    \label{table::baseline::adaptations}
\end{table*}

\annotate{C1.2 \& C2.3}\addedtext{\subsubsection{Baselines}\label{sub::baselines} As noted in \autoref{Introduction}, to the best of our knowledge, our technique is the first to perform FL on the test code. Therefore, a direct comparison with tools that also operate on test code is not feasible. To enable the evaluation of our method against existing baselines, we instead selected three established tools that perform FL on the SUT code: FauxPy\cite{PythonFL-FauxPy-Tool}, which implements seven established FL techniques for Python code, the information-retrieval-based FL technique BoostNSift (IRFL)\cite{9610655}, and FlexFL\cite{10.1109/TSE.2025.3553363}, the most recent LLM-guided FL approach for the SUT. FlexFL extends the state-of-the-art LLM-based SUTFL tool AutoFL~\cite{DBLP:journals/pacmse/KangAY24} by integrating traditional and LLM-based FL methods and defining an expanded set of agent function calls.

Given the distinct nature of the SUTFL and TCFL problems, as well as the differences between the baseline datasets and our industrial dataset, we had to consider certain restrictions and apply several adaptations. \newadded{These adaptations are minimal and necessary to make the SUTFL baselines applicable to the TCFL problem, and their implementations can be found in our tool repository.\footnote{\underline{\url{https://github.com/Ahmadreza-SY/TCFL}}}
In this section, we discuss these adaptations in detail. In \autoref{adaptation::analysis}, we analyze their effects on the baselines' original behavior and outline the measures taken to mitigate potential biases introduced by these adaptations.} \newdeleted{as follows:}

\newannotate{C1.1}\newadded{\autoref{table::baseline::adaptations} presents a detailed overview of the adaptations for each baseline, along with the rationale behind them. While the implementations of these adaptations are available in our tool repository, the table also includes a column that provides guidance for replicating them. 

The adaptations can be grouped into four high-level categories:
\begin{enumerate}
    \item Excluding execution-based information from the FL process
    \item Expanding FL support across multiple granularity levels beyond the method level
    \item Excluding unsupported artifacts or replacing them with comparable ones
    \item Modifying the set of agent function calls and revising the prompt terminology
\end{enumerate}
}

Below, we detail the reasoning behind each of these decisions while presenting the specifics of \newadded{implementing them for}each baseline.

\newannotate{C1.1}\noindent\newadded{\emph{1. Excluding execution-based information from the FL process.}}
\newdeleted{Most importantly, o}Our choice of FL tools is limited to those that do not require code execution. This constraint arises from two main reasons. First, many SUT-based tools, such as those implementing spectrum-based fault localization (SBFL)\cite{DBLP:journals/ieicetd/ZhengHCYFX24, DBLP:journals/jss/RaselimoF24, DBLP:journals/access/SarhanB22,de2016spectrum,zakari2020spectrum}, execute multiple test cases and subsequently run parts of the SUT code to assign a suspiciousness score to each code element based on the frequency of its presence in passing and failing test runs. Since there are no test cases for the test code, the same concept cannot be directly applied to our setting.}
\annotate{Part of C2.1 \& C3.2} \addedtext{Second, executing test cases is costly in practical scenarios, for our industry partner and in many other contexts, due to the complexity of system-level testing, which involves numerous interactions among the test code, the SUT, and intermediate layers. As mentioned in \autoref{Introduction} and \autoref{subsub::exec::cost}, this high execution cost is common in system-level testing setups and is not specific to our case study. Although the average number of lines per test in our benchmark is 244, the large number of components involved in each execution\newadded{---from both the SUT and the test codebase---}makes execution inefficient and costly. For example, as shown in \autoref{table::benchmark::additional::characteristics}, the high-level SUT interaction logic (i.e., the closest layer to the test script in \autoref{fig:problem_def_overview}) receives an average of 479 calls per test execution in our dataset. As we move to lower, more complex API layers, the number of calls to the SUT increases significantly.}

\newadded{Like our technique, IRFL does not rely on execution for fault localization. It solely relies on textual information embedded in bug reports and
source code to link bug descriptions to the corresponding faulty
code elements.}This implies that a single failed execution is sufficient to be used as input to the FL technique, eliminating the need to run the code during the FL process itself. \newadded{Therefore, this baseline is not affected by this adaptation. In contrast, FlexFL and FauxPy leverage execution in their FL process; however, they also support execution-free FL.}In particular, among the seven FL techniques implemented by FauxPy, the stack-trace-based approach can be adapted without requiring multiple code executions. This technique relies solely on the error traceback (i.e., the stack trace associated with the faulty test's execution) produced during a single failed run. Given an error traceback, FauxPy produces a list of error locations in reverse order, prioritizing the most recent calls (i.e., those closer to the exception) as the most likely faulty locations. \newadded{We retain only the stack-trace-based approach from the FL techniques implemented in FauxPy.\footnote{\underline{\url{https://github.com/atom-sw/fauxpy}}}} \newdeleted{The original implementation of FauxPy}\newdeleted{ performs FL only at the function level. We extend its capability to the block and line levels by extracting each error line from the traceback and mapping it to its corresponding line, block, and function. This enhancement enables us to compare the performance of FauxPy with our technique across all three code-granularity levels.}

\newdeleted{The BoostNSift IRFL approach can also identify faulty code by leveraging textual information embedded in bug reports and source code to link bug descriptions to the corresponding faulty code elements, without requiring code execution. However, because our dataset consists of commercial code
and is subject to role-based access control}

\newdeleted{, we do not have access to internal, proprietary bug-tracking systems or to historical bug reports. FlexFL faced the same limitation and modified the original BoostNSift accordingly, producing a variant called \textit{BoostN}. Following the same rationale, we also adopt FlexFL's BoostN and substitute bug reports with error tracebacks, which serve as our fault-related query information.}

FlexFL consists of two core components: Space Reduction (SR) and Localization Refinement (LR). The SR phase integrates IRFL \newdeleted{(BoostN), Agent4SR, }and other traditional FL techniques, \newadded{along with a component named Agent4SR,}to narrow the search space and produce an initial set of suspicious methods. Agent4SR leverages an LLM equipped with code-exploration functions to construct part of this set. The LR phase then employs Agent4LR, which follows a similar LLM-based approach to Agent4SR but focuses exclusively on identifying the most likely faulty methods from the SR candidates.

FlexFL also offers the flexibility to avoid code execution by leveraging certain design considerations. In particular, selecting IRFL from the traditional FL techniques integrated into FlexFL allows it to operate without code execution.
\newdeleted{To this end, in}\newadded{To replicate this adaptation, during}the SR phase of the FlexFL framework, \newdeleted{we retained}only the IRFL and the Agent4SR components \newadded{are retained, while techniques that rely on code execution, including SBFL and HybridFL, are excluded.} As a result, the hybrid candidate list produced in the SR phase consists of the top 15 suspicious elements identified by IRFL and the top 5 elements identified by Agent4SR. 

\newannotate{C1.1}\noindent\newadded{\emph{2. Expanding FL support across multiple granularity levels
beyond the method level.}}\newadded{As shown in \autoref{sub::problem::importance}, method-level FL does not always provide sufficiently detailed information, making a more fine-grained analysis necessary. To address this, our FL technique supports two additional granularity levels---code block and line---beyond the default function-level granularity offered by the three baseline techniques. We extend the FL support of these baselines accordingly, enabling a more comprehensive analysis and allowing a like-for-like comparison with our technique in terms of granularity. To replicate this adaptation in FauxPy, error lines should be extracted from the traceback
and mapped to their corresponding code blocks. In IRFL, separate document corpora should be constructed for each program element type from the test code---functions, blocks, or lines---corresponding to the desired level of granularity. In FlexFL, only minor adjustments are needed to the function-level pipeline, primarily replacing function-level agent calls with block- and line-level equivalents. Details are presented in \autoref{table::baseline::adaptations}.
}

\newannotate{C1.1}\noindent\newadded{\emph{3. Excluding unsupported artifacts or replacing them with
comparable ones.}}We use the exact Java implementation of BoostN,\footnote{\underline{\url{https://zenodo.org/records/11524997}}} with a modification to the query and document contents \newadded{when applying IRFL, either as a traditional FL technique within FlexFL or as a standalone approach. Specifically, in TCFL, the document corpora naturally consist of program elements from the test code rather than from the SUT's source code. Furthermore, our dataset consists of commercial code and is subject to role-based access control\cite{sandhu1998role}.
Therefore, we do not have access to internal, proprietary bug-tracking systems and replace bug reports in BoostN---including titles, descriptions, and comments---with error tracebacks, which serve as our fault-related query information.

To replicate this adaptation in the LLM-based components of FlexFL, the bug report component should be removed from the prompt
template, leaving the rest unchanged. This variant, which excludes bug reports and relies solely on the trigger test (i.e., the error traceback), has already been introduced in FlexFL's work.} \newdeleted{In BoostN,
the query is constructed from the bug report title, description,
and comments, while the documents represent the SUT's source code methods. However, in our setting, the query corresponds to the error traceback, and the documents consist of program elements from the test code---functions, blocks, or lines---
depending on the desired level of granularity.}

\newannotate{C1.1}\noindent\newadded{\emph{4. Modifying the set of agent function calls and revising the prompt terminology.}}We also modified the set of agent function calls used in FlexFL to better align with our problem context. Specifically, we replaced \textit{get\_classes\_of\_path} and \textit{get\_methods\_of\_class} with a single function, \textit{get\_methods\_of\_path}, and removed the \textit{find\_class} function. These changes reflect the structure of our industrial benchmark, which is composed of files (paths) and methods, without a class hierarchy. \newadded{Moreover, we revised the prompt terminology by updating the programming language from Java to Python and explicitly specifying the FL granularity (function, block, or line) in both the provided input and the expected output.}

\newdeleted{FlexFL originally supports FL only at the function level. To enable comparison across multiple granularity levels, we extended its capabilities to include block-level and line-level FL. This required only minor adjustments to the function-level pipeline, primarily involving replacement of the function-level agent calls (e.g., \textit{find\_method}, \textit{get\_methods\_of\_path}, \textit{get\_code\_snippet\_of\_method}) with block-level and line-level equivalents such as \textit{find\_block}, \textit{get\_blocks\_of\_path}, \textit{get\_code\_snippet\_of\_block}, and corresponding line-level variants.}We use the same maximum number of agent function calls as FlexFL (10 calls), applied uniformly across all granularity levels and for both the SR and LR phases.

\newdeleted{Finally, we adapted the LLM prompts used within FlexFL. Since we do not have access to bug reports, we supply the LLM with the error traceback. In addition, we revised the prompt terminology to better align with our context. For example, we updated the programming language from Java to Python and explicitly specified the FL granularity (function, block, or line) in both the provided input and the expected output.}

We conduct our FL experiments on a system with a 72-core CPU, 128 GB of RAM, and four 32GB GPUs.

\subsection{RQ1: Accuracy and Impact of Estimated Execution Traces}\label{rq1}
To address RQ1, we generate a set of estimated execution traces using the estimation algorithms described in \autoref{approach::estimation}, and evaluate each one using three metrics: \emph{masked prediction F1 score (MPF1)}, \emph{pruning rate}, and \emph{fault preservation ratio}, all of which will be defined later in this section. The first column of \autoref{table2} presents 12 different estimated execution traces. \(\tau_f\) considers all lines from the original faulty test code as executed, without applying any pruning. Variants of \(T_1\) are generated using the fill-in-the-gaps algorithm (see \autoref{algorithm1}), where doubtful statements are assumed to be executed in \(T_{1,dbt=\text{EXE}}\) and non-executed in \(T_{1,dbt=\text{NEXE}}\). The \(T_2\) variants are produced by applying control flow graph (CFG) pruning on \(\tau_f\) (see \autoref{algorithm2}). \(T_{1,dbt=\text{EXE}} \cap T_2\) represents the intersection of \(T_{1,dbt=\text{EXE}}\) and \(T_2\), including only the statements shared by both traces. Similarly, \(T_{1,dbt=\text{NEXE}} \cap T_2\)  includes statements common to both \(T_{1,dbt=\text{NEXE}}\) and \(T_2\). These variants allow us to analyze the combined effect of both estimation algorithms. Finally, the CSR trace variants apply call site refinement (see \autoref{algorithm3}) to remove any function definitions from a trace that are not associated with a call site within that trace. In the following, we define the metrics used to evaluate these traces and present the corresponding results.

\begin{table}[tbp]
    \centering
    \resizebox{\columnwidth}{!}{
    \begin{tabular}{l c c cc}
        \toprule
         \multirow{2.5}{*}{Estimated Trace} & \multirow{2.5}{*}{MPF1 (\%)} & \multirow{2.5}{*}{Pruning Rate (\%)} & \multicolumn{2}{c}{Fault Preservation (\%)} \\ 
        \cmidrule(lr){4-5} 
        &  & &  Partial &  Full \\  
        \midrule
        
        \(\text{CSR}(T_{1, dbt = \texttt{EXE}}\cap T_2)\) & 89.6 & 26.1 & 87.4 &  59.0 \\
         \(T_{1, dbt = \texttt{EXE}} \cap T_2\) & 89.4& 25.7 & 87.4 &  59.0\\
         \midrule

          \(\text{CSR}(T_{1, dbt = \texttt{EXE}})\) & 86.2 & 24.0 & 87.8 & 60.9  \\
           \(T_{1, dbt = \texttt{EXE}}\) & 86.1 & 23.8 & 87.8 & 60.9 \\
            \(\text{CSR}(T_2)\) & 85.1 & 16.7 & 97.3 &  80.1 \\
               \(T_2\) & 84.9 & 16.5 & 97.3 & 80.1 \\
        \midrule

           \(\text{CSR}(T_{1, dbt = \texttt{NEXE}}\cap T_2)\) & 73.5 & 35.7 & 83.6 &  48.5 \\
           \(T_{1, dbt = \texttt{NEXE}} \cap T_2\) & 73.5 & 35.4 & 83.6 & 48.7 \\
           \(\text{CSR}(T_{1, dbt = \texttt{NEXE}})\) & 73.0 & 35.0 & 83.6 & 48.9 \\
   \(\text{CSR}(\tau_f)\) & 73.0 & 0.04 & 100.0 & 100.0  \\
       \(\tau_f\) & 73.0 & 0.0 & 100.0 & 100.0  \\
        \(T_{1, dbt = \texttt{NEXE}}\) & 72.9 & 34.7 & 83.6 & 49.0 \\
        \bottomrule
    \end{tabular}
    }
    \caption{Average accuracy, pruning rate, and fault preservation ratio of estimated execution traces across 785 faulty test cases.}
    \label{table2}
\end{table}

\subsubsection{Masked Prediction F1 Score (MPF1)}\label{subsub::mpf1} 
In the absence of the possibility to execute test cases due to high costs, irreproducibility, or missing artifacts from the time of failure (as described in \autoref{Problem_Def}), we propose a \emph{masking strategy}, inspired by the token masking approach used in LLMs\cite{DBLP:conf/naacl/DevlinCLT19}, to evaluate the accuracy of our trace estimation algorithms. Specifically, we use statements with known execution states, as described in \autoref{approach::estimation} (i.e., \(L_{exe}\) and \(L_{nexe}\)). To measure each estimator's accuracy, we mask (hide) the execution states of a subset of statements in \(L_{exe} \cup L_{nexe}\) and evaluate whether the estimation algorithm can accurately infer their states using the remaining known information.
We then compute F1 score to evaluate each estimator's accuracy on the masked statements. A higher F1 score indicates greater accuracy in determining execution states by the estimated trace.

The second column of \autoref{table2} presents the average MPF1 score for each estimated trace, computed across 785 faulty test cases and sorted in descending order. To compute this value, we use four masking rates: 0\%, 20\%, 50\%, and 80\%. Each masking rate, except for the 0\% rate, is applied five times, with a different random subset of statements from \(L_{exe} \cup L_{nexe}\) masked in each run. A 0\% masking rate means that no statements are masked, while an 80\% rate masks a large portion of the set. Since applying a 0\% rate results in an empty masked set regardless of repetition, multiple runs produce the same outcome and are therefore unnecessary. We perform random sampling using stratified selection\cite{Singh1996} to preserve the overall distribution of executed and non-executed statements. 
The values in the second column represent the average MPF1 across all 785 test cases in our dataset, computed at four masking rates with five repetitions each, when applicable. For all masking rates above 0\%, the F1 score is calculated over the set of masked statements, as the goal is to evaluate how accurately the estimators can infer the execution states of statements hidden from them. In contrast, for the 0\% masking rate, where the masked set is empty, computing accuracy over this empty set is not meaningful. Instead, we compute the F1 score over all statements with known execution states. This special case allows us to assess whether a trace may alter the execution state of a statement whose status is already precisely known. Interestingly, such changes can occur. For example, \(T_2\) may overestimate execution by performing estimation at the node level, treating an entire node as executed if it contains at least one executed statement. Thus, some truly non-executed lines within the node (e.g., those following an error line) may be incorrectly marked as executed by \(T_2\), reducing the MPF1 at a 0\% masking rate and thereby lowering \(T_2\)'s overall MPF1.

The MPF1 results show the highest accuracy for \(\text{CSR}(T_{1,dbt=\text{EXE}} \cap T_2)\) and \(T_{1,dbt=\text{EXE}} \cap T_2\), with scores of 89.6\% and 89.4\%, respectively. In contrast, the lowest MPF1 values are observed for the original \(\tau_f\) and for traces that assign doubtful statements as non-executed. This discrepancy may be attributed to execution overestimation, as seen in \(\tau_f\), or underestimation, as in the \(dbt=\text{\footnotesize{NEXE}}\) variants, both of which can adversely affect accuracy. The higher MPF1 scores obtained from combining \(T_{1,dbt=\text{EXE}}\) and \(T_2\) traces, compared to using each trace individually, further highlight the advantage of integrating these two trace types. 

\begin{table*}[tbp]
    \centering
    \resizebox{\linewidth}{!}{
    \annotate{C1.4}\addedtext{
    \begin{tabular}{l ccccc ccccc ccccc ccccc}
        \toprule
         \multirow{2.5}{*}{Estimated Trace} & \multicolumn{5}{c}{MPF1 (\%)} & \multicolumn{5}{c}{Pruning Rate (\%)} & \multicolumn{5}{c}{Partial Fault Preservation (\%)} &  \multicolumn{5}{c}{Full Fault Preservation (\%)}\\ 
         \cmidrule(lr){2-6} \cmidrule(lr){7-11}  \cmidrule(lr){12-16}  \cmidrule(lr){17-21} 
        &  0 & 20 & 50 & 80 & 100 & 0 & 20 & 50 & 80 & 100 & 0 & 20 & 50 & 80 & 100 & 0 & 20 & 50 & 80 & 100 \\  
        \midrule
        
        \(\text{CSR}(T_{1, dbt = \texttt{EXE}}\cap T_2)\) & 99.1 & 90.5 & 90.3 & 87.1 & 72.5 & 25.9 & 25.7 & 24.9 & 23.0 & 0.04 & 87.0 & 86.6 & 87.5 & 87.3 & 100.0 & 58.9 & 59.1 & 60.6 & 64.8 & 100.0 \\
         \(T_{1, dbt = \texttt{EXE}} \cap T_2\) & 99.4 & 89.7 & 89.9 & 87.1 & 72.5 & 25.5 & 25.3 & 24.4 & 22.6 & 0.0 & 86.9 & 86.2 & 86.5 & 87.1 & 100.0 & 58.9 & 58.9 & 60.4 & 62.9 & 100.0 \\
         \midrule

          \(\text{CSR}(T_{1, dbt = \texttt{EXE}})\) &  99.8 & 87.1 & 86.1 & 83.3 & 72.5 & 24.0 & 23.6 & 21.7 & 18.1 & 0.04 & 87.8 & 87.5 & 88.7 & 90.2 & 100.0 & 60.9 & 61.5 & 63.9 & 69.7 & 100.0 \\
           \(T_{1, dbt = \texttt{EXE}}\) & 99.8 & 87.3 & 86.2 & 83.0 & 72.5 & 23.8 & 23.2 & 21.5 & 18.1 & 0.0 & 87.8 & 87.8 & 88.5 & 90.3 & 100.0 & 60.9 & 61.4 & 64.6 & 71.3 & 100.0 \\
            \(\text{CSR}(T_2)\) & 86.8 & 85.5 & 85.7 & 84.1 & 72.5 & 16.9 & 17.0 & 16.9 & 16.9 & 0.04 & 97.2 & 97.1 & 96.1 & 94.0 & 100.0 & 79.4 & 79.1 & 77.7 & 75.9 & 100.0 \\
               \(T_2\) & 86.4 & 84.9 & 84.8 & 83.8 & 72.5 & 16.5 & 16.6 & 16.6 & 16.4 & 0.0 & 97.3 & 96.7 & 95.8 & 92.9 & 100.0 & 80.1 & 79.6 & 78.0 & 75.8 & 100.0 \\
               \midrule

           \(\text{CSR}(T_{1, dbt = \texttt{NEXE}}\cap T_2)\) & 99.2 & 72.7 & 74.1 & 70.4 & 72.5 & 35.3 & 36.6 & 39.2 & 41.7 & 0.04 & 83.7 & 79.4 & 68.4 & 52.2 & 100.0 & 48.4 & 45.5 & 35.9 & 25.0 & 100.0 \\
           \(T_{1, dbt = \texttt{NEXE}} \cap T_2\) & 99.2 & 70.9 & 73.4 & 70.4 & 72.5 & 34.9 & 36.6 & 38.6 & 41.1 & 0.0  & 83.4 & 77.3 & 66.8 & 53.9 & 100.0 & 48.5 & 43.6 & 36.3 & 26.9 & 100.0 \\
           \(\text{CSR}(T_{1, dbt = \texttt{NEXE}})\) & 99.8 & 72.3 & 73.2 & 69.8 & 72.5 & 34.9 & 36.3 & 39.1 & 41.4 & 0.04 &  83.7 & 78.1 & 66.2 & 51.8 & 100.0 & 49.0 & 43.9 & 34.4 & 24.8 & 100.0 \\
   \(\text{CSR}(\tau_f)\) & 72.5 & 73.4 & 72.8 & 72.8 & 72.5 & 0.04 & 0.04 & 0.04 & 0.04 & 0.04 & 100.0 & 100.0 & 100.0 & 100.0 & 100.0 & 100.0 & 100.0 & 100.0 & 100.0 & 100.0 \\
       \(\tau_f\)& 72.5 & 73.4 & 72.8 & 72.8 & 72.5 & 0.0 & 0.0 & 0.0 & 0.0 & 0.0 & 100.0 & 100.0 & 100.0 & 100.0 & 100.0 & 100.0 & 100.0 & 100.0 & 100.0 & 100.0 \\
        \(T_{1, dbt = \texttt{NEXE}}\) & 99.8 & 70.6 & 71.2 & 70.3 & 72.5 & 34.6 & 36.0 & 38.8 & 40.7 & 0.0 & 83.7 & 78.0 & 66.0 & 55.5 & 100.0 & 49.2 & 44.5 & 35.2 & 28.8 & 100.0 \\
        \bottomrule
    \end{tabular}
    }
    }
    \caption{\protect\addedtext{Ablation study on the impact of masking static log statements on the performance of estimated execution traces. Results are reported at masking fractions of 0\%, 20\%, 50\%, 80\%, and 100\%, averaged over five repetitions on our dataset of 785 faulty test cases.}}
    \label{table::trace::ablation}
\end{table*}

We further apply the non-parametric Friedman statistical test\cite{friedman1937use} to determine whether at least one estimated trace is statistically different from other traces in terms of MPF1. The test result confirms this hypothesis ($\text{p-value} < 0.001$). Subsequently, we conduct a post-hoc analysis by comparing every pair of traces using the paired Wilcoxon Signed-Rank test\cite{wilcoxon1945test}. This analysis reveals three distinct groups of traces, visually separated by horizontal lines in \autoref{table2}, with traces within each group showing no statistically significant differences in MPF1.

\subsubsection{Pruning Rate} For each estimated trace, we define the pruning rate as the proportion of lines excluded from the trace relative to the total number of lines. The pruning rates are presented in the third column of \autoref{table2}. As expected, variants of \(T_1\) traces that set the execution state of doubtful statements to non-executed prune the most. For example, \(\text{CSR}(T_{1,dbt=\text{NEXE}} \cap T_2)\) achieves a pruning rate of 35.7\%. 
Moreover, the traces generated by intersecting the executed states of \(T_{1,dbt=\text{EXE}}\) and \(T_2\) prune more lines than either of them individually. The lowest pruning rates are observed for \(\tau_f\) and \(\text{CSR}(\tau_f)\), with no or minimal pruning, respectively. 

\subsubsection{Fault Preservation Ratio} The final metric we define and compute in this section is the fault preservation ratio, which measures the overlap between the faulty lines and the lines retained in the estimated traces. Ideally, the estimated traces should retain most of the faulty lines, as these represent the ground truth and must remain within the search space of the FL technique. To quantify how well our traces preserve faulty lines, we define two types of fault preservation ratios: partial and full. The partial ratio reflects whether \emph{at least one} faulty line is present in the estimated trace, while the full ratio provides a stricter criterion, requiring that \emph{all} faulty lines be included. 
The last two columns of \autoref{table2} present the average values of this metric across all test cases for each trace. Excluding \(\tau_f\) and \(\text{CSR}(\tau_f)\), which either include all lines or prune only negligibly, \(T_2\) and \(\text{CSR}(T_2)\) achieve the highest fault preservation ratios. They retain at least one faulty line in 97.3\% of the test cases and all faulty lines in 80.1\% of the cases. Although this increased fault preservation comes at the cost of a lower pruning rate for \(T_2\) by about 7\% compared to \(T_1\) variants in the same statistical group, it delivers a notable 20\% improvement for \(T_2\) over \(T_1\) in the full fault preservation ratio.

\annotate{C1.4 \& part of C2.1}\addedtext{\subsubsection{Ablation Study}\label{trace::ablation} Since our execution trace estimation algorithms rely on a set of static log statements as input, variations in this set can directly influence the accuracy of the estimated traces. We conducted an ablation study in which 0\%, 20\%, 50\%, 80\%, and 100\% of the static log statements were removed from the test code in different rounds. This ablation study simulates varying levels of log message quality and availability, as fewer static messages reduce the number of matches, revealing how our trace estimation performs under different logging conditions. \newannotate{C2.3}\newadded{Note that this study does not demonstrate the feasibility of log manipulation, as this is beyond its scope. It is strictly based on data collected from a single execution using the initial logging configuration. It varies the amount of known execution information provided to the trace estimation algorithms, but does not simulate or validate the effects of modifying logging behavior across runs. In other words, log manipulation in this study involves only masking (hiding) existing logs rather than adding new ones, which is impractical due to the execution costs or failure irreproducibility.}Similar to computing MPF1 in \autoref{subsub::mpf1}, we employed random sampling with stratified selection in each round to maintain the overall distribution of executed and non-executed static log statements, repeating each round five times.
The evaluation of the 0\% masking rate, in particular for computing MPF1, differs from the other masking rates, as fully explained in \autoref{subsub::mpf1}.

Results of this study are presented in \autoref{table::trace::ablation}. As expected, when all static log statements are present, the trace estimation algorithms achieve the highest accuracy (MPF1). As log statements are progressively removed, this accuracy decreases. At the 80\% masking rate---the condition under which static log statements are available in the smallest proportion---the accuracy of our best estimation algorithm in terms of MPF1 declines by only about 12 percentage points, from 99\% to 87\%. Notably, even when no static log statements are available in the test code, our trace estimation algorithms still achieve 72.5\% accuracy. This is because, in the absence of log information, all test code lines are estimated as executed (see \autoref{algorithm1}, lines 8--10, and \autoref{algorithm2}, lines 5--7), and due to this over-approximation, no executed line is missed by our algorithms. \newannotate{C1.2}\newadded{In addition to completely static log-free test cases represented by 100\% static log statement removal in this study, this setting also captures other scenarios, such as cases where no log message can be \emph{uniquely} matched to a static log statement (e.g., due to repetitive generic messages like ``Test failed" in the log). From the perspective of our approach, these two scenarios are fundamentally the same, as both introduce ambiguity in log matching. The former represents ambiguity in terms of quantity, where no static log messages are available, while the latter represents ambiguity in terms of log quality, where no log message can be \emph{uniquely} matched to a static log statement. Our technique handles log-matching ambiguity in the same manner, regardless of its cause, by considering only log messages that are either uniquely matched to a single static log statement, marking that statement as executed, or not matched at all, yielding the set of non-executed statements. When no such log message is available, the approach conservatively assumes that all lines are executed.}

The pruning rate remains relatively consistent across different removal fractions of static log statements, except for the 100\% removal case, where the pruning rate is either zero because the entire test code is estimated as executed, or negligible due to the removal of helper function definitions that are not invoked anywhere in the test code.
For traces of the \(T_{1,dbt=\texttt{EXE}}\) variant, less pruning occurs at higher masking rates. In contrast, for traces of the \(T_{1,dbt=\texttt{NEXE}}\) variant, more pruning is observed at higher masking rates. This behavior aligns with the design goals of the two algorithms: the former adopts a more conservative approach, tending to classify lines as executed when information is missing, whereas the latter tends to classify lines as non-executed under the same conditions. 
Fault preservation ratios are inversely related to the pruning rate, with higher pruning rates generally resulting in lower fault preservation. 
}

\subsubsection{Subset of Estimated Traces}\label{trace::selection}
Evaluating all 12 estimated traces to measure their impact on FL accuracy and efficiency is impractical. Therefore, we select a representative subset of four traces for the experiments presented in the remainder of the paper. To ensure diversity, we include \(\tau_f\) and one trace from each of the three trace groups identified as statistically distinct based on the MPF1 metric, as shown in our statistical analysis and in \autoref{table2}. These selected traces differ not only in MPF1 but also in pruning rates and fault preservation ratios. We select the following four estimated traces. 
\begin{enumerate}
    \item \(\tau_f\): This unpruned version of the faulty test case serves as the baseline for evaluating FL performance without pruning. Given its zero pruning rate, we refer to it as \(T_0\). We exclude \(\text{CSR}(\tau_f)\) from our options due to its similarity to \(\tau_f\) and negligible pruning.
    \item \(T_2\): With the \textbf{highest fault preservation ratio}, this variant---selected from the second statistically distinct group---achieves an MPF1 score comparable to the top-performing group and has a lower pruning rate than other traces. Given its minimal pruning rate, we refer to it as \(T_{min}\). 
    \item \(\text{CSR}(T_{1,dbt=\text{EXE}} \cap T_2)\): With the \textbf{highest MPF1 score}, this variant---selected from the first statistically distinct group--- shows a higher fault preservation ratio than traces from the third group. Its pruning rate is moderate, greater than that of \(T_{min}\) but lower than the third group's traces. Given its moderate pruning rate, we refer to it as \(T_{mid}\).
    \item \(\text{CSR}(T_{1,dbt=\text{NEXE}} \cap T_2)\): With the \textbf{highest pruning rate}, this trace---selected from the third statistically distinct group---has the lowest MPF1 and fault preservation ratio among the selected traces. Given its maximal pruning rate, we refer to it as \(T_{max}\).
\end{enumerate}

The traces described above are used in the experiments to address the second and third research questions.

\Finding{
Our findings for RQ1 demonstrate that our estimated trace \(\text{CSR}(T_{1,dbt=\text{EXE}} \cap T_2)\) achieves the highest MPF1 accuracy, reaching nearly 90\%, while preserving at least one faulty location in the FL search space for more than 87\% of the test cases. In terms of pruning effectiveness, \(\text{CSR}(T_{1,dbt=\text{NEXE}} \cap T_2)\) yields the best performance by eliminating nearly one-third of the test code and still maintaining a strong ratio of faulty location preservation ($\approx$ 84\%). These traces offer a viable alternative to the cost and complexity of executing test cases directly.
}

\annotate{C2.2}\addedtext{\subsection{RQ2: The Impact of Estimated Execution Traces on the Effectiveness, Efficiency, and Scalability of LLM-Based TCFL}}\label{rq2}

\addedtext{Using the prompt template outlined in \autoref{approach::prompt}, we query the LLM to localize faults across our dataset of 785 faulty test cases at three levels of code granularity: function, block, and line. 
To evaluate the impact of different input code on LLM-based TCFL, we select the four estimated traces described in \autoref{trace::selection}. We evaluate how pruning the faulty test code affects the FL accuracy, the total number of input tokens processed by the LLM, and its inference time.}

\begin{table*}[tbp]
    \centering
    \resizebox{\textwidth}{!}{
    \begin{tabular}{c c cccc cccc cccc cccc cccc}
        \toprule
        \multirow{3}{*}{\shortstack{FL \\ Granularity}}  & 
        \multirow{3}{*}{k}  & 
        \multicolumn{4}{c}{Precision@k (\%)}  & 
        \multicolumn{4}{c}{Recall@k (\%)}  & 
        \multicolumn{4}{c}{Hit@k (\%)}  & 
        \multicolumn{4}{c}{MAP@k (\%)}  & 
        \multicolumn{4}{c}{MRR@k (\%)} \\
        \cmidrule(lr){3-6} \cmidrule(lr){7-10} \cmidrule(lr){11-14} 
        \cmidrule(lr){15-18} \cmidrule(lr){19-22} 
        & & 
        \(T_0\) & \(T_{min}\) & \(T_{mid}\) & \(T_{max}\) & 
        \(T_0\) & \(T_{min}\) & \(T_{mid}\) & \(T_{max}\) & 
        \(T_0\) & \(T_{min}\) & \(T_{mid}\) & \(T_{max}\) & 
        \(T_0\) & \(T_{min}\) & \(T_{mid}\) & \(T_{max}\) & 
        \(T_0\) & \(T_{min}\) & \(T_{mid}\) & \(T_{max}\) \\
        \midrule
        \multirow{2}{*}{Function} 
        & 1  & 
          81.6  & 80.9  & 82.1  & 81.9  & 
          76.3  & 75.5  & 76.7  & 76.4  & 
          81.6  & 80.9  & 82.1  & 81.9  & 
          81.6  & 80.9  & 82.1  & 81.9  & 
          81.6  & 80.9  & 82.1  & 81.9 \\
        & 3  & 
          37.1  & 36.5  & 36.4  & 36.0  & 
          95.7  & 94.7  & 94.0  & 92.2  & 
          97.3  & 96.3  & 95.9  & 94.5  & 
          88.6  & 88.7  & 88.7  & 87.8  & 
          88.8  & 88.9  & 89.0  & 88.0 \\
        \hline
        \multirow{4}{*}{Block} 
        & 1  & 
          65.4  & 65.6  & 66.2  & 65.4  & 
          46.1  & 46.4  & 46.6  & 46.4  & 
          65.4  & 65.6  & 66.2  & 65.4  & 
          65.4  & 65.6  & 66.2  & 65.4  & 
          65.4  & 65.6  & 66.2  & 65.4 \\
        & 3  & 
          33.8  & 34.0  & 32.9  & 32.9  & 
          60.7  & 60.9  & 59.5  & 59.5  & 
          79.3  & 81.0  & 79.2  & 78.8  & 
          71.9  & 72.4  & 71.8  & 70.6  & 
          72.1  & 72.5  & 71.9  & 70.8 \\
        & 5  & 
          24.8  & 25.0  & 24.4  & 24.1  & 
          67.2  & 68.6  & 67.5  & 66.3  & 
          83.8  & 86.2  & 85.5  & 83.8  & 
          71.2  & 72.5  & 71.9  & 71.5  & 
          72.3  & 73.9  & 73.2  & 72.6 \\
        & 10  & 
          17.4  & 16.5  & 16.1  & 15.6  & 
          77.6  & 75.9  & 75.4  & 74.0  & 
          89.9  & 91.2  & 90.8  & 89.9  & 
          70.2  & 71.4  & 71.2  & 71.2  & 
          73.8  & 74.3  & 74.4  & 74.1 \\
        \hline
        \multirow{4}{*}{Line} 
        & 1  & 
          33.7  & 34.1  & 33.5  & 33.2  & 
          13.1  & 13.1  & 13.0  & 12.9  & 
          33.7  & 34.1  & 33.5  & 33.2  & 
          35.2  & 35.6  & 35.1  & 34.7  & 
          35.2  & 35.6  & 35.1  & 34.7 \\
        & 3  & 
          22.8  & 23.0  & 21.0  & 20.3  & 
          21.7  & 21.0  & 19.8  & 18.6  & 
          45.9  & 45.9  & 43.5  & 43.5  & 
          40.0  & 39.6  & 38.9  & 38.6  & 
          40.1  & 40.0  & 39.2  & 38.9 \\
        & 5  & 
          18.6  & 17.8  & 18.0  & 16.2  & 
          26.0  & 25.2  & 25.7  & 23.3  & 
          52.7  & 51.1  & 52.3  & 48.7  & 
          40.8  & 40.4  & 40.6  & 39.3  & 
          41.5  & 41.2  & 41.3  & 40.0 \\
        & 10  & 
          14.2  & 14.6  & 12.6  & 11.9  & 
          32.8  & 33.6  & 29.8  & 29.6  & 
          57.8  & 60.8  & 56.6  & 57.1  & 
          40.3  & 40.5  & 39.9  & 39.5  & 
          42.1  & 42.6  & 41.4  & 41.2 \\
        \bottomrule
    \end{tabular}
    }
    \caption{The average top-k performance of fault localization at different test code granularity levels using Qwen2.5 72B.}
    \label{table3}
\end{table*}

\begin{table*}[tbp]
    \centering
    \resizebox{\linewidth}{!}{
    \begin{tabular}{cc cccccccc cccccccc}
    \toprule
    \multirow{4}{*}{\shortstack{FL \\ Granularity}} & \multirow{4}{*}{k} 
    & \multicolumn{8}{c}{Average Token Counts} 
    & \multicolumn{8}{c}{Inference Time} \\
    \cmidrule(lr){3-10} \cmidrule(lr){11-18}
    & 
    & \multicolumn{2}{c}{\(T_0\)} 
    & \multicolumn{2}{c}{\(T_{min}\)} 
    & \multicolumn{2}{c}{\(T_{mid}\)} 
    & \multicolumn{2}{c}{\(T_{max}\)} 
    & \multicolumn{2}{c}{\(T_0\)} 
    & \multicolumn{2}{c}{\(T_{min}\)} 
    & \multicolumn{2}{c}{\(T_{mid}\)} 
    & \multicolumn{2}{c}{\(T_{max}\)} \\
    \cmidrule(lr){3-4} \cmidrule(lr){5-6} \cmidrule(lr){7-8} \cmidrule(lr){9-10}
    \cmidrule(lr){11-12} \cmidrule(lr){13-14} \cmidrule(lr){15-16} \cmidrule(lr){17-18}
    & 
    & In & Out 
    & In & Out 
    & In & Out 
    & In & Out
    & Avg (s) & Sum 
    & Avg (s) & Sum 
    & Avg (s) & Sum 
    & Avg (s) & Sum \\
    \midrule
	\multirow{2}{*}{Function}
		& 1 & 5.31k & 24 & 4.49k & 24 & 4.16k & 24 & 3.77k & 24 & 17.9 & 3h 54m & 15.0 & 3h 15m & 14.2 & 3h 5m & 12.6 & 2h 44m \\ 
		& 3 & 5.31k & 38 & 4.49k & 37 & 4.16k & 37 & 3.77k & 37 & 19.1 & 4h 9m & 16.1 & 3h 29m & 15.0 & 3h 15m & 13.6 & 2h 57m \\ 
		\hline
		\multirow{4}{*}{Block}
		& 1 & 6.43k & 15 & 5.35k & 15 & 4.92k & 15 & 4.42k & 16 & 21.5 & 4h 41m & 17.3 & 3h 46m & 15.8 & 3h 26m & 14.2 & 3h 5m \\ 
		& 3 & 6.43k & 27 & 5.35k & 27 & 4.92k & 27 & 4.42k & 27 & 22.6 & 4h 54m & 18.3 & 3h 59m & 16.8 & 3h 39m & 15.8 & 3h 26m \\ 
		& 5 & 6.43k & 38 & 5.35k & 38 & 4.92k & 38 & 4.42k & 41 & 23.6 & 5h 8m & 19.5 & 4h 14m & 17.8 & 3h 52m & 16.3 & 3h 32m \\ 
		& 10 & 6.43k & 66 & 5.36k & 68 & 4.93k & 71 & 4.42k & 73 & 26.2 & 5h 41m & 21.9 & 4h 46m & 20.9 & 4h 32m & 18.9 & 4h 6m \\ 
		\hline
		\multirow{4}{*}{Line}
		& 1 & 6.18k & 24 & 5.16k & 24 & 4.75k & 24 & 4.26k & 22 & 21.4 & 4h 39m & 17.3 & 3h 46m & 16.0 & 3h 28m & 14.1 & 3h 4m \\ 
		& 3 & 6.18k & 73 & 5.16k & 70 & 4.75k & 69 & 4.26k & 67 & 25.7 & 5h 35m & 21.3 & 4h 38m & 19.8 & 4h 18m & 18.0 & 3h 54m \\ 
		& 5 & 6.18k & 113 & 5.16k & 112 & 4.75k & 112 & 4.26k & 108 & 29.3 & 6h 23m & 25.0 & 5h 26m & 23.8 & 5h 11m & 21.9 & 4h 46m \\ 
		& 10 & 6.18k & 216 & 5.16k & 212 & 4.75k & 210 & 4.26k & 203 & 38.7 & 8h 25m & 33.7 & 7h 20m & 32.1 & 6h 59m & 29.6 & 6h 26m \\ 
        \bottomrule
    \end{tabular}
    }
    \caption{Average token count along with average and cumulative inference time using Qwen2.5 72B.}
    \label{table4}
\end{table*}

To prevent the LLM from generating excessively long or endless outputs, which can be a sign of hallucination, we set the maximum output length to 2,048 tokens.
We conduct our initial TCFL experiments using \texttt{Qwen2.5} with 72 billion parameters\footnote{\underline{\url{https://huggingface.co/Qwen/Qwen2.5-72B}}}.
To assess the effectiveness of LLM-based TCFL, we employ the metrics Precision@\(k\) and Recall@\(k\), which evaluate performance based on the counts of true positives (faulty elements correctly identified), false positives (non-faulty elements incorrectly flagged as faulty), and false negatives (faulty elements missed by the model) among the top-\(k\) elements predicted as faulty. We select different
\(k\) values of 1, 3, 5, and 10 for block-level and line-level FL experiments based on the advice of test engineers from our industry partner. Since the total number of functions per test case is smaller---averaging only six in our benchmark---we limit the function-level FL experiments to \(k\) values of 1 and 3.

Similar to previous FL techniques\cite{DBLP:journals/jss/MiryeganehHH21, DBLP:journals/tse/WenCTWHHC21, DBLP:conf/compsac/LiBWL20}, in addition to the precision and recall rates, we also evaluate the performance using three additional metrics: Hit@\(k\), MAP@\(k\), and MRR@\(k\).
The Hit@\(k\) metric represents the proportion of test cases in which the LLM successfully ranks at least one faulty element among its top-\(k\) predicted outputs. MAP@\(k\) (Mean Average Precision at \(k\)) computes the mean of the average precision scores for correctly identified faulty elements among the top-\(k\) results across all test cases, while MRR@\(k\) (Mean Reciprocal Rank at \(k\)) calculates the average reciprocal rank of the first correctly identified faulty element within the top-\(k\) results across test cases. Formally, these metrics are defined as follows.

\begin{equation}
\text{AP@}k_i = \frac{\sum_{j=1}^{k} P_i(j) \cdot \text{rel}_i(j)}{m_i} \quad
\text{MAP@}k = \frac{\sum_{i=1}^{N} \text{AP@}k_i}{N}
\end{equation}

\begin{equation}
RR_i = \frac{1}{\text{rank}_i} \quad\quad
\text{MRR@}k = \frac{\sum_{i=1}^{N} RR_i}{N}
\end{equation}

Here, \(N\) denotes the total number of faulty test cases. For the \(i\)th test case, \(m_i\) is the number of faulty elements correctly identified by the LLM among its top-\(k\) outputs. \(P_i(j)\) represents the precision at position \(j\), and \(\text{rel}_i(j)\) is 1 if the element at position \(j\) is faulty, and 0 otherwise. 
\(\text{rank}_i\) denotes the rank of the first correctly identified faulty element in the top-\(k\) results, and \annotate{C2.4 (c)}\addedtext{its \(RR_i\)} is set to 0 if no such element is found. Higher values of Hit@\(k\), MAP@\(k\), and MRR@\(k\) indicate better performance of the LLM-based TCFL technique.

\autoref{table3} presents the evaluation results of our LLM-based TCFL approach using different values included in the prompt template. The first column lists three levels of TCFL granularity: function, block, and line. The second column denotes \(k\), the number of faulty elements the LLM is prompted to identify, ranked in descending order of suspiciousness. Each evaluation metric is accompanied by four columns, corresponding to different variants of the input code pruned based on the estimated traces described earlier. 

As shown, our FL technique achieves the highest performance at the function level across all evaluation metrics, followed by the block level, with the line level showing lower performance than the other two. These results can be attributed to the increase in the LLM's search space (total number of elements) at finer levels of granularity, as also visible in \autoref{table::benchmark}. On average, our dataset contains approximately 6 functions, 40 blocks, and 244 lines, with 24.5\%, 9.5\%, and 3.2\% of these being faulty, respectively. These proportions indicate a higher likelihood of detecting faulty functions, since nearly one in every four functions is faulty, compared to about one in 10 blocks and one in 33 lines. 

Increasing the value of \(k\) leads to a decrease in precision and an increase in recall. This trade-off is expected because requesting the LLM to identify more faulty elements increases the likelihood of capturing actual faults, thereby improving recall by reducing false negatives caused by missing faulty elements. However, precision declines as \(k\) increases, since more elements are flagged, increasing the chance of including non-faulty elements in the output.

In terms of the test code, we observe no significant differences across the variants \(T_0\), \(T_{min}\), \(T_{mid}\), and \(T_{max}\). This suggests that our approach effectively prunes unnecessary parts of the test code that are not executed, while preserving fault-identification capability and maintaining the performance of the FL technique. 

While the optimal configuration for running LLM-based TCFL may vary depending on the specific needs and datasets of each debugging team, the overall findings indicate that the top-3 block-level setting using \(T_{min}\) strikes an effective balance between precision and recall for FL.
Compared to top-1, it achieves higher recall, and unlike top-5 and top-10, it prevents a drop in precision. Furthermore, using code blocks provides a focused and informative scope for fault investigation, as they cover a smaller range of statements than functions while offering more context than individual lines. Finally, selecting \(T_{min}\) as the input trace for this setting yields slightly better performance across all metrics.

\annotate{C2.2}\addedtext{To evaluate how pruning the input test code affects the scalability and efficiency of our technique, we measure both the token usage and the LLM's inference time.}
The reported inference times exclude the model-loading time, which is typically high during the first inference when the model has not yet been loaded into the GPU.
Results are presented in \autoref{table4}. On average, \(T_{max}\) at the function level results in the fewest total input tokens, with approximately 3,770 tokens per test case. 
As expected, the total output token count rises as the number of requested outputs (\(k\)) increases, with line-level outputs producing the largest number of tokens due to the inclusion of both line numbers and their content.

Inference time increases with larger output and input token counts. For example, at the line level, asking the LLM to identify one faulty line in \(T_{max}\) takes around 14.1 seconds per test case. This time nearly doubles to 29.6 seconds when requesting 10 faulty lines, even though the input token count remains constant at an average of 4,260 tokens. This illustrates that the LLM's FL efficiency is directly affected by the value of \(k\), with line-level granularity most affected. 

Our estimated traces also significantly reduce inference time by lowering the input token count, with the most noticeable improvement observed when using \(T_{max}\). The isolated effect of input token count on inference time becomes evident when comparing different trace variants at the same \(k\) value. For each \(k\), a clear decreasing trend in inference time is observed from \(T_0\) to \(T_{max}\), with traces having higher pruning rates resulting in lower inference times. Specifically, \(T_{max}\) leads to an inference time reduction of 24\% to 34\% on average per test case compared to the unpruned version \(T_0\). 
As demonstrated in the performance results presented in \autoref{table3}, there is no significant difference in effectiveness across pruning levels. This indicates that using pruned traces can achieve similar performance to the original test case while requiring less time.

In addition to reporting average inference times, we also include the total time required for each FL setting across the full dataset of 785 test cases. This provides a better sense of the time savings achievable at scale. For example, identifying three faulty blocks using the original test case \(T_0\) takes 294 minutes, while using \(T_{max}\) reduces this to 206 minutes, representing a 30\% reduction in time. Similar improvements are observed across different variants. The more the test case is pruned, the faster the localization process becomes, and the fewer tokens are required. 

\Finding{Our findings for RQ2 demonstrate that our proposed LLM-based approach, which, to the best of our knowledge, is the first TCFL technique applied to industrial test cases, is capable of identifying faults effectively at various levels of code granularity. Our approach prunes unnecessary test code without execution while preserving FL performance. \annotate{C2.2}\addedtext{While we do not indeed observe substantial differences in FL accuracy across the variants of our estimated traces, they provide significant speedups (up to 34\%) and significantly reduce token usage (up to 31\%), mitigating the risk of exceeding the LLM's context window limit and decreasing the cost for commercial LLMs. In particular, \(\text{CSR}(T_{1,dbt=\text{NEXE}} \cap T_2)\), which applies the most extensive pruning, achieves the greatest reduction in inference time---up to 34\%--- thereby enhancing localization efficiency and mitigating the risk of exceeding the LLM's context window limit.} As explained above, we propose top-3 block-level FL using our CFG-based pruning technique as a pragmatic FL setting, achieving 34\% precision, 60.9\% recall, and 81\% hit rate. The practical implications of these results will be discussed in \autoref{Discussion}. }

\begin{table*}[tbp]
    \centering
    \resizebox{\textwidth}{!}{
     \annotate{C1.2 \& C2.3}\addedtext{
    \begin{tabular}{c c cccc cccc cccc cccc cccc}
        \toprule
        \multirow{3}{*}{\shortstack{FL \\ Granularity}}  & 
        \multirow{3}{*}{k}  & 
        \multicolumn{4}{c}{Precision@k (\%)}  & 
        \multicolumn{4}{c}{Recall@k (\%)}  & 
        \multicolumn{4}{c}{Hit@k (\%)}  & 
        \multicolumn{4}{c}{MAP@k (\%)}  & 
        \multicolumn{4}{c}{MRR@k (\%)} \\
        \cmidrule(lr){3-6} \cmidrule(lr){7-10} \cmidrule(lr){11-14} 
        \cmidrule(lr){15-18} \cmidrule(lr){19-22} 
        &  
         & TCFL & FauxPy & IRFL & FlexFL 
         & TCFL & FauxPy & IRFL & FlexFL 
         & TCFL & FauxPy & IRFL & FlexFL 
         & TCFL & FauxPy & IRFL & FlexFL 
         & TCFL & FauxPy & IRFL & FlexFL\\
        \midrule
        \multirow{3}{*}{Function} 
        & 1  & 
          80.9  & 82.7  & 86.2 & 83.2
             &  75.5 &  77.0 & 80.6 & 77.7  
             &   80.9  &  82.7 & 86.2 & 83.2
             & 80.9  & 82.7  & 86.2 &  83.2
             & 80.9  & 82.7  &  86.2 & 83.2 \\
        & 3  & 
           36.5  & 33.8  & 35.8 & 33.9 & 
           94.7  &  82.2 &  96.2 & 91.3 &
           96.3  &  87.4 & 97.8  & 93.8 &
           88.7  & 84.9  & 91.2  & 87.6 &
           88.9  &  84.9 & 91.4 & 87.9 \\
     & 5 & 21.9 & 21.4 & 22.3 & 21.8
        & 97.1 & 82.2 & 99.0 & 96.4
           & 97.8 & 87.4 & 99.4 & 97.3
           & 88.7 & 84.9 & 91.1 & 87.6
           & 89.4 & 84.9 & 91.7 & 88.8\\
        \hline
        \multirow{3}{*}{Block} 
        & 1  & 
          65.6  & 64.9  &   58.3 & 58.3 &
           46.4  &  46.4 &  41.3 & 42.0 &
           65.6  &  64.9 & 58.3   & 58.3 &
           65.6  & 64.9  &  58.3 & 58.3 &
           65.6  &  64.9 &  58.3 & 58.3 \\
        & 3  & 
          34.0  & 26.1  &  31.2 & 31.9 &
          60.9  &  49.0 & 57.6 & 56.7 &
           81.0 &  68.1 & 73.4  & 71.5 &
           72.4  &  66.2 &  64.1 & 63.6 &
           72.5  & 66.2  & 64.9 & 64.0 \\
        & 5  & 25.0 & 16.5 & 24.6  & 23.4
        & 68.6 &  49.1 & 68.3  & 63.6
        &  86.2 & 68.1  & 82.8 & 77.1
        &  72.5 & 66.2 & 64.8 & 64.1
        & 73.9  &  66.1 & 67.1 & 65.3\\
        \hline
        \multirow{3}{*}{Line} 
        & 1  & 
          34.1  &  34.4 &  11.1 & 31.3 &
          13.1  & 13.8  &  2.7  & 11.8 &
           34.1  &  34.4 & 11.1  & 31.3 &
          35.6  &  34.4 & 11.5  & 32.7 &
          35.6  & 34.4  & 11.5  & 32.7\\
        & 3  & 
        23.0  &  13.4 & 11.8 & 24.3 &
        21.0  &  14.8 &  8.8 & 22.3 &
          45.9  &  36.7 & 24.1 & 45.5 &
           39.6  &  35.4 & 16.8  & 38.7 &
         40.0  &  35.4 & 16.8  & 39.0 \\
        & 5  & 
           17.8  &  8.3 & 11.2 & 19.7 &
          25.2 &  14.9 & 12.6  & 26.4 &
        51.1  &  37.0 & 29.6  & 50.8 &
        40.4  &  35.5 &  18.2 & 39.4 &
         41.2  &  35.5 &  18.1 & 40.1\\
        \bottomrule
    \end{tabular}
    }
    }
    \caption{\protect\addedtext{Performance comparison of our test code fault localization using the estimated trace \(T_{min}\) (TCFL) against FauxPy, IRFL, and FlexFL across different test code granularity levels. The LLM employed in both TCFL and FlexFL is Qwen2.5 72B.}}
    \label{table::baseline::compare}
\end{table*}

\begin{table}[tbp]
    \centering
    \scriptsize
    \resizebox{0.95\linewidth}{!}{
    \annotate{C1.2 \& C2.3}\addedtext{
    \begin{tabular}{c cccc cccc}
    \toprule
    \multirow{4}{*}{\shortstack{FL \\ Granularity}}  
    & \multicolumn{4}{c}{Average Token Counts} 
    & \multicolumn{4}{c}{Inference Time} \\
    \cmidrule(lr){2-5} \cmidrule(lr){6-9}
    &  \multicolumn{2}{c}{TCFL} 
    & \multicolumn{2}{c}{FlexFL} 
    & \multicolumn{2}{c}{TCFL} 
    & \multicolumn{2}{c}{FlexFL} \\
    \cmidrule(lr){2-3} \cmidrule(lr){4-5} \cmidrule(lr){6-7} \cmidrule(lr){8-9}
    & In & Out
    & In & Out
    & Avg (s) & Sum 
     & Avg (s) & Sum\\
    \midrule
	Function & 4.49k & 49 & 63.03k & 913 & 17.4 & 3h 47m & 121.2 & 26h 26m  \\ 
		Block
		  & 5.35k & 38 & 54.85k &   859  & 19.5 & 4h 14m & 105.3 & 22h 58m \\ 
		Line
		  & 5.16k & 112 & 45.98k & 871 & 25.0 & 5h 26m & 97.7 & 21h 18m \\ 
        \bottomrule
    \end{tabular}
    }
    }
    \caption{\protect\addedtext{Comparison of the average token count, as well as the average and cumulative inference times for function, block, and line-level FL at \(k=5\), between FlexFL and our technique using the estimated trace \(T_{min}\) with Qwen2.5 72B.}}
    \label{table::baseline::efficiency}
\end{table}

\annotate{C1.2 \& C2.3}\addedtext{\subsection{RQ3: LLM-Based TCFL vs. State-of-the-Art Techniques}\label{rq3}
We evaluate our TCFL technique against the three baselines introduced in \autoref{sub::baselines} using the estimated trace \(T_{min}\) identified in \autoref{rq2} as a practical setting. Similar to FlexFL, for each baseline, we collect FL results for \(k=5\) and compute the accuracy metrics for \(k=1\) and \(k=3\) in a post-processing phase. We use the \texttt{Qwen 2.5 72B} model for both TCFL and FlexFL. 
}

\addedtext{The performance results are presented in \autoref{table::baseline::compare}. At the function level, TCFL achieves competitive accuracy, with precision comparable to all three baselines while outperforming FauxPy in recall and hit rates.
For block-level FL with \(k=1\), our technique performs on par with FauxPy while improving upon IRFL and FlexFL by more than 7 percentage points in precision, hit, MAP, and MRR. The advantage of TCFL over FauxPy becomes more pronounced at larger \(k\) values. This outcome is expected, as FauxPy is limited to identifying only error locations as faults and cannot capture all faults when the root causes extend beyond the error locations. TCFL maintains its advantage over IRFL and FlexFL at higher \(k\) values for block-level FL, with a MAP@5 of 72.5\% versus roughly 64\% for these two baselines.
The same trend is observed at the line-level, with IRFL exhibiting the lowest accuracy. This stems from the small document size in line-level IRFL, as each line of code is treated as an individual document, offering limited useful information when matched against the larger error traceback query. These results highlight the limitations of information-retrieval-based techniques and underscore the advantage of leveraging LLMs for the FL task.}

\addedtext{In addition to measuring the FL accuracy for TCFL and the baselines, we compare the token usage and LLM's inference time between our technique and FlexFL. Results are shown in \autoref{table::baseline::efficiency}. As shown, FlexFL consumes a substantially larger number of tokens and consequently takes much longer to run (e.g., an average of 17.4 seconds per test case for TCFL versus 121.2 seconds for FlexFL at the function level). This heavy token usage and slow processing also appear at the block and line levels. This outcome is expected given FlexFL's chain of LLM invocations resulting from its interactive function-calling workflow.}

\Finding{\annotate{C1.2 \& C2.3}\addedtext{Our findings for RQ3 demonstrate that TCFL offers significantly greater scalability and efficiency compared to the latest LLM-guided FL technique, FlexFL. Notably, TCFL achieves this while either outperforming or being on par with FlexFL at all granularity levels and across all \(k\) values.}}

\section{Discussion}\label{Discussion}
Beyond addressing the main research questions, we conduct further analyses to provide additional insights that complement the results presented above. For these analyses, we use top-3 block-level FL for \(T_{min}\), established as a practical and effective setting in \autoref{rq2}.

\subsection{Applicability}
Let us revisit the practicality of our previously defined FL setting: top-3 block-level FL on test code pruned using control flow graph (CFG) analysis. Defining an optimal FL setting that outperforms all others across every evaluation metric (e.g., precision and recall) is unrealistic. However, we believe that \textbf{pruning test code using CFG analysis} and prompting the LLM to identify \textbf{three} faulty \textbf{blocks} is more effective than other settings. This setting maximizes the benefits of the three components, FL granularity, \(k\), and code pruning, while minimizing their disadvantages. We explain the impact of each component below.

\subsubsection{Block-Level Granularity}
\label{sec:discu:blk-lvl}
As noted earlier, code blocks cover fewer statements than functions but offer more surrounding context than individual lines. The impact of this granularity difference becomes more evident in large datasets with many faulty test cases, such as the dataset provided by our industry partner. For example, if an FL system returns \(k\) faulty functions rather than \(k\) blocks, the scope of investigation for a developer increases significantly. In our dataset of 785 faulty test cases, each test case contains an average of 6 functions, 40 blocks, and 244 lines. This means that function-level FL requires examining approximately 40\(\times k\) lines of code, many of which may not be faulty. In contrast, block-level FL reduces this scope to around 6\(\times k\) lines, a reduction of about 85\%. With a large number of test cases, this leads to substantial time savings.

Conversely, line-level FL outputs only \(k\) lines, which is smaller than the block-level investigation scope. However, given the larger search space in line-level FL and the low ratio of faulty lines, it is less likely to localize all actual faulty lines. In our dataset, about 3\% of lines are faulty, compared to 10\% of blocks and 25\% of functions. As a result, the chance of localizing all faulty elements in the top-\(k\) results is lower for lines. This is also reflected in the precision@\(k\) scores: while function and block-level FL achieve up to 82\% and 65\% precision at top-1, respectively, line-level FL reaches only 34\%. Finally, examining individual lines in isolation rather than entire code blocks can impede effective debugging and repair due to a lack of surrounding code context. In practice, developers rarely limit their focus to a few lines; instead, they analyze entire blocks or functions to trace variables and understand behavior across a broader scope. While blocks may not always provide perfect contextual relevance, they facilitate debugging.

\subsubsection{\texorpdfstring{Choice of \(k\)}{Choice of k}}
When selecting code blocks as the level of granularity for FL, another important variable to determine is the number of elements \(k\) to request from the LLM. Based on input from our industry partner, we experimented with values of 
\(k=\)1, 3, 5, and 10, with a preference for lower values to align with industrial needs. Our findings suggest that, at the block level, \(k=\)3 (i.e., an average of 18 lines out of more than 240 lines in our dataset) is a near-optimal choice. It offers a reasonable trade-off between precision and recall: larger \(k\) values entail lower precision, while smaller values tend to miss faulty blocks, reducing recall, as discussed in \autoref{rq2}.

This choice is also supported by the statistics of our dataset, where the average number of blocks per test case is around 40, and the average faulty block ratio is about 10\%, resulting in about four faulty blocks per test case. Thus, selecting \(k=\)3 is justified among the tested values. 
We believe that choosing a very low \(k\), such as 1 or 2, is overly restrictive, particularly in complex test cases where multiple faulty blocks may exist. This restriction could prevent the TCFL technique from presenting all relevant faulty candidates, forcing developers to manually search for remaining faults, hence increasing their burden.
Conversely, \(k>\)3, and especially values greater than 5, can broaden the investigation scope and reduce productivity by including non-faulty blocks in the scope. Therefore, we conclude that \(k=\)3 strikes a practical balance between completeness and efficiency in our industrial context. However, we acknowledge that the ideal value of \(k\) can vary depending on factors such as the number of lines, blocks, and fault statistics of the codebase. In practice, engineers should analyze these metrics to determine the most suitable \(k\) for their context.

\subsubsection{Pruned Test Code}
The larger a faulty test case is, the broader its FL search space becomes, thereby decreasing the effectiveness and efficiency of identifying faulty locations. Therefore, a smaller pruned test case that still preserves the faults is desirable.
To this end, we proposed three novel algorithms to estimate execution traces and applied them, both individually and in combination, to prune faulty test cases. As thoroughly discussed in \autoref{Evaluation}, each of these traces had an impact on the test case size and, as a result, on the LLM-based FL time, ranging from minor to significant. Despite the pruning, their performance remained comparable to FL on the original, unpruned test cases. These findings generally support the use of our estimated traces to speed up FL by reducing the test code size and complexity. This speed-up can directly enhance developers' workflows, boosting their productivity and efficiency\cite{DBLP:journals/sqj/ParninR11}. Additionally, it can accelerate the development of emerging FL applications, including those focused on models, neural networks, and hardware specifications\cite{DBLP:journals/tim/MohammedMCS21, DBLP:conf/icse/WardatLR21, DBLP:conf/iccd/Wu0YMHML22}.

To illustrate the impact of this speed-up, consider a developer applying automated FL to a batch of test cases. This practice is often both common and necessary, particularly when a large number of test cases are generated automatically, for example by using LLMs or deep learning techniques\cite{DBLP:journals/corr/abs-2305-04764, DBLP:journals/infsof/AlagarsamyTA24, lemieux2023codamosa}, or when the system's code coverage has significantly degraded due to many faulty test cases existing within the test suite, resulting in a large backlog of failing test cases to debug\cite{IMTIAZ20191}.

In our dataset, the FL search space consists of 785 test cases, each averaging 244 lines of code. Without pruning, this amounts to over 190,000 lines, which, despite being processed by an automated LLM-based technique, can considerably prolong inference time. In contrast, applying our estimated traces can reduce this search space by up to 36\%, bringing it down to fewer than 125,000 lines. Eliminating these extra 65,000 lines can significantly decrease FL time, resulting in a reduction of total LLM inference time of up to 2 hours.

Regarding FL performance on precision, recall, and other accuracy metrics, all versions of our estimated traces yielded comparable results. Consequently, we believe that any of our trace estimations, with accuracies (MPF1s) ranging from 73\% to 90\%, are viable without negatively affecting FL accuracy. We selected the trace inferred through CFG analysis (i.e., \(T_{min}\)) because it preserves faults more effectively than the other estimates and showed slightly better overall performance, with approximately a 1\%--2\% improvement across all accuracy metrics in the top-3 block-level results reported in \autoref{table3}. However, this marginal difference does not constitute a significant improvement, suggesting that other traces may also be suitable for this setting.

\begin{table}[tbp]
    \centering
    \resizebox{\columnwidth}{!}{
    \begin{tabular}{ccc c c c c c}
        \toprule
         \multicolumn{3}{c}{Hyperparameter} & \multirow{3}{*}{\shortstack{Precision@k\\(\%)}} & \multirow{3}{*}{\shortstack{Recall@k\\(\%)}} & \multirow{3}{*}{\shortstack{Hit@k\\(\%)}} & \multirow{3}{*}{\shortstack{MAP@k\\(\%)}} & \multirow{3}{*}{\shortstack{MRR@k\\(\%)}}\\  
        \cmidrule(lr){1-3} 
        temperature & top\_k & top\_p &  &  &  &  &  \\
        \midrule
        0.3 & 10 & 0.5 &33.7 & 60.6 & 80.1 & 72.5 & 72.7  \\ 
        0.3 & 40 & 0.9 &33.6 & 60.6 & 80.2 & 72.5 & 72.7  \\ 
        0.3 & 100 & 0.95 &33.7 & 60.6 & 80.3 & 72.4 & 72.6 \\ 
        0.8 & 10 & 0.5 &33.6 & 60.6 & 80.1 & 72.6 & 72.8  \\ 
        0.8 & 40 & 0.9 &33.7 & 60.7 & 80.3 & 72.1 & 72.3  \\ 
        0.8 & 100 & 0.95 &33.7 & 60.9 & 80.6 & 72.1 & 72.2  \\ 
        1.0 & 10 & 0.5 &33.6 & 60.7 & 80.1 & 72.6 & 72.8  \\ 
        1.0 & 40 & 0.9 &33.6 & 60.7 & 80.5 & 72.2 & 72.3  \\ 
        1.0 & 100 & 0.95 &33.3 & 60.5 & 80.4 & 72.1 & 72.3  \\ 
        \bottomrule
    \end{tabular}
    }
    \caption{Results of hyperparameter tuning for top-3 block-level fault localization on \(T_{min}\) using Qwen2.5 72B (averaged across three repetitions).}
    \label{table8}
\end{table}

\begin{table*}[tbp]
    \centering
    \resizebox{\textwidth}{!}{
    \begin{tabular}{l c c c c c c c c c}
        \toprule
         \multirow{-2}{*}{Model} & {\shortstack{Precision@k \\ (\%)}} & {\shortstack{Recall@k\\ (\%)}} & {\shortstack{Hit@k \\ (\%)}} & {\shortstack{MAP@k \\ (\%)}} & {\shortstack{MRR@k \\ (\%)}} & \shortstack{Avg. \# \\Input Tokens} & \shortstack{Avg. \# \\Output Tokens} & \shortstack{Avg. Inference \\Time (Seconds)} & \shortstack{Total Inference \\Time} \\  
        \midrule
	Qwen2.5-72B & 34.0 & 60.9 & 81.0 & 72.4 & 72.5 & 5.35k & 27 & 18.3 & 3h 59m \\ 
	DeepSeek-R1-Distill-Llama-70B & 27.2 & 51.0 & 67.9 & 56.9 & 57.1 & 5.06k & 880 & 89.3 & 19h 26m \\ 
    Qwen2.5-Coder-32B & 32.6 & 58.6 & 78.1 & 68.8 & 69.2 & 5.35k & 27 & 11.2 & 2h 25m \\ 
	Qwen2.5-7B & 26.5 & 46.9 & 66.2 & 54.5 & 54.8 & 5.35k & 26 & 2.9 & 37m 24s \\ 
        \bottomrule
    \end{tabular}
    }
    \caption{The average performance of top-3 block-level fault localization on \(T_{min}\) using different models.}
    \label{table7}
\end{table*}

\subsection{Hyperparameter Tuning}\label{tuning}
To examine how varying LLM inference hyperparameter values affect FL performance, we selected three parameters: \emph{temperature}, \emph{top\_p}, and \emph{top\_k}. We tested nine distinct combinations of their values, each repeated three times. Lower values for these parameters result in more deterministic LLM outputs, whereas higher values increase randomness and diversity. 
We paired top\_p and top\_k at low, medium, and high values to evaluate how jointly adjusting both affects output diversity, given their close interaction.
The temperature parameter was assigned three values independently. \autoref{table8} presents the TCFL results using these combinations. The results show minimal variation across different combinations, suggesting consistent outputs. Thus, hyperparameter tuning appears to have little effect within the ranges explored.

\subsection{Generalizability}
To assess the generalizability of our TCFL technique, we extend our experiments with cross-model validation using three additional LLMs: the smaller \emph{Qwen2.5 7B},\footnote{\underline{\url{https://huggingface.co/Qwen/Qwen2.5-7B}}} the code-oriented \emph{Qwen2.5-Coder-32B},\footnote{\underline{\url{https://huggingface.co/Qwen/Qwen2.5-Coder-32B}}} and \emph{DeepSeek-R1-Distill-Llama-70B},\footnote{\underline{\url{https://huggingface.co/deepseek-ai/DeepSeek-R1-Distill-Llama-70B}}} a recent open-source and popular reasoning model. The results are shown in \autoref{table7}.

As shown, the lowest inference time is achieved by Qwen2.5 7B, the model with the fewest parameters. Interestingly, DeepSeek has the longest inference time and lower performance than Qwen2.5 72B. Specifically, DeepSeek takes more than 19 hours to complete FL across all instances, about 5 times longer than Qwen2.5 72B. Moreover, DeepSeek yields lower accuracy across all metrics: its precision is 27.2\% compared to 34\%, recall is 51\% versus 60.9\%, and it identifies at least one faulty block in 67.9\% of test cases (Hit@3), compared to 81\% for Qwen2.5 72B. Our analysis reveals that DeepSeek's extended reasoning process is the primary cause of its significantly longer inference time. This is evident in its average output token count of 880, compared to only 26--27 tokens generated by the other models. These large outputs also negatively impacted DeepSeek's performance: In 66 test cases, DeepSeek's reasoning output exceeded our predefined 2,048-token limit (see \autoref{rq2}), preventing it from producing the final answer and thereby reducing its FL accuracy.

Another notable observation is the comparable performance of Qwen2.5-Coder 32B to that of Qwen2.5 72B. Qwen2.5-Coder 32B required 145 minutes to perform FL, while Qwen2.5 72B took 239 minutes, representing a 39\% reduction in inference time. Despite this, the differences in their accuracy metrics were marginal, ranging from 1\% to 4\%. This highlights the advantage of using Qwen2.5-Coder 32B to lower computational costs while maintaining similar performance. 

Finally, it is worth noting that the average number of input tokens is the same across all Qwen2.5 models, whereas it differs for DeepSeek. This difference arises from the distinct tokenizers used by these models, with DeepSeek breaking down the same input into fewer tokens. Different tokenizers can also affect the number of tokens in the output. However, the primary reason for the significantly larger outputs in DeepSeek is the additional reasoning tokens included its responses.

\subsection{Direct Prompting (DP) vs. Line-to-Block Mapping (LM)}\label{direct::map}
The inherent challenges of line-level FL, which are discussed in \autoref{sec:discu:blk-lvl}, underscore the need to compare methods that leverage or mitigate these limitations. To this end, we introduce two approaches for performing block-level FL. The first, named \emph{Direct Prompting} (DP), is our existing block-level FL and directly asks the LLM to identify faulty blocks. An alternative approach, named \emph{Line-to-Block Mapping} (LM), prompts the LLM to identify faulty lines and then maps each line to its corresponding block in a post-processing step. Comparing these approaches helps us understand the trade-offs between granular information and broader context, guiding the choice of FL strategy in practical debugging scenarios. To explore potential differences between these two methods, we compared the results of our top-\(k\) block-level FL using \(T_{min}\) with those of line-level FL, where the predicted lines were mapped to blocks, for all values of \(k\). The results of this comparison are presented in \autoref{table5}.

\begin{table}[tbp]
    \centering
    \resizebox{\columnwidth}{!}{
    \begin{tabular}{c *{12}{c}}
        \toprule
         \multirow{2.5}{*}{\(k\)} & \multicolumn{2}{c}{Precision@k (\%)} 
            & \multicolumn{2}{c}{Recall@k (\%)} 
            & \multicolumn{2}{c}{Hit@k (\%)} 
            & \multicolumn{2}{c}{MAP@k (\%)}
            & \multicolumn{2}{c}{MRR@k (\%)}
            & \multicolumn{2}{c}{Avg. \# Blocks} \\
        \cmidrule(lr){2-3} \cmidrule(lr){4-5} \cmidrule(lr){6-7} \cmidrule(lr){8-9} \cmidrule(lr){10-11} \cmidrule(lr){12-13}
         & DP & LM & DP & LM & DP & LM & DP & LM & DP & LM & DP & LM \\
        \midrule
	1 & 65.6 & 67.0 & 46.4 & 47.5 & 65.6 & 67.0 & 65.6 & 67.0 & 65.6 & 67.0 & 1.0 & 1.0 \\ 
	3 & 34.0 & 54.0 & 60.9 & 57.9 & 81.0 & 77.7 & 72.4 & 71.2 & 72.5 & 71.4 & 3.0 & 2.0 \\ 
	5 & 25.0 & 48.1 & 68.6 & 61.8 & 86.2 & 80.6 & 72.5 & 71.9 & 73.9 & 72.3 & 5.0 & 2.7 \\ 
	10 & 16.5 & 42.2 & 75.9 & 67.4 & 91.2 & 85.3 & 71.4 & 72.4 & 74.3 & 73.7 & 9.9 & 3.9 \\ 
        \bottomrule
    \end{tabular}
    }
    \caption{Top-\(k\) performance of block-level fault localization on \(T_{min}\) using Qwen2.5 72B, comparing Direct Block Prompting (DP) and Line-to-Block Mapping (LM).}
    \label{table5}
\end{table}

As shown, LM achieves higher precision than DP, whereas the latter outperforms in recall and hit rate. The performance gap between the two methods on these three metrics widens as \(k\) increases. For the other two metrics, MAP@\(k\) and MRR@\(k\), the difference between the methods is negligible. To investigate the cause of the variation in precision and recall, we analyzed the average number of blocks generated by each method, as shown in the last column of \autoref{table5}. Except for \(k=\)1, where both techniques produce a single block, LM yields fewer blocks on average for \(k=\)3, 5, and 10. We found that the LM approach can lead to repeated identification of the same block, thereby reducing the total number of distinct blocks. This is mainly due to the narrower scope of exploration in line-level FL compared to block-level. In essence, LM functions as a more focused, less exploratory variant of block-level FL: it tends to prioritize precision by pinpointing a smaller set of likely faulty blocks, but this comes at the cost of reduced recall. This balance between precision and recall mirrors the trade-offs discussed earlier in \autoref{rq2} across different values of \(k\).

\subsection{Limitations}\label{limit}

In this section, we acknowledge certain limitations of our technique, outline potential threats to validity, and describe the measures we implemented to address them.

\newannotate{C1.1}\subsubsection{Analysis of Adaptations' Effects on Baselines' Original Behavior}\label{adaptation::analysis}
\newdeleted{While both the system's source code and its test code are typically accessible in SUTFL, black-box TCFL does not assume access to the SUT's source code.}As discussed in \autoref{Problem_Def}, SUTFL and TCFL differ in scope and present distinct challenges, making a direct, like-for-like comparison between them inappropriate.
While prior work has explored the use of LLMs in SUTFL\cite{DBLP:journals/corr/abs-2403-16362, DBLP:conf/icse/YangGMH24, 10.1109/TSE.2025.3553363}, to the best of our knowledge, our approach is the first to apply LLMs to TCFL. \annotate{C1.2 \& C2.3}\deletedtext{Consequently, a comparison with existing state-of-the-art techniques was not feasible.}\addedtext{To enable a comparison with state-of-the-art techniques, we employed the latest LLM-guided SUTFL method, FlexFL\cite{10.1109/TSE.2025.3553363}, along with two traditional SUTFL techniques, IRFL\cite{9610655} and FauxPy\cite{PythonFL-FauxPy-Tool}. We adapted these techniques to fit our problem setting (see \autoref{table::baseline::adaptations}). \newdeleted{Although this does not allow an evaluation against each baseline in its original form, it provides a practical and reasonable framework for comparison, which would not be feasible otherwise.}  \newdeleted{We believe our technique can serve as a foundational baseline for future research in LLM-based TCFL, paving the way for further exploration in this area.}}

\newadded{In this section, we discuss the potential biases or changes in the original methods' behavior introduced by these adaptations. For each adaptation, we either provide a detailed explanation of why it does not raise such concerns or describe the measures we took to mitigate them.}

\begin{table*}[tbp]
    \centering
    \resizebox{\textwidth}{!}{
    \newannotate{C1.1}\newadded{
    \begin{tabular}{c c ccc ccc ccc ccc ccc}
        \toprule
        \multirow{3}{*}{\shortstack{FL \\ Granularity}}  & 
        \multirow{3}{*}{k}  & 
        \multicolumn{3}{c}{Precision@k (\%)}  & 
        \multicolumn{3}{c}{Recall@k (\%)}  & 
        \multicolumn{3}{c}{Hit@k (\%)}  & 
        \multicolumn{3}{c}{MAP@k (\%)}  & 
        \multicolumn{3}{c}{MRR@k (\%)} \\
        \cmidrule(lr){3-5} \cmidrule(lr){6-8} \cmidrule(lr){9-11} 
        \cmidrule(lr){12-14} \cmidrule(lr){15-17} 
        & & 
        c=10 & c=20 & c=30 & c=10 & c=20 & c=30 & c=10 & c=20 & c=30 & c=10 & c=20 & c=30 & c=10 & c=20 & c=30 \\
        \midrule
        \multirow{3}{*}{Block} 
        & 1  & 58.3 & 59.6 & 58.1
          & 42.0  & 42.7 & 41.3
          & 58.3  & 59.6 & 58.1 
          &  58.3 & 59.6 & 58.1
          &  58.3 & 59.6 & 58.1 \\
        & 3  & 31.9 & 33.5 & 33.2 
          &  56.7 & 58.5 & 58.5
          &  71.5 & 74.3 & 74.5
          &  63.6 & 65.5 & 64.6
          &  64.0 & 65.9 & 65.3 \\
        & 5  & 23.4 & 25.2  & 24.5
          &  63.6 & 66.4 & 65.8
          &  77.1 & 79.6 & 80.5
          &   64.1 & 65.8 & 65.1
          &  65.3 & 67.1 & 66.6\\
        \hline
        \multirow{3}{*}{Line} 
        & 1   & 31.3 & 32.4 & 32.6
          &  11.8 & 12.3  & 12.5
          & 31.3  & 32.4 & 32.6
          &  32.7 & 33.9 & 34.0
          &  32.7 & 33.9 & 34.0 \\
        & 3  & 24.3 & 24.5  & 24.8
          &  22.3 & 22.2 & 22.6
          & 45.5  & 46.0 & 46.9
          &  38.7 & 39.1 & 39.7
          &  39.0 & 39.5 & 40.1 \\
        & 5  & 19.7 & 20.1 & 19.8
          &  26.4 & 26.1 & 27.0
          &  50.8 & 50.1 & 51.8
          & 39.4  & 39.4  & 40.4
          &  40.1 & 40.4 & 41.2 \\
        \bottomrule
    \end{tabular}
    }
    }
    \caption{\protect\newadded{FlexFL performance for top-k block- and line-level FL using a maximum of 10, 20, and 30 agent function calls (c).}}
    \label{table::adaptation::func::call}
\end{table*}

\begin{table*}[tbp]
    \centering
    \resizebox{\textwidth}{!}{
    \newannotate{C1.1}\newadded{
    \begin{tabular}{c c cc cc cc cc cc}
        \toprule
        \multirow{3}{*}{Artifact}  & 
        \multirow{3}{*}{k}  & 
        \multicolumn{2}{c}{Precision@k (\%)}  & 
        \multicolumn{2}{c}{Recall@k (\%)}  & 
        \multicolumn{2}{c}{Hit@k (\%)}  & 
        \multicolumn{2}{c}{MAP@k (\%)}  & 
        \multicolumn{2}{c}{MRR@k (\%)} \\
        \cmidrule(lr){3-4} \cmidrule(lr){5-6} \cmidrule(lr){7-8} 
        \cmidrule(lr){9-10} \cmidrule(lr){11-12} 
        & & 
        FlexFL & IRFL & FlexFL & IRFL & FlexFL & IRFL & FlexFL & IRFL & FlexFL & IRFL \\
        \midrule
        \multirow{3}{*}{Error Trace} 
        & 1  & 83.2 & 86.2 & 77.7 & 80.6 & 83.2 & 86.2 & 83.2 & 86.2 & 83.2 & 86.2 \\
        & 3  & 33.9 & 35.8 & 91.3 & 96.2 & 93.8 & 97.8 & 87.6 & 91.2 & 87.9 & 91.4 \\
        & 5  & 21.8 & 22.3 & 96.4 & 99.0 & 97.3 & 99.4 & 87.6 & 91.1 & 88.8 & 91.7 \\
        \hline
         \multirow{3}{*}{Gen Bug Report} 
        & 1  & 80.8 & 81.3 & 75.6 & 75.5 & 80.8 & 81.3 & 80.8 & 81.3  & 80.8 & 81.3\\
        & 3  & 33.5 & 36.0 & 90.0 & 96.2 & 92.6 & 97.5 & 85.8 & 88.5 & 86.0 & 88.8 \\
        & 5  & 21.4 & 22.3 & 94.9 & 98.6 & 96.2 & 99.1 & 86.1 & 88.6 & 86.9 & 89.2\\
        \hline
        \multirow{3}{*}{\shortstack{Error Trace + \\Gen Bug Report}} 
        & 1  & 82.7 & 85.7 & 77.3 & 80.1 & 82.7 & 85.7 & 82.7 & 85.7 & 82.7 & 85.7 \\
        & 3  & 33.2 & 36.1 & 90.0 & 96.6 & 92.7 & 97.8 & 86.9 & 91.1 & 87.1 & 91.4\\
        & 5  & 21.7 & 22.3 & 96.6 & 98.8 &  98.0 &  99.2 & 87.3 & 91.2 & 88.4 & 91.7\\
        \bottomrule
    \end{tabular}
    }
    }
    \caption{\protect\newadded{FlexFL and IRFL performance in top-k function-level FL using error tracebacks, generated bug reports, and their combination.}}
    \label{table::adaptation::gen::bug::report}
\end{table*}

\noindent\newadded{\emph{1. Excluding execution-based FL methods such as SBFL (affected baselines: FlexFL and FauxPy).} Excluding execution-based methods removes the additional information they provide, which can affect the FL accuracy of the baseline, but it does not alter its core behavior. In particular, FlexFL operates in two main phases: (1) generating a candidate list of potentially faulty methods using both traditional non-LLM-based techniques and the LLM-guided approach, and (2) using only the LLM-guided approach to finalize the ranking of suspicious methods. Our adaptation preserves both phases: the candidate list is still generated using traditional techniques (restricted to IRFL in our setting), and the LLM-guided phase is unchanged. The only difference is that the set of traditional FL techniques is reduced---original FlexFL uses IRFL, SBFL, and HybridFL, whereas the adapted baseline uses only IRFL---but the baseline's overall reasoning and final ranking mechanism remain intact. Importantly, this reduction does not affect the size of the candidate list in the first phase: we maintain a total of 20 candidates, with 15 generated by IRFL and 5 by the first-phase LLM (Agent4SR), following a variant already used and discussed in the original FlexFL paper.}

\newadded{The same reasoning applies to FauxPy. Since the input (i.e., the error traceback) is fully preserved, removing execution-based methods affects only the information available for ranking, without altering the technique's original behavior.

Although we preserve the original baselines' intended behavior, we acknowledge a reduction in their FL accuracy due to the absence of execution-based information, a limitation inherent to the TCFL problem rather than to our technique.}

\noindent\newadded{\emph{2. Expanding FL support across multiple granularity levels beyond the method level (affected baselines: FlexFL, IRFL, and FauxPy).} This adaptation is unlikely to alter the original behavior of the baselines. The main reason is that the original granularity level (i.e., method-level FL) is preserved in the adapted versions, and its results are directly compared against our technique. Extending the baselines to support finer-granularity levels is an additional exploration to evaluate their behavior when more detailed analysis is required (see \autoref{sub::problem::importance} for an example).}

\newadded{While this justification mitigates concerns of bias in IRFL and FauxPy, further analysis is required for FlexFL. In particular, at finer granularity, the code snippets available to FlexFL are smaller than at the method level. Therefore, the original setting of a maximum of 10 agent function calls may not provide FlexFL with sufficient context, potentially leading to an unfair comparison with our technique. To address this, in addition to the original limit of 10, we also collected FlexFL's results for block- and line-level FL using increased limits of 20 and 30 function calls before the LLM invokes the \texttt{exit()} function. This enables the LLM to collect additional context at finer granularity levels before reaching the limit, if required. 

Results are presented in \autoref{table::adaptation::func::call}. As shown, increasing the limit to 2 or 3 times the default does not significantly affect FL accuracy. All metrics remain within the same range as the original setting of 10 maximum calls, with only a 1--4\% variation. These results suggest that comparing our technique with FLexFL at finer granularity levels, while retaining the default setting, does not introduce meaningful bias. This is because FlexFL already captures the essential context needed for fault localization at both block and line levels, and providing additional opportunities to gather context does not improve performance. In other words, extending FL support to finer-grained levels in FlexFL does not hinder its ability to collect the necessary context, thereby preserving its original behavior.}

\noindent\newadded{\emph{3. Excluding bug reports from the prompt and data corpus queries (affected baselines: FlexFL and IRFL).}
As described in \autoref{table::baseline::adaptations}, we do not have access to the internal bug-tracking system of test cases in our dataset, which is a common limitation caused by role-based access control in large organizations\cite{sandhu1998role}. Consequently, we exclude bug reports from the prompt template in FlexFL and use error tracebacks instead of bug reports when querying the data corpora in IRFL.

To assess the impact of bug reports on the behavior of these two baselines and reduce the potential bias introduced by their exclusion, we generate synthetic bug reports from error tracebacks and incorporate them into the baselines in a separate set of experiments. For this purpose, we employ the Claude Code agentic coding assistant tool,\footnote{\underline{\url{https://code.claude.com/docs/en/overview}}} enabled via Claude Code Router\footnote{\underline{\url{https://github.com/musistudio/claude-code-router}}} to support open-source models for data privacy. In particular, we use the \texttt{gpt-oss-120b} model\footnote{\underline{\url{https://ollama.com/library/gpt-oss:120b}}}.

To generate bug reports, we run the agentic assistant within a directory containing the failed execution log file and the corresponding failing test scripts. The assistant is instructed to analyze the log file, including error messages and the source code of the test scripts, to produce a structured bug report with a title, description, steps to reproduce, and expected and actual results. Although the resulting reports may differ from user-written bug reports in some cases, particularly in terms of the level of detail or the specific steps provided, this setup represents the closest feasible approximation of a realistic scenario in which a user files a bug report after encountering a failure during test execution.}

\newadded{Once the bug reports are generated, we incorporate them into the FlexFL prompt while keeping all other prompt components unchanged. We also use these reports to query the data corpora in IRFL. The results for function-level FL (i.e., the original configuration of FlexFL and IRFL) are presented in \autoref{table::adaptation::gen::bug::report}. The first three rows of the table present the FL results for FlexFL and IRFL when the bug report is not used: for FlexFL, it is excluded from the prompt, while for IRFL, it is replaced with error tracebacks during corpus querying. Note that, in this case, error tracebacks remain part of the FlexFL prompt, as in the original configuration, where they represent the trigger test component. The next three rows show the FL results when generated bug reports are used in both the FlexFL prompt and the IRFL query. The final three rows combine these inputs by concatenating generated bug reports and available error tracebacks in both the prompt and the corpus queries. As shown, incorporating bug reports, either alone or combined with error tracebacks, does not improve the FL performance of the baselines compared to using error tracebacks alone. The results are very similar, with some cases even showing slight metric degradation when bug reports are included. These findings suggest that error tracebacks in our adapted versions of FlexFL and IRFL provide sufficient information for the baselines to reason about faults, thereby preserving their original behavior and ensuring a fair comparison with our technique.}

\noindent\newadded{\emph{4. Deleting class-related agent function calls and revising the prompt terminology (affected baseline: FlexFL).} These are minor adjustments required when applying FlexFL to a new dataset, particularly when the code is written in a programming language other than Java or follows a different structure. As FlexFL is a generalizable approach that can be applied across diverse datasets and models, such modifications are minimal, necessary to maintain its generalizability, and do not affect its core functionality.}

\subsubsection{Open-Source vs. Commercial LLMs}
To protect data privacy, we conducted our FL experiments using open-source LLMs, which are freely available and effective but generally slower due to limited computational resources. Commercial LLMs, on the other hand, typically achieve higher efficiency due to their greater computational power. As a result, while our technique reduces prompt tokens to lower LLM API call costs, it may not significantly reduce inference time when applied to commercial LLMs. This can be explored in future work using datasets that do not involve privacy concerns, unlike most industrial datasets. However, using such datasets introduces the risk of data leakage, potentially leading to inaccurate evaluations. This concern, which is not typically present in confidential industrial datasets, refers to situations in which LLMs may have been exposed to portions of the data during training, as discussed in \autoref{Problem_Def}.

\begin{table*}[tbp]
    \centering
    \resizebox{\textwidth}{!}{
     \newannotate{C1.3}\newadded{
    \begin{tabular}{c cc cc cc cc cc c}
        \toprule 
        \multirow{2.5}{*}{k}  & 
        \multicolumn{2}{c}{Precision@k (\%)}  & 
        \multicolumn{2}{c}{Recall@k (\%)}  & 
        \multicolumn{2}{c}{Hit@k (\%)}  & 
        \multicolumn{2}{c}{MAP@k (\%)}  & 
        \multicolumn{2}{c}{MRR@k (\%)} & \multirow{2.5}{*}{\shortstack{Avg. Mismatch \\Rate (\%)}} \\
        \cmidrule(lr){2-3} \cmidrule(lr){4-5} \cmidrule(lr){6-7} 
        \cmidrule(lr){8-9} \cmidrule(lr){10-11}   
         & Enabled & Disabled & Enabled & Disabled 
         & Enabled & Disabled & Enabled & Disabled 
         & Enabled & Disabled & \\
        \midrule
        1 & 33.9 & 34.1 & 12.9 & 13.1 & 33.9 & 34.1 & 35.5 & 35.6 & 35.5 & 35.6 & 7.9 \\
        3 & 23.0 & 23.0 & 21.2 & 21.0 & 46.2 & 45.9 & 39.9 & 39.6 & 40.2 & 40.0 & 18.2 \\
        5 & 18.1 & 17.8 & 25.3 & 25.2 & 51.4 & 51.1 & 40.5 & 40.4 & 41.3 & 41.2 & 20.5\\
        10 & 14.7 & 14.6 & 33.9 & 33.6 & 60.8 & 60.8 & 40.5 & 40.5 & 42.6 & 42.6 & 21.5 \\
        \bottomrule
    \end{tabular}
    }
    }
    \caption{\protect\newadded{Line-level FL performance of our test code fault localization using the estimated trace \(T_{min}\), with mismatch mitigation enabled and disabled, across different values of \(k\), along with the corresponding average mismatch rates.}}
    \label{table::mismatch}
\end{table*}

\subsubsection{Output Validity}\label{discuss::output::validity}
LLM-based FL techniques are generally prone to producing invalid outputs. For example, the LLM may return results that are out of range, such as a function name not present in the search space or a block ID \newadded{or line number}that exceeds the available range. \newannotate{C1.3}\newdeleted{In line-level FL, it may generate inconsistent outputs, such as a line number within the valid range that refers to content that does not exist at that position. We refer to such inconsistencies as mismatches.}Repeated outputs or duplicates also occur, which provide no additional useful information.
Although these issues are common across all LLM-based techniques, we attempted to mitigate them by incorporating clear instructions in our prompt template to steer the LLM away from generating invalid outputs. Our LLM-based TCFL technique produces fewer than 1\% out-of-range or duplicate outputs across all levels of granularity.

\newdeleted{In line-level FL, it may generate inconsistent outputs, such as a line number within the valid range that}\newannotate{C1.3}\newadded{In line-level FL, an additional type of inconsistent output, specific to this level of granularity, may arise. In such cases, the predicted line number falls within the valid range but}refers to content that does not exist at that position. We refer to such inconsistencies as mismatches. \newdeleted{Repeated outputs or duplicates also occur, which provide no additional useful information. The mismatch rate---observed only at the line level---varied between 8\% and 22\%, with \(k=\)1 yielding the lowest rate and \(k=\)10 the highest.}To resolve mismatches in line-level FL and evaluate their impact on performance, we \newadded{applied a mitigation strategy leveraging}the \emph{Levenshtein distance}\cite{levenstein1966} to measure content similarity and combined it with line number distance to identify the code line closest to the LLM-predicted output in both position and content. When no mismatch occurred, the identified line number matched the LLM's prediction exactly (distance = 0).
Otherwise, we replaced the mismatched line with the closest match. \newadded{This strategy is well-suited to TCFL, as it accounts for both test code proximity and content similarity.} 

\newdeleted{In a post-processing step, we evaluated FL performance using the updated outputs.}\newannotate{C1.3}\newadded{\autoref{table::mismatch} reports the ratio of mismatches to the total number of predicted lines and performance results of top-\(k\) line-level FL for different values of \(k\), with the mitigation strategy enabled and disabled. As shown, the accuracy metrics differ by less than 0.4\% between the two cases, with an average mismatch rate ranging from 7.9\% for \(k\)=1 to 21.5\% for \(k\)=10. Even the highest mismatch rate (21.5\%) has little to no effect on line-level FL performance, including precision and recall.}\newdeleted{Interestingly, no significant change was observed in the evaluation metrics, including precision and recall.}Further analysis showed that \newadded{the difference is minimal because}mismatches in our outputs typically involve line numbers differing by only one, with both lines usually sharing the same faulty or non-faulty status. Thus, \newdeleted{the updated lines}\newadded{updating mismatches based on our mitigation strategy}rarely affected the accuracy metrics.

\newadded{Although our mitigation strategy effectively handles mismatches and their impact on FL performance is minimal, such mismatches occur only at the line level and are generally easy to identify, as both line numbers and content are available in the LLM outputs and the input test code. This is particularly beneficial in sensitive scenarios where even low error rates are costly.}

Finally, the use of LLMs in code-related tasks can lead to other forms of semantic hallucination. However, these hallucinations are more commonly observed in code generation and automated program repair tasks, where the generated code or patch does not align with the intended requirements or fails to address the underlying bug, even though the code may be syntactically correct and successfully pass execution.

\subsubsection{Executed Trace Estimation}\label{trace::discussion}
While we use precise data to uniquely match test code statements to log messages, our trace estimation algorithms have limitations. For example, the fill-in-the-gaps algorithm assumes the code contains no branches. Although branchless code is often used to optimize execution\cite{DBLP:conf/pldi/MuellerW95} and reduce CPU branch mispredictions\cite{DBLP:conf/uemcom/HealyGE23}, most code, including that in our dataset, contains branches. In such cases, fill-in-the-gaps provides a trade-off between accuracy and efficiency. This was particularly evident when statements with a doubtful execution state were treated as non-executed. Although this approach reduced accuracy, it resulted in faster FL.

Pruning test code using CFGs accounts for branches and yields higher accuracy than the above fill-in-the-gaps algorithm. However, it also has limitations due to its node-level granularity. It does not consider the execution state of individual lines within a node, which can lead to overestimation. For example, if a node contains an error line that terminates execution, any subsequent lines in that node should be marked as non-executed, which is not captured by the CFG-based method. To address this, we generated a new trace by intersecting the executed lines from both the fill-in-the-gaps and CFG-based traces. The improved accuracy of these intersected traces highlights the benefit of combining CFG-based estimation with the line-level precision of fill-in-the-gaps. \annotate{Part of C2.1 \& C1.4}\addedtext{Finally, our test-to-log matching algorithm and, subsequently, our trace estimation algorithms depend on the availability of \newadded{unambiguous}static log statements by construction. \newdeleted{Although this limits our approach to datasets that include printed logs, it}\newannotate{C1.2}\newadded{However, this does not compromise the robustness of our trace estimation algorithms. First, as demonstrated in our ablation study in \autoref{trace::ablation}, even without such log statements, our approach maintains high accuracy by conservatively treating all statements as executed, thereby avoiding the loss of execution information. Second, the dependence on unambiguous static log statements} is consistent with standard software engineering practices. Logging is an integral and inseparable part of software systems, providing essential observability\newadded{, diagnostic information,}and debugging signals. Because \newadded{unambiguous}static log statements are maintained to support these purposes, their presence is highly likely in real-world systems, whereas fully dynamic \newadded{ambiguous}logging would undermine effective debugging. Consequently, many downstream analysis tasks inherently depend on the stability and availability of such static logs~\cite{app14188421, 9778893, 10.1145/506378.506421}.}

\subsection{Threats to Validity}\label{threats}
Several factors may influence the results of our study. Key variables include the size, complexity, and nature of the faulty test cases, which can impact the effectiveness of our techniques. To alleviate this threat, we evaluated our approach on an industrial dataset comprising 785 test cases with an average length of 244 lines of code, \newannotate{C2.4}\newadded{a test codebase comprising 5,159,790 lines,}\annotate{Part of C2.1 \& C3.2}\addedtext{and a significant number of interactions with the SUT}. These test cases are actively used by our industrial partner for product validation and represent realistic scenarios, supporting the practical relevance of our findings.

Fault types also directly impact the credibility of our results. We focus on localizing real faults rather than artificial ones. Using a diff-based approach to identify and annotate faulty locations may inadvertently include refactoring changes as faults, expanding the fault set. To address this, we applied a 3-\(\sigma\) rule\cite{pukelsheim1994} to detect outliers and refined fault labels for a smaller subset using domain knowledge. Although this manual inspection helps, it is not feasible for all test cases, which means some false positives may remain in our ground truth.

While our trace estimation and FL techniques are designed to be generally applicable across programming languages, the dataset used for our evaluation consists solely of Python test cases. This limitation may affect the generalizability of our results to other languages and domains. Future work should investigate the applicability of our methods on more diverse datasets to further validate their robustness.

\subsection{Future Work}\label{future}
For future work, researchers can enhance our estimation algorithms by leveraging more information from call sites to mitigate overestimation. For instance, if the first statement in a function is non-executed and the function has call sites, those call sites can also be marked as non-executed to support further pruning. This can, for instance, help eliminate branches that lack log-based lines but contain such call sites.
Conversely, if a function contains an executed log statement and has only one call site, we can infer that the call site was executed, allowing us to confidently determine the execution state of more statements. \annotate{C2.2}\addedtext{In this work, we demonstrated that estimated traces improve the efficiency of FL. Further exploration of such enhancements could help identify approaches for estimating traces that not only boost FL efficiency but also enhance its accuracy.}

Another potential direction is to incorporate data flow graphs (DFGs)\cite{DBLP:journals/cacm/AllenC76} to capture data dependencies through techniques such as program slicing\cite{DBLP:journals/ac/BinkleyG96}, and integrate this information into our trace estimation algorithms. Additionally, combining DFGs with blocks in FL can offer a richer context, leading to a better understanding of the fault and more effective repair.

A two-phase FL approach is another avenue for research. For instance, one could compare the effectiveness of first identifying faulty functions and then asking the LLM to localize faulty blocks within them, versus directly localizing faulty blocks in a single phase.

Another promising direction is to investigate how the syntactic and semantic validity of pruned test code influences LLM performance. At present, we do not assess code validity after pruning. Evaluating its effect on LLM-based FL could offer important insights. Additionally, examining various prompt templates, including the impact of different component orderings, may further enhance our understanding of how to optimize LLM performance.

Additionally, one can explore applications of our estimated traces beyond FL, including tasks such as test code repair\cite{yaraghi2025automated} or vulnerability identification\cite{DBLP:conf/kbse/ShafiuzzamanDSB24}. Considering the execution \emph{order} of functions or statements is another potential direction that can help further reduce the LLM's search space, particularly by addressing ambiguities associated with multi-matched statements. 
Finally, examining whether the results of trace estimation accuracy and FL experiments generalize to other datasets can further validate the applicability and practicality of our approach.

\section{Related Work}\label{Related Work}
In this section, we discuss state-of-the-art research related to our work from two perspectives: fault localization (FL) and source-to-log matching for estimating execution traces.

\subsection{Fault Localization}
To the best of our knowledge, our work is the first study of system-level black-box TCFL within an industrial context. Thus, we cannot make a direct, like-for-like comparison between our work and the state of the art. However, we review some SUTFL approaches in this section, highlighting the innovations of our technique and differentiating it from existing FL methods, while acknowledging that SUTFL and TCFL have distinct scopes.

Existing SUTFL approaches can be broadly classified into several categories: Spectrum-Based Fault Localization (SBFL)~\cite{DBLP:journals/ieicetd/ZhengHCYFX24, DBLP:journals/jss/RaselimoF24, DBLP:journals/access/SarhanB22,de2016spectrum,zakari2020spectrum}, Machine Learning and Deep Learning-Based FL~\cite{DBLP:journals/corr/abs-2411-17101, DBLP:conf/ieem/FangH22, DBLP:journals/ieicet/ZhangLTMZC17,DBLP:conf/sigsoft/LouZDLSHZZ21,DBLP:conf/issta/LiLZZ19,DBLP:conf/icse/Li0N21a,DBLP:conf/icse/MengW00022}, and FL using LLMs~\cite{DBLP:conf/icse/YangGMH24,DBLP:journals/corr/abs-2403-16362,10.1109/TSE.2025.3553363,jambigi2025fault}.

SBFL techniques~\cite{DBLP:journals/ieicetd/ZhengHCYFX24, DBLP:journals/jss/RaselimoF24, DBLP:journals/access/SarhanB22,de2016spectrum,zakari2020spectrum} use execution spectra to identify suspicious code regions by statistically analyzing coverage information, prioritizing program elements that are executed more often by failing tests and less often by passing tests.
Ochiai~\cite{DBLP:conf/icse/PearsonCJFAEPK17} and DStar~\cite{DBLP:journals/tr/WongDGL14} are among the widely used SBFL formulas.
They assign a suspiciousness score to each program element, with elements deemed more likely to be faulty receiving higher scores.
While SBFL may be effective, its performance is highly dependent on the quality of the test suite and code coverage information, which can be difficult to collect in large-scale systems due to the high computational cost~\cite{fatima2022flakify}.

With the advances in machine learning (ML) and deep learning (DL), data-driven FL techniques have emerged~\cite{DBLP:journals/corr/abs-2411-17101, DBLP:conf/ieem/FangH22, DBLP:journals/ieicet/ZhangLTMZC17,DBLP:conf/sigsoft/LouZDLSHZZ21,DBLP:conf/issta/LiLZZ19,DBLP:conf/icse/Li0N21a,DBLP:conf/icse/MengW00022}. These approaches leverage features such as code complexity, code history, and test coverage to predict faulty code locations. To this end, DeepFL~\cite{DBLP:conf/issta/LiLZZ19} combines SBFL, mutation-based FL, and code complexity metrics using neural networks~\cite{DBLP:books/el/17/Suk17}. Similarly, GRACE~\cite{DBLP:conf/sigsoft/LouZDLSHZZ21} integrates coverage information with graph neural networks (GNNs) to improve FL accuracy.  
FLUCCS~\cite{DBLP:conf/issta/SohnY17} enhances SBFL accuracy by incorporating code change metrics, such as the frequency and history of code modifications and DeepRL4FL~\cite{DBLP:conf/icse/Li0N21a} employs reinforcement learning (RL)~\cite{DBLP:journals/tse/ChakrabortyAN24} to analyze code coverage information, enhancing FL.
While models can offer significant benefits, techniques using coverage information often demand extensive preprocessing and are computationally costly~\cite{DBLP:conf/sigsoft/IvankovicPJF19}. Similarly, employing RL or other resource-intensive ML techniques can be costly. As a result, both categories of FL techniques encounter scalability challenges.

The emergence of LLMs has provided new opportunities for FL. LLMs like ChatGPT~\cite{liu2024gpt} and Codex~\cite{DBLP:journals/corr/abs-2107-03374} have shown promise in code comprehension and bug detection~\cite{DBLP:conf/icse-apr/PrennerBR22, DBLP:journals/corr/abs-2403-14734}. AGENTFL~\cite{DBLP:journals/corr/abs-2403-16362} employs a multi-agent system based on ChatGPT to localize faults at the project level. LLMAO~\cite{DBLP:conf/icse/YangGMH24} fine-tunes LLMs with adapters to predict buggy lines without requiring test coverage information.

Our work distinguishes itself from existing FL techniques in several key aspects, addressing unique challenges and advancing the state of the art in FL. Unlike most existing FL techniques, such as AGENTFL~\cite{DBLP:journals/corr/abs-2403-16362} and LLMAO~\cite{DBLP:conf/icse/YangGMH24}, which focus on localizing faults in the SUT (SUTFL), our approach targets faults in the system-level test code. SUTFL typically relies on the SUT's source code, execution logs, and code coverage information, which are unavailable in our black-box context.

Further, our approach eliminates the need for coverage information by estimating execution traces through inferring test behavior patterns directly from the test code. This makes our method more applicable in scenarios where coverage data is unavailable or impractical to collect~\cite{DBLP:conf/icse/PanGB23}.
Moreover, existing FL techniques are often evaluated on widely used benchmarks like Defects4J ~\cite{DBLP:conf/icse/PearsonCJFAEPK17} and BugsInPy~\cite{widyasari2022real}, which primarily contain SUT faults. These datasets may not reflect the complexity and diversity of industrial test code faults.
Our approach is evaluated on a unique industrial Python dataset comprising thousands of lines of test code from an actual product line. Since LLMs are not exposed to this dataset during pre-training, it further strengthens the credibility of our results. 
Additionally, unlike techniques such as AutoFL~\cite{DBLP:journals/pacmse/KangAY24} and AgentFL~\cite{DBLP:journals/corr/abs-2403-16362}, our approach utilizes open-source LLMs, making it freely accessible, while ensuring data privacy. This is particularly crucial for industrial applications, where proprietary models may be impractical due to licensing constraints.
Lastly, techniques such as LLMAO~\cite{DBLP:conf/icse/YangGMH24} fine-tune LLMs on specific datasets. Fine-tuning is resource-intensive, especially compared to prompt engineering for large-scale datasets like ours. Investigating the trade-offs between the costs and benefits of fine-tuning is an interesting avenue for future work.

\subsection{Source-to-Log Matching and Execution Trace Estimation}
We estimate the execution traces of faulty test cases using information obtained by matching log statements in the test source code to log messages in the execution log. In this section, we discuss related state-of-the-art work and highlight how our approach differs. 

Schipper et al.~\cite{DBLP:conf/msr/SchipperAD19} propose a technique that parses the source code into an AST to locate statements corresponding to log messages. Like our method, they treat each log message as having constant (referred to as static in our approach) and dynamic parts, and generate its regex accordingly. Additionally, we extend our regexes to automatically handle domain-specific log statements alongside standard logging constructs. Their method also leverages additional information about each object by utilizing overridden methods such as \texttt{toString()} in Java. However, such information is not commonly used in test code due to differences in code architecture, where fewer objects override \texttt{toString()} or similar methods. 
Xu et al.~\cite{xu2009detecting} parse console logs by combining source code analysis with information retrieval to construct features for models capable of detecting operational problems. They also leverage information provided by methods such as \texttt{toString()}. Shang~\cite{shang2012bridging} defines regexes for larger source code units, such as classes and methods, rather than for individual source lines. Iprof~\cite{10.5555/2685048.2685099}, a non-intrusive request flow profiler for distributed systems, performs matching using more complex relationships based on data flow analysis to distinguish individual requests. In contrast, our approach focuses on a different domain and relies on simpler relationships, such as the static and dynamic components of log statements in test code. Exploring the use of data flow analysis for matching and trace estimation in the context of test code presents an interesting future direction, as discussed in~\autoref{future}.

We further leverage information obtained from source-to-log matching to estimate the test execution trace and reduce the FL search space. A well-known state-of-the-art technique for this purpose is program slicing~\cite{DBLP:journals/ac/BinkleyG96}. Typically, program slicing starts at an endpoint, such as an error line, and performs a backward traversal using data and control flow analysis to identify the statements that led to that point. Many techniques leverage program slicing to some degree, along with other techniques such as SAT solvers~\cite{yuan2010sherlog}, stack traces~\cite{jiang2010debugging, sinha2009fault, zhang2011improved}, and call graphs~\cite{sui2017use} to estimate execution traces.
While program slicing is effective for tracing information back to its origin in the source code of the SUT, it can become complex and less informative for system-level test code. This complexity arises because system-level test code involves multiple layers of logic and interactions with the SUT, and it typically lacks direct access to the internal origins of data within the SUT. To address this, our trace estimation algorithms rely solely on the information available in the test code.

\section{Conclusions}\label{Conclusion}
In this paper, we introduced a novel black-box technique for system-level test code fault localization (TCFL) that leverages Large Language Models (LLMs) in a fully static manner, avoiding the need for repeated test executions. Our approach estimates the execution trace of a failing test case from a single failure log and prunes irrelevant statements to guide the LLM in identifying likely faulty locations. We presented three estimation algorithms and a masking-based evaluation strategy to assess trace accuracy.
Through a comprehensive evaluation on an industrial dataset, we demonstrated that our estimated traces achieve high fidelity, enabling effective and efficient fault localization (FL) across different levels of code granularity. Our findings highlight that coarse-grained function-level prompting yields high precision, while block-level TCFL offers a favorable balance between context and granularity.

Our TCFL technique is highly beneficial across different parameter settings, particularly for identifying faults in system-level test code when test engineers lack access to the system's source code. This is a common scenario in industrial development environments, where large numbers of complex faulty test cases are often encountered.
This work opens a new direction for practical, execution-free FL in complex system-level test code. It sets the stage for future enhancements in trace estimation and static LLM-guided debugging.

\section*{Data Availability}\label{Availability}
The data used in this study were provided by our industrial partner under a non-disclosure agreement and cannot be made publicly available due to confidentiality restrictions.

\section*{Acknowledgement}
This work was supported by a research grant from Huawei Technologies Canada Co., Ltd., as well as the Canada Research Chair and Discovery Grant programs of the Natural Sciences and Engineering Research Council of Canada (NSERC). Lionel C. Briand's contribution was partially funded by the Research Ireland grant 13/RC/209.
\newpage

\bibliographystyle{IEEEtran} 
\bibliography{references}

\end{document}